\documentclass[twoside,twocolumn,9pt]{article}
\usepackage{extsizes}
\usepackage[super,sort&compress,comma]{natbib} 
\usepackage[version=3]{mhchem}
\usepackage[left=1.5cm, right=1.5cm, top=1.785cm, bottom=2.0cm]{geometry}
\usepackage{balance}
\usepackage{mathptmx}
\usepackage{sectsty}
\usepackage{graphicx} 
\usepackage{lastpage}
\usepackage[format=plain,justification=justified,singlelinecheck=false,font={stretch=1.125,small,sf},labelfont=bf,labelsep=space]{caption}
\usepackage{float}
\usepackage{fancyhdr}
\usepackage{fnpos}
\usepackage[english]{babel}
\addto{\captionsenglish}{%
  
}
\usepackage{array}
\usepackage{droidsans}
\usepackage{charter}
\usepackage[T1]{fontenc}
\usepackage[usenames,dvipsnames]{xcolor}
\usepackage{setspace}
\usepackage[compact]{titlesec}
\usepackage{hyperref}

\usepackage{chemformula} 
\usepackage{rotating} 
\usepackage{amsmath} 
\usepackage{multirow} 
\usepackage{amssymb} 
\usepackage{numprint} 
\usepackage{enumitem} 

\usepackage{epstopdf}

\definecolor{cream}{RGB}{222,217,201}

\begin{document}

\pagestyle{fancy}
\thispagestyle{plain}
\fancypagestyle{plain}{
\renewcommand{\headrulewidth}{0pt}
}

\makeFNbottom
\makeatletter
\renewcommand\LARGE{\@setfontsize\LARGE{15pt}{17}}
\renewcommand\Large{\@setfontsize\Large{12pt}{14}}
\renewcommand\large{\@setfontsize\large{10pt}{12}}
\renewcommand\footnotesize{\@setfontsize\footnotesize{7pt}{10}}
\makeatother

\renewcommand{\thefootnote}{\fnsymbol{footnote}}
\renewcommand\footnoterule{\vspace*{1pt}%
\color{cream}\hrule width 3.5in height 0.4pt \color{black}\vspace*{5pt}} 
\setcounter{secnumdepth}{5}

\makeatletter 
\renewcommand\@biblabel[1]{#1}            
\renewcommand\@makefntext[1]%
{\noindent\makebox[0pt][r]{\@thefnmark\,}#1}
\makeatother 
\renewcommand{\figurename}{\small{Fig.}~}
\sectionfont{\sffamily\Large}
\subsectionfont{\normalsize}
\subsubsectionfont{\bf}
\setstretch{1.125} 
\setlength{\skip\footins}{0.8cm}
\setlength{\footnotesep}{0.25cm}
\setlength{\jot}{10pt}
\titlespacing*{\section}{0pt}{4pt}{4pt}
\titlespacing*{\subsection}{0pt}{15pt}{1pt}

\fancyfoot{}
\fancyfoot[LO,RE]{\vspace{-7.1pt}\includegraphics[height=9pt]{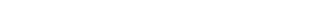}}
\fancyfoot[CO]{\vspace{-7.1pt}\hspace{13.2cm}\includegraphics{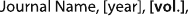}}
\fancyfoot[CE]{\vspace{-7.2pt}\hspace{-14.2cm}\includegraphics{head_foot/RF}}
\fancyfoot[RO]{\footnotesize{\sffamily{1--\pageref{LastPage} ~\textbar  \hspace{2pt}\thepage}}}
\fancyfoot[LE]{\footnotesize{\sffamily{\thepage~\textbar\hspace{3.45cm} 1--\pageref{LastPage}}}}
\fancyhead{}
\renewcommand{\headrulewidth}{0pt} 
\renewcommand{\footrulewidth}{0pt}
\setlength{\arrayrulewidth}{1pt}
\setlength{\columnsep}{6.5mm}
\setlength\bibsep{1pt}

\makeatletter 
\newlength{\figrulesep} 
\setlength{\figrulesep}{0.5\textfloatsep} 

\newcommand{\topfigrule}{\vspace*{-1pt}%
\noindent{\color{cream}\rule[-\figrulesep]{\columnwidth}{1.5pt}} }

\newcommand{\botfigrule}{\vspace*{-2pt}%
\noindent{\color{cream}\rule[\figrulesep]{\columnwidth}{1.5pt}} }

\newcommand{\dblfigrule}{\vspace*{-1pt}%
\noindent{\color{cream}\rule[-\figrulesep]{\textwidth}{1.5pt}} }

\makeatother

\twocolumn[
  \begin{@twocolumnfalse}
{\includegraphics[height=30pt]{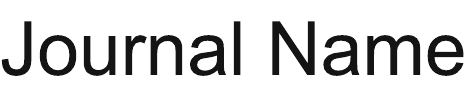}\hfill\raisebox{0pt}[0pt][0pt]{\includegraphics[height=55pt]{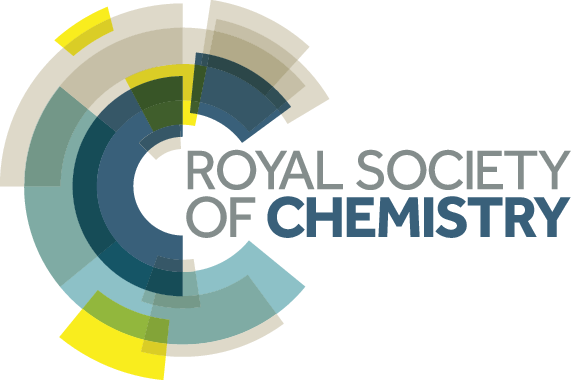}}\\[1ex]
\includegraphics[width=18.5cm]{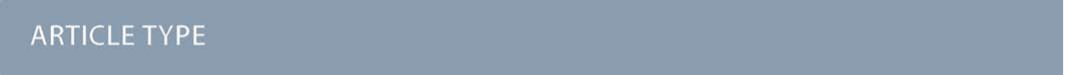}}\par
\vspace{1em}
\sffamily
\begin{tabular}{m{4.5cm} p{13.5cm} }

\includegraphics{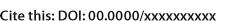} & \noindent\LARGE{\textbf{Towards Standardized Grid Emission Factors: Methodological Insights and Best Practices$^\dag$}} \\
\vspace{0.3cm} & \vspace{0.3cm} \\

 & \noindent\large{Malte Schäfer,$^{\ast}$\textit{$^{a}$} Felipe Cerdas,\textit{$^{a}$} and Christoph Herrmann\textit{$^{a}$}} \\

\includegraphics{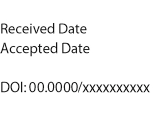} & \noindent\normalsize{Inconsistent calculation of grid emission factors (EF) can result in widely divergent corporate greenhouse gas (GHG) emissions reports. We dissect this issue through a comprehensive literature review, identifying nine key aspects—each with two to six methodological choices—that substantially influence the reported EF. These choices lead to relative effect variations ranging from 1.2\% to 69\%. Using Germany's 2019-2022 data as a case study, our method yields results that largely align with prior studies, yet reveal relative effects from 0.4\% to 34.6\%. This study is the first to methodically unpack the key determinants of grid EF, quantify their impacts, and offer clear guidelines for their application in corporate GHG accounting. Our findings hold implications for practitioners, data publishers, researchers, and guideline-making organizations. By openly sharing our data and calculations, we invite replication, scrutiny, and further research.} \\

\end{tabular}

 \end{@twocolumnfalse} \vspace{0.6cm}

  ]

\renewcommand*\rmdefault{bch}\normalfont\upshape
\rmfamily
\section*{}
\vspace{-1cm}


\footnotetext{\textit{$^{a}$~Institute of Machine Tools and Production Technology (IWF), Technische Universität Braunschweig, 38106 Braunschweig, Germany. Tel: +49 (0)531 391-7650; E-mail: malte.schaefer@tu-braunschweig.de}}

\footnotetext{\dag~Electronic Supplementary Information (ESI) available online: document, code and data. Document: ADD DOI OF THIS ARTICLE, code: \url{https://doi.org/10.24355/dbbs.084-202309131139-0}, data: \url{https://doi.org/10.24355/dbbs.084-202309111514-0} }



\section{Introduction}
\label{sec:intro}

In the European Union, companies will be legally obliged to report on sustainability in the near future. According to the  \textit{Corporate Sustainability Reporting Directive (CSRD)}, all large and many small and medium-sized companies have to start doing so, beginning with the financial year 2024 \cite{eu2022directive}. Meanwhile, in California, the \textit{Climate Corporate Data Accountability Act} requires businesses with a revenue of USD 1 billion or more to disclose their greenhouse gas (GHG) emissions by the year 2026 \cite{california2023sb253}. In addition, sustainability reporting is not only important for meeting legal requirements, but it can also increase an organization's credibility towards its stakeholders and help legitimize its business operations towards society.

One of the key tasks in preparing a sustainability report is the calculation of the company's annual GHG emissions. Given the often substantial electricity usage of companies, understanding the emissions from this sector is crucial for the company and its stakeholders. Many organizations rely on the Greenhouse Gas Protocol for guidelines on GHG accounting \cite{WRI.2004}, and more specifically, its Scope 2 Guidelines for electricity-related emissions \cite{WRI.2015, Hickmann.2017}.

A vital part of these calculations involves emission factors (EF), which quantify the amount of emissions (e.g. \ch{CO2}) generated per unit of electricity consumed (e.g. kWh). For example, to assess the company's annual electricity-based GHG emissions, its total annual electricity consumption is multiplied with the EF.

\begin{figure}[h!]
\centering
    \includegraphics[keepaspectratio, width=\linewidth]{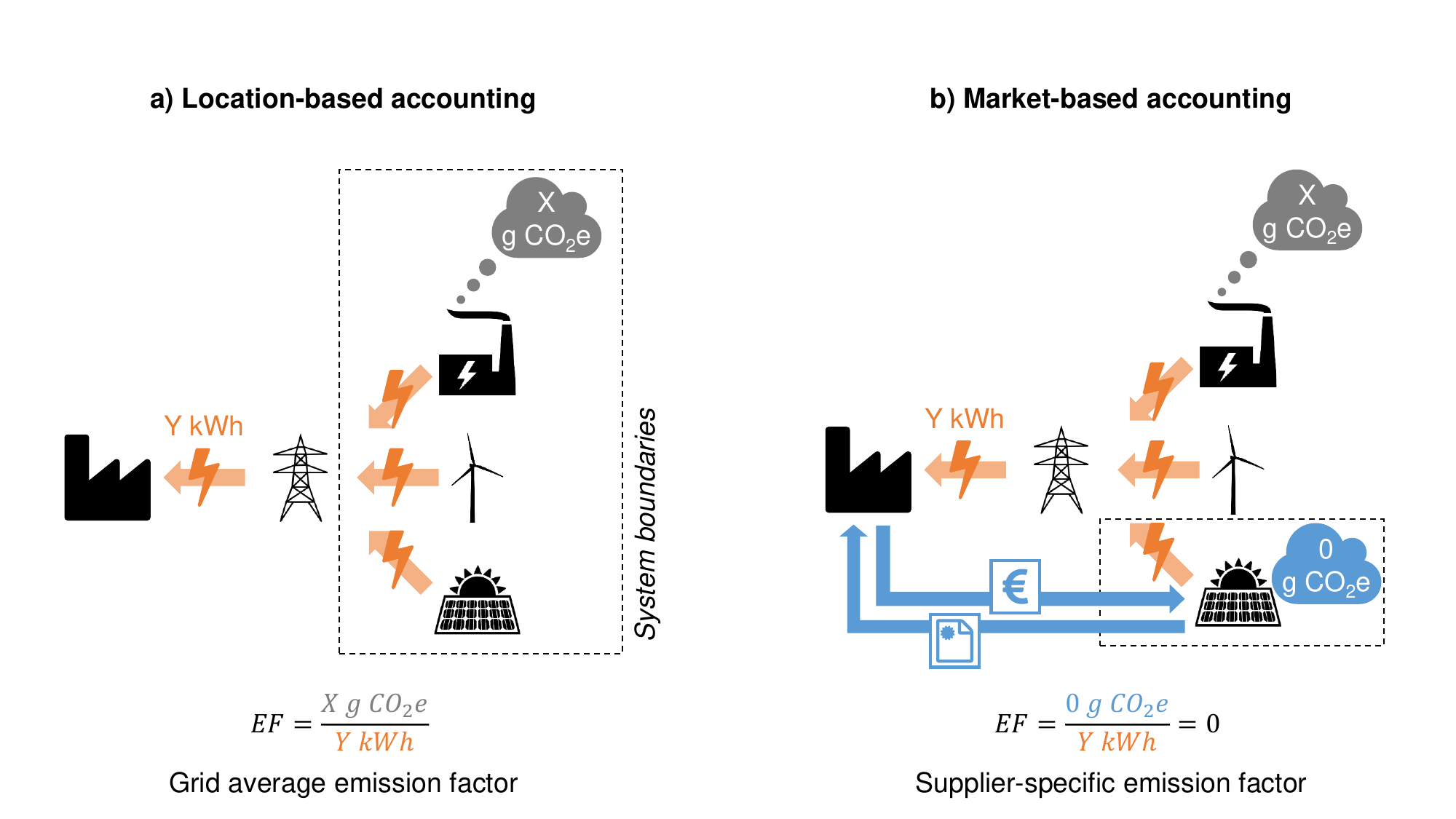}
    \caption{a) Location-based and b) market-based accounting, as described in the GHG Protocol Scope 2 Guidelines \cite{WRI.2015}. Location-based accounting relies on a grid-average EF (focus of this study), which reflects the emissions from all generators feeding into a grid. Market-based accounting relies on a supplier-specific EF, which reflects the emissions from the energy supplier that the electricity consumer has a contractual agreement with.}
    \label{fig:mb_lb}
\end{figure}

The EF value depends on the mix of primary energy sources used for electricity generation. If a company procures electricity through a specific supplier, then the EF should correspond to that source, known as the market-based approach (cf. Figure \ref{fig:mb_lb}b). The market-based approach may take into consideration instruments such as guarantees of origin (GOs), which allow consumers to claim electricity from a specific source. In addition to the market-based approach, a grid-average EF should also be calculated, termed the location-based approach \cite{WRI.2015} (cf. Figure \ref{fig:mb_lb}a). The location-based approach does not take into account GOs.

\begin{figure}[hbt]
    \includegraphics[keepaspectratio, width=\linewidth]{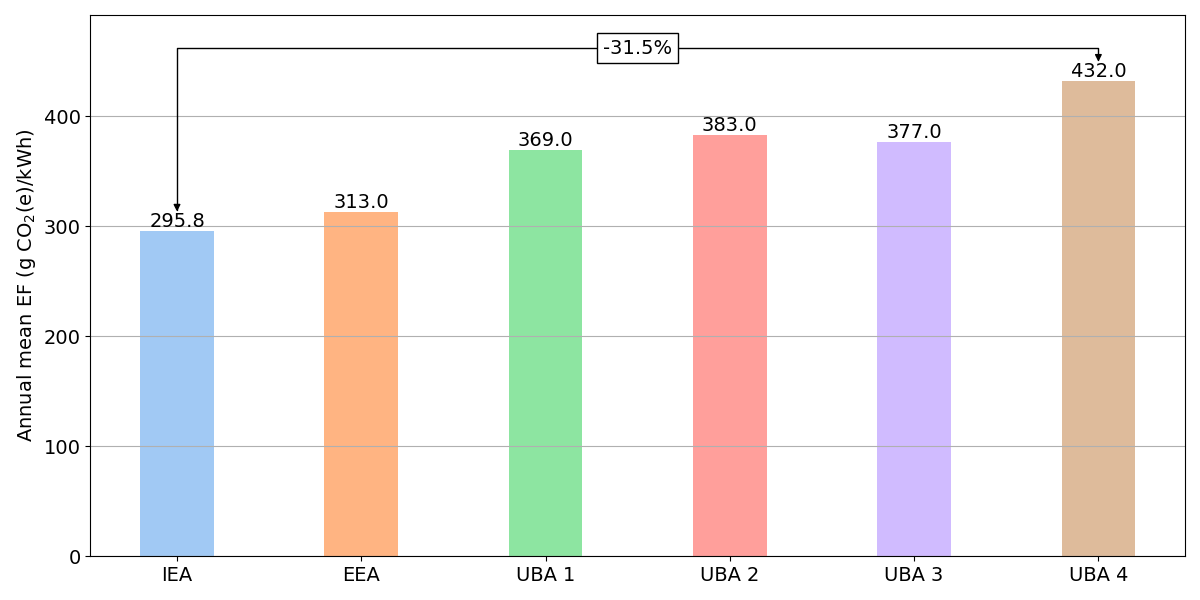}
    \caption{Annual mean grid emission factor for Germany in 2020, according to different sources (IEA \cite{IEA.2020, IEA.2021}, EEA \cite{EEA.2023} and UBA \cite{UBA.2023}). \textit{UBA 1}-\textit{UBA 4} represent four different approaches to calculating a grid EF, varying the aspects impact metric (\ch{CO2}, \ch{CO2}-equivalents), inclusion of electricity trade (with, w/o) and system boundaries (direct, life-cycle emissions). \textit{UBA 1}: \ch{CO2}, w/o trade, direct emissions; \textit{UBA 2}: \ch{CO2}, with trade, direct emissions; \textit{UBA 3}: \ch{CO2}e, w/o trade, direct emissions; \textit{UBA 4}: \ch{CO2}e, w/o trade, life-cycle emissions.}
    \label{fig:motivation}
\end{figure}

One of the challenges for determining a grid-average EF lies in selecting suitable data sources. To highlight this issue, Figure \ref{fig:motivation} presents the 2020 grid-average EF for Germany, as reported by diverse organizations such as the International Energy Agency (IEA) \cite{IEA.2020, IEA.2021}, the European Environmental Agency (EEA) \cite{EEA.2023}, and the German Federal Environmental Agency (UBA) \cite{UBA.2023}.

As illustrated in the figure, the disparity in reported grid EF values is significant, with the lowest being 31.5\% smaller than the highest. At least part of this divergence stems from variations in calculation methodologies. For instance, the UBA differentiates between an electricity production (w/o trade) and consumption (with trade) perspective, operational (direct/combustion) versus life-cycle (including upstream and downstream) emissions, and \ch{CO2} versus \ch{CO2}-equivalents (including multiple GHG instead of only \ch{CO2}). The result are UBA values ranging from 369 to 432 g CO$_2$(e)/kWh.

The GHG Protocol Scope 2 Guidance provides limited advice on these methodological aspects, suggesting only that electricity trade across borders should not be factored into the EF \cite{WRI.2015}. It falls short in offering guidance on other aspects or recommending specific data sources. Consequently, an organization aiming to report lower Scope 2 emissions could technically achieve a one-third reduction simply by choosing an EF from the IEA over one from the UBA---without altering its electricity supply or consumption.

Given this landscape, and the increasing importance of reliable data on grid emissions, there is a clear need to scrutinize how grid EFs are calculated. Thus, the question arises: What constitutes a methodology for calculating a grid EF that best represents the emissions caused by the electricity consumer, and should therefore be used in Scope 2 emission accounting? Understanding the methodological aspects and choices involved in determining grid EFs, their impact on the outcomes, and issuing recommendations related to these choices is crucial.

The need for scrutinization leads to three research questions (RQs) guiding this study, each aimed at dissecting the complexities of grid emission factors (EF):
\begin{itemize}
    \setlength\itemsep{0em}
    \item[\textbf{RQ1:}] Which methodological aspects impact the final grid-average electricity EF?
    \item[\textbf{RQ2:}] How significant is the effect of various choices within these aspects on the outcome?
    \item[\textbf{RQ3:}] Which methodological choices best represent the emissions from an organization's electricity consumption?
\end{itemize}

To address RQ1, we conduct a literature review of studies that calculate grid-average electricity EFs, focusing on key methodological aspects. This review also informs RQ2 as we compile insights from studies that quantify the influence of these methodological aspects. We supplement these findings with our own analysis, examining the impact of various choices within these aspects on Germany's grid EF for the years 2019-2022. Lastly, for RQ3, we offer recommendations on which choices best reflect the emissions of an organization drawing electricity from the grid.

The remaining paper is structured as follows: Section \ref{sec:literature} dives into the existing literature to identify and assess the methodological aspects and choices that affect the grid EF calculations. Section \ref{sec:methodology} outlines the methodology and data used for our own calculations, guided by insights from the literature review. Section \ref{sec:results} presents the results of the analysis. In Section \ref{sec:discussion}, we compare our results to prior studies and official grid EF data sources, and offer recommendations based on our findings. Finally, Section \ref{sec:conclusion} contains our conclusions.

\section{State of Research}
\label{sec:literature}
To address RQ1, we undertake a comprehensive literature review, aiming to pinpoint the methodological decisions that influence grid EF calculations. Section \ref{sec:literature_results} presents the results of the review, summarized in Section \ref{sec:literature_summary}. For more information on the scope and search process, the reader is referred to the ESI\dag.

\subsection{Key Methodological Aspects}
\label{sec:literature_results}
The review produced 48 primary research articles \cite{Jiusto.2006, Soimakallio.2012, Li.2013, Louis.2014, Maurice.2014, Messagie.2014, Stoll.2014, Tamayao.2015, Colett.2016, Ji.2016, Roux.2016, Kono.2017, KopsakangasSavolainen.2017, Mills.2017, Nilsson.2017, Qu.2017, Clauss.2018, Collinge.2018, Fiorini.2018, Khan.2018, Milovanoff.2018, Qu.2018, Vuarnoz.2018, Baumann.2019, Baumgartner.2019, BeloinSaintPierre.2019, Chalendar.2019, Clauss.2019, Donti.2019, MunneCollado.2019, Rupp.2019, Schram.2019, Schwabeneder.2019, Tranberg.2019, Walzberg.2019, Worner.2019, Braeuer.2020, Neirotti.2020, Noussan.2020, Papageorgiou.2020, Pereira.2020, Chalendar.2021, Fleschutz.2021, Mehlig.2021, Peters.2022, Scarlat.2022, Unnewehr.2022, Blizniukova.2023}. The nine aspects that most frequently appeared in these articles and were found to have an impact on the resulting grid EF are:

\begin{itemize}
    \setlength\itemsep{0em}
    \item Choice of \textbf{impact metric} (e.g. \ch{CO2} vs. multiple GHG)
    \item Choice of \textbf{system boundaries} (e.g. operational vs. life-cycle)
    \item Allocation for \textbf{co-generated heat} (e.g. by energy vs. by exergy)
    \item Treatment of \textbf{auto-producers} (e.g. inclusion vs. exclusion)
    \item Treatment of \textbf{auxiliary consumption} (incl./excl.)
    \item Treatment of \textbf{electricity trading} (incl./excl.)
    \item Treatment of \textbf{storage cycling} losses (incl./excl.)
    \item Treatment of \textbf{transformation \& distribution} (T\&D) losses (incl./excl.)
    \item Choice of \textbf{temporal resolution} (e.g. annual vs. hourly)
\end{itemize}

In addition to the aspects listed above, there are additional ones that are relevant. These include the spatial and technological resolution, both of which are not covered in this study. The primary reason for excluding these aspects is data availability. The rationale behind this decision is further discussed in the ESI\dag.

\subsection{Summary of the State of Research}
\label{sec:literature_summary}
None of the reviewed studies covers all nine methodological aspects, but each study addresses at least one. Notably, only five studies delve into the role of auto-producers (also referred to as \textit{self-generation} or \textit{distributed generation}), whereas 31 consider the impact of electricity trading on grid EF calculations. Table \ref{tbl:lit_aspects_effect} details the magnitude of each aspect's effect, specifically focusing on data from Germany.

\begin{table}[h]
\small
    \caption{Effect of key methodological aspects in primary research articles. The table displays the range of changes in grid emission factors when different aspects are considered, in both absolute and relative terms. All values pertain to the German grid (except for \textit{Temporal resolution}, where no German data is available).}
    \label{tbl:lit_aspects_effect}
    \begin{tabular*}{0.48\textwidth}{lll}
        \hline
        Aspect & Abs. effect (g/kWh) & Rel. effect (\%) \\
        \hline
        Impact metric & +9...+33 &  +1.9...+5.9 \\
        System boundaries & +14...+69 &  +2.2...+13.2 \\
        Co-generation of heat & +54...+60 & +9.9...11.4 \\
        Auto-producers & -- &  -- \\
        Auxiliary consumption & +20 & +5.1 \\
        Electricity trade & -22...+12 & -4.0...+2.9 \\
        Storage cycling & +5...+6 & +1.2...+1.3 \\
        Transformation \& distribution & -- & +3.9...+4.2 \\
        Temporal resolution* & -- & -28...+69 \\
        \hline
    \end{tabular*}   
    * Countries other than Germany
\end{table}

One can observe that changing the impact metric (e.g. from \ch{CO2} to one that includes multiple GHG) increases the EF by 9-33 g/kWh in absolute terms, which is equivalent to 1.9-5.9\% in relative terms. For auto-producers the effect has not been quantified before, while for T\&D losses it has only been quantified in relative terms. For the temporal resolution, the effect has only been quantified for countries other than Germany.

The literature review covered in this section addresses RQ1, and to some extent also RQ2: nine methodological aspects influencing the grid EF have been identified, and for most of them, the effect that these aspects have on the grid EF have been quantified. However, no study provides a comprehensive analysis using consistent assumptions and data across all aspects, which is the focus of the subsequent sections.

\section{Methodology and Data}
\label{sec:methodology}
\begin{figure}[hbt]
\centering
    \includegraphics[keepaspectratio, width=\linewidth]{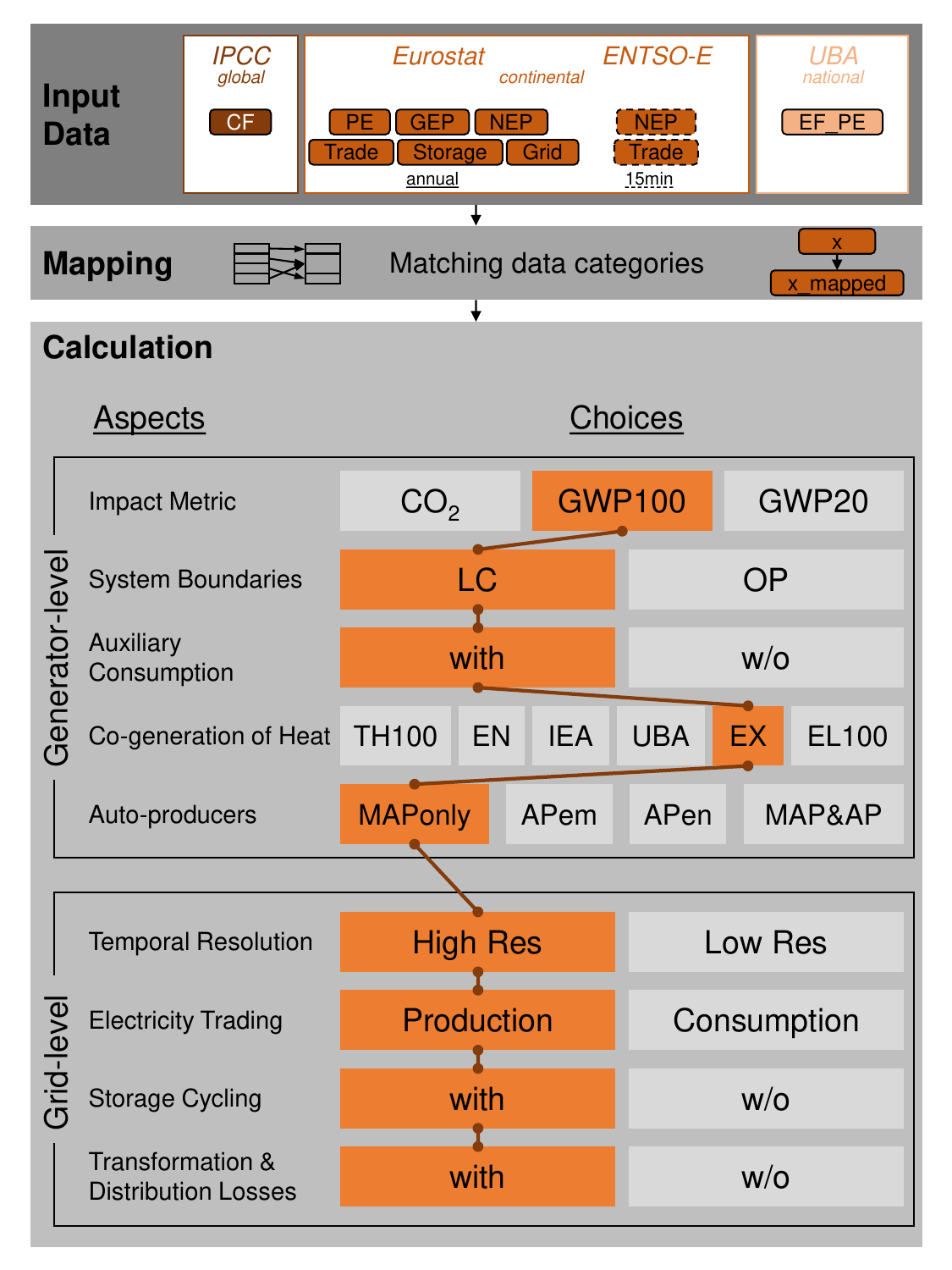}
    \caption{Grid EF calculation methodology, grouped into the layers \textit{Input Data}, \textit{Mapping} and \textit{Calculation}. The Calculation layer consists of two parts, comprising \textit{generator-level} and \textit{grid-level} calculations. The boxes highlighted in orange illustrate one example set of choices.}
    \label{fig:methodology_overview}
\end{figure}

In this section, we describe how we calculate grid EF, considering each of the aspects mentioned in Section \ref{sec:literature}. The methodology outlined here serves the purpose of calculating a grid EF at a temporal resolution of 15 minutes, while providing multiple choices for each of the methodological aspects reviewed in the previous section. An example of a methodological aspect is \textit{Impact metric}, and an exemplary choice with respect to that aspect is \textit{GWP100} (the global warming potential observed over a time period of 100 years). Figure \ref{fig:methodology_overview} depicts the calculation procedure. The link to the code and data used in this article can be found in the ESI\dag.

The four primary input data sources are the IPCC (characterization factors), Eurostat (low resolution energy data), ENTSO-E (high resolution energy data) and the UBA (primary energy referenced EF). The input data does not match in all cases with respect to the categories used to describe fuels/energy carriers (e.g. \textit{Fossil Gas} is used by ENTSO-E, \textit{Natural gas} by Eurostat). Thus, mapping is required to match the different types of categories. Finally, in multiple calculation steps, the input data is combined and transformed.

The first part of these calculations are conducted at the generator level, i.e. separate EF exist for individual production types (e.g. \textit{Hard coal}, \textit{Wind onshore}). The second part occurs at the grid level, where individual fuels/energy carriers cannot be not distinguished anymore. The following sections describe in more detail each of the three layers of the methodology depicted in Figure \ref{fig:methodology_overview}.

\subsection{Input Data}
\label{sec:methodology_input}
The input data layer encompasses all data necessary for calculating grid EFs. The selection criteria for choosing the input datasets are as follows:

\begin{itemize}
    \setlength\itemsep{0em}
    \item Comprehensive
    \item Relevant to the German context
    \item Available for/applicable to the years 2019-2022
    \item Consistent with all methodological aspects
\end{itemize}

The ESI\dag provides more information on each data source and any necessary pre-mapping adjustments.

\subsection{Mapping}
\label{sec:methodology_map}
The mapping layer aligns disparate data categories from the raw datasets. This harmonization is essential, given that the datasets originate from varied sources with inconsistent categorization. Without mapping, some production types may be over- or underrepresented, or in some cases not counted towards the grid EF at all. This would lead to distorted results.

\subsection{Calculation}
\label{sec:methodology_calc}
The calculation layer transforms the mapped input data into emission factors through a series of steps. Initial calculations are made at the generator level, producing individual EFs for each production type (fuel/energy carrier). As electricity flows into the grid, subsequent EF calculations are generalized to the grid level. Table \ref{tbl:method_calc_aspects} summarizes the methodological considerations incorporated into our calculations.

\begin{table}[htb]
\small
    \caption{Summary of methodological aspects and choices addressed in this study}
    \label{tbl:method_calc_aspects}
    \begin{tabular*}{0.48\textwidth}{ll}
        \hline
        Aspect & Choices \\
        \hline
        Impact metric &  \ch{CO2}, GWP100, GWP20 \\
        System boundaries & OP, LC \\
        Co-generation of heat & TH100, EN, IEA, UBA, EX, EL100 \\
        Auto-producers & MAPonly, APem, APen, MAP\&AP \\
        Auxiliary consumption & with, w/o \\
        Electricity trade & with, w/o \\
        Storage cycling & with, w/o \\
        Transformation \& distribution & with, w/o \\
        Temporal resolution & high (15 min), low (1 year) \\
        \hline
    \end{tabular*} 

    TH100: all emissions allocated to heat; EN: emissions allocated by energy;
    IEA: IEA allocation method;
    UBA: UBA allocation method;
    EX: allocation by exergy;
    EL100: all emissions allocated to electricity;
    MAPonly: emissions and electricity from main-activity producers only;
    APem: emissions from all generators (main-activity producers and auto-producers), electricity from main-activity producers only;
    APen: emissions from main-activity producers only, electricity from all generators;
    MAP\&AP: emissions and energy from all generators.
    
\end{table}

The choices outlined in Table \ref{tbl:method_calc_aspects} represent a broad spectrum found in the literature. For \textit{Co-generation of heat}, we introduce two new choices not previously found in the literature reviewed in this study. \textit{EX}, or allocation by exergy, is commonly used in CHP units \cite{Choi.2022, Mollenhauer.2016}, even though it was not featured in the literature review. \textit{TH100}, which allocates all emissions to heat, serves as a counterpoint to \textit{EL100}, which allocates all emissions to electricity. The ESI\dag contains a section breaking down the calculation steps from mapped input data to finalized grid EFs in detail.

\section{Results}
\label{sec:results}
This section presents the calculated grid emission factors (EF) for Germany for the years 2019-2022. After an overview of the whole dataset, two methodological aspects' influence on the grid EF are explored in detail.

\subsection{Overview}
\label{sec:results_over}
\begin{figure}[hbt]
\centering
    \includegraphics[keepaspectratio, width=\linewidth]{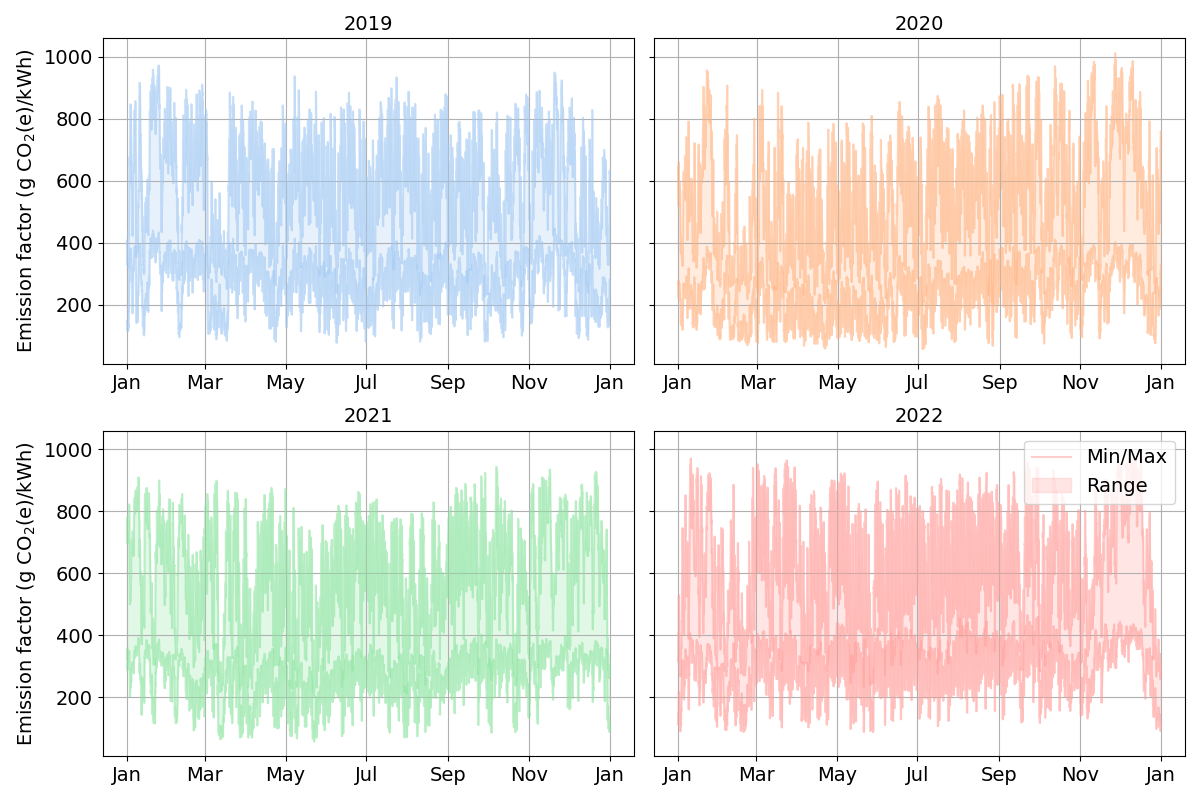}
    \caption{Temporal summary of 2304 unique grid EFs for Germany from 2019 to 2022. The plot captures the minimum, maximum and the range in between for each time point.}
    \label{fig:result_overview_line}
\end{figure}

The entire dataset comprises \numprint{323149824} data points. This number represents 2304 grid EF configurations, measured every 15 minutes for four years (equivalent to \numprint{140256} time steps). Figure \ref{fig:result_overview_line} plots the temporal evolution of these grid EFs.

\begin{figure}[hbt]
\centering
    \includegraphics[keepaspectratio, width=\linewidth]{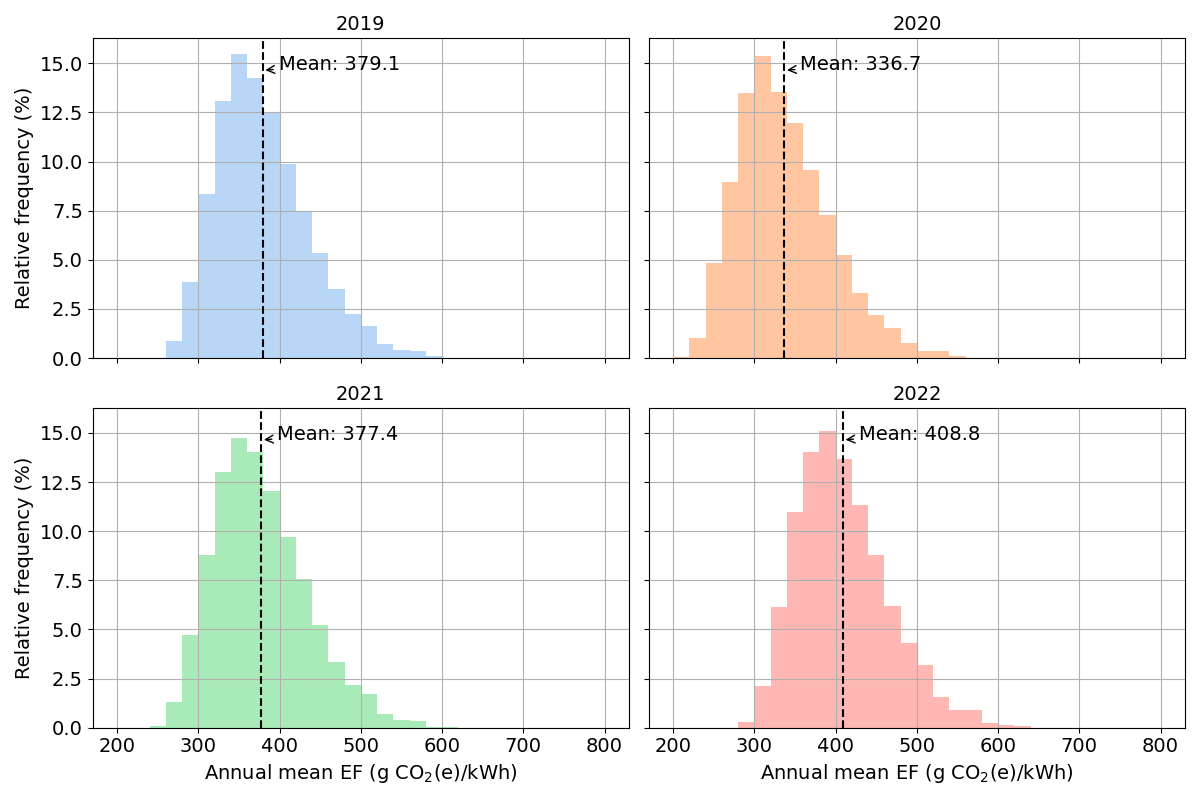}
    \caption{Frequency distribution of 2304 distinct annual mean grid EFs for Germany between 2019 and 2022. The relative frequency denotes the share of 2304 calculated EFs falling within a given bin, with a bin width of 20 g \ch{CO2}(e)/kWh.}
    \label{fig:result_overview_histogram}
\end{figure}

The figure differentiates by year, revealing noticeable temporal variability. Extreme values range from approximately 100 to nearly 1000 g \ch{CO2}(e)/kWh. However, it is difficult to perceive other temporal trends, e.g. how the EF has evolved over the years or how the different EF configurations are distributed around the mean. For an alternative view, Figure \ref{fig:result_overview_histogram} presents a histogram of the annual mean grid EFs for these configurations.

This histogram is based on the same data as Figure \ref{fig:result_overview_line}, but depicts the annual average instead of 15-minute values. The plot indicates the share of all 2304 grid EF configurations falling into a certain bin. For example, for 2020, most configurations (> 15\%) fall into the bin ranging from 300 to 320 g \ch{CO2}(e)/kWh. Additionally, one can observe that the mean of all configurations shifts over the years, reaching its lowest point in 2020 with 336.7 g \ch{CO2}(e)/kWh. The data further reveal that the smallest and largest annual mean grid EFs can differ by a factor of three, e.g. ranging from about 200 to 600 g \ch{CO2}(e)/kWh for the year 2020.

\subsection{Influence of Individual Methodological Aspects}
\label{sec:results_ind}

This part analyzes the sensitivity of the grid EF to two out of nine aspects: \textit{Impact metric} and \textit{Temporal resolution}. The remaining aspects are investigated in the ESI\dag.

\subsubsection{Impact Metric.}
\label{sec:results_ind_imp}

\begin{figure}[hbt]
\centering
    \includegraphics[keepaspectratio, width=\linewidth]{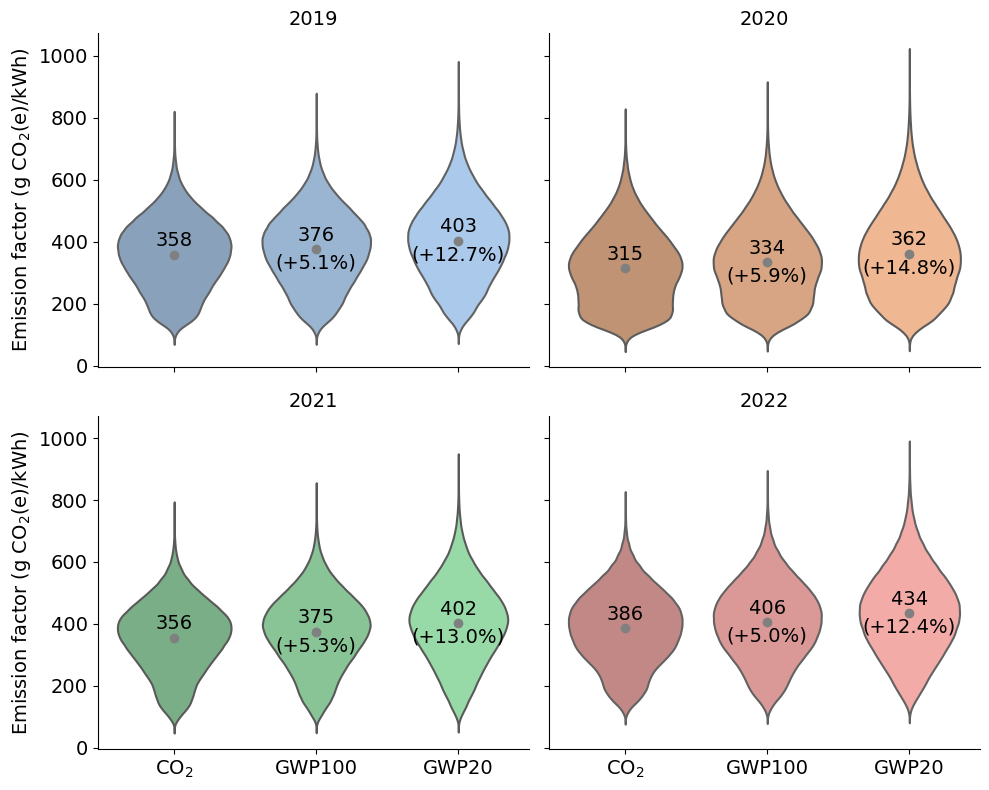}
    \caption{Grid EF variation due to impact metric, disaggregated by year. Three metrics are considered: \textit{\ch{CO2}}, \textit{GWP100}, and \textit{GWP20}. The data labels indicate the mean values for each metric and the relative differences compared to the reference metric (\textit{\ch{CO2}}).}
    \label{fig:result_effect_metric_multi}
\end{figure}

Figure \ref{fig:result_effect_metric_multi} illustrates the variation in grid EF attributable to different impact metrics: \textit{\ch{CO2}}, \textit{GWP100}, and \textit{GWP20}. The plot showcases the mean values associated with each choice, in addition to their relative difference when compared to a reference metric (here, \textit{\ch{CO2}}).

The analysis reveals that, when broken down by year, a \textit{GWP100}-based EF tends to be 5.0-5.9\% higher than a \textit{\ch{CO2}}-based EF. Similarly, a \textit{GWP20}-based EF exhibits an average increase of 12.4-14.8\% over a \textit{\ch{CO2}}-based EF. The trend across years is consistent with figures \ref{fig:result_overview_line} and \ref{fig:result_overview_histogram}: the mean values are lowest for the year 2020 and highest for the year 2022. The fact that \textit{GWP20} values are consistently higher than \textit{GWP100} values, which are again higher than \textit{\ch{CO2}} values, aligns with our expectations. \textit{GWP} covers multiple climate-change-relevant substances, while \textit{\ch{CO2}} represents only one. \textit{GWP20} has higher characterization factors for methane (\ch{CH4}) than \textit{GWP100}, which explains the difference between these two metrics.

\subsubsection{Temporal Resolution.}
\label{sec:results_ind_temp}

To investigate the effect of the temporal resolution on the resulting emissions, it is not sufficient to study only the grid EF. Additional data on an electricity consumer's grid electricity load profile is required to quantify how a change in the temporal resolution affects the consumer's electricity-related emissions. This section first describes the temporal trends that can be observed in Germany's grid EF, before applying the grid EF to a case study load profile.

\paragraph{Grid EF Temporal Trends.}

\begin{figure}[hbt]
\centering
    \includegraphics[keepaspectratio, width=\linewidth]{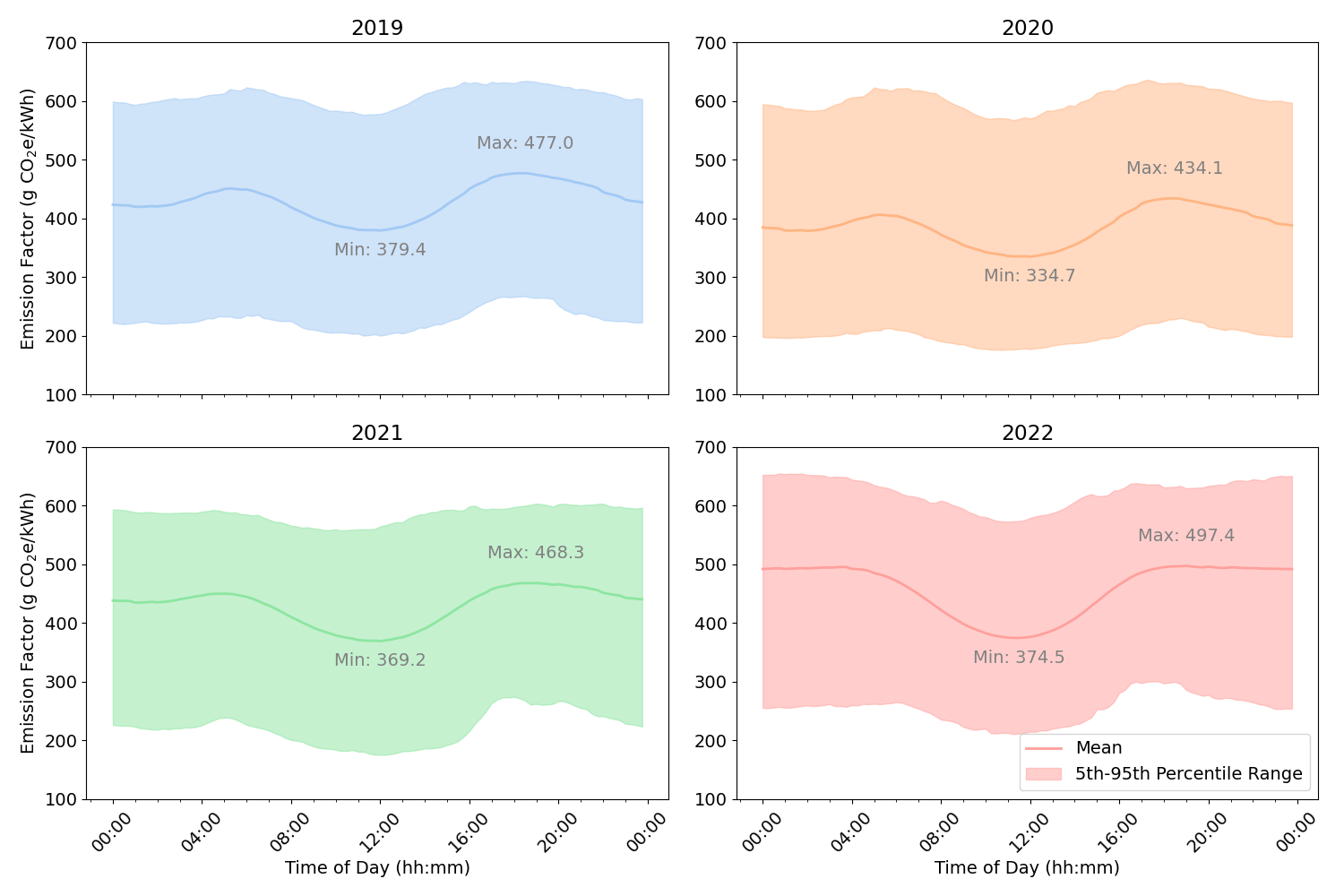}
    \caption{Grid EF by year and time of day. The line represents the mean value for specific time points (e.g. 12:00 h) and years (e.g. 2019). The shaded area delineates the range between the 5th and 95th percentiles of the data, highlighting the distribution's variability and indicating where 90\% of the values lie for the given time and year. The values `Min' and `Max' indicated the minimum and maximum value of the line representing the mean.}
    \label{fig:result_effect_temp_tod}
\end{figure}

Germany's grid EF exhibits some typical temporal patterns, depicted in figure \ref{fig:result_effect_temp_tod}. The plot illustrates how the grid EF varies between years and throughout a typical day. The grid EF configuration is the recommended configuration described in section \ref{sec:discussion_rec_calc}.

It is apparent that while the grid EF changes from year to year, reaching a low point in 2020, the pattern throughout a typical day remains relatively stable. The grid EF is typically highest in the morning and in the evening, and lowest at night and around midday. However, the `dip' at night becomes less pronounced and is barely detectable for the year 2022.

\begin{figure}[hbt]
\centering
    \includegraphics[keepaspectratio, width=\linewidth]{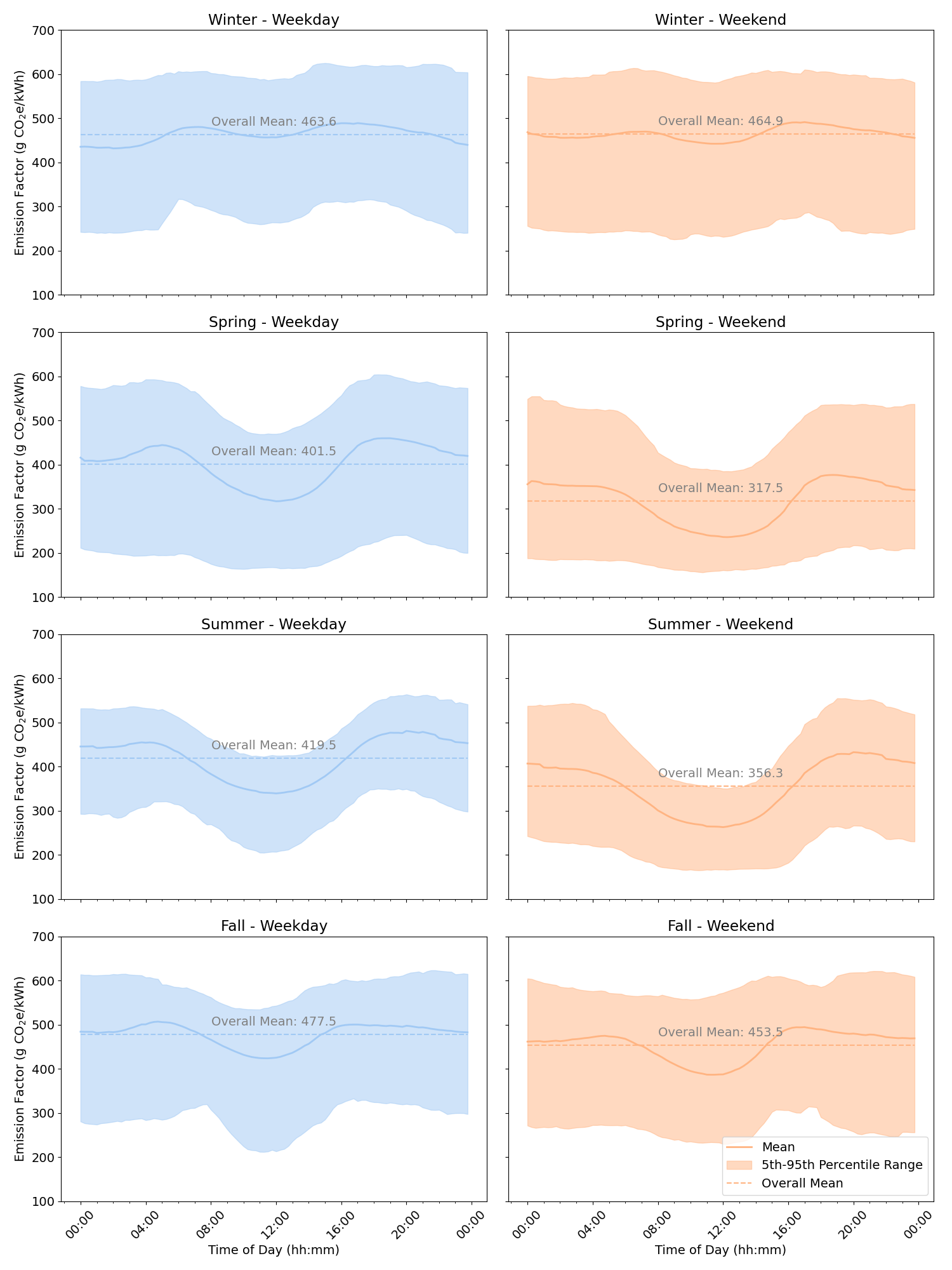}
    \caption{Grid EF by day type, season, and time of day for the year 2021. The solid line represents the mean value for specific time points (e.g. 12:00 h), day type (e.g. weekday), and season (e.g. Summer). The shaded area delineates the range between the 5th and 95th percentiles of the data, highlighting the distribution's variability and indicating where 90\% of the values lie for the given time and year. The dashed line represents the overall mean, i.e., the daily mean for a given day type and season.}
    \label{fig:result_effect_temp_day_season}
\end{figure}

Other temporal patterns besides inter-annual and intraday changed in the grid EF can be observed as well. Figure \ref{fig:result_effect_temp_day_season} illustrates how the grid EF varies throughout the day, distinguishing between weekdays and weekends, as well as between seasons.

The plot demonstrates how the overall mean grid EF tends to be lower on weekends than on weekdays, with the exception of the Winter season. The overall mean grid EF further tends to be lowest in the spring and highest in the Fall and in the Winter. The grid EF variation throughout the day is most pronounced in the Spring and in the Summer, and least pronounced in the Winter. Finally, the range between the 5th and the 95th percentile is notably narrower in the Summer than in the Winter.

A more detailed analysis of temporal trends, including possible explanations for the patterns described above, and a correlation analysis with overall generation, can be found in the ESI\dag.

\paragraph{Case Study.}

To investigate the effect of the temporal resolution on the emissions of an electricity consumer, we calculate emissions at two different temporal resolutions and compare the results. The two resolutions are one year and 15 minutes. The resolution applies to both the grid EF and the load profile of an exemplary consumer.

\begin{figure}[hbt]
\centering
    \includegraphics[keepaspectratio, width=\linewidth]{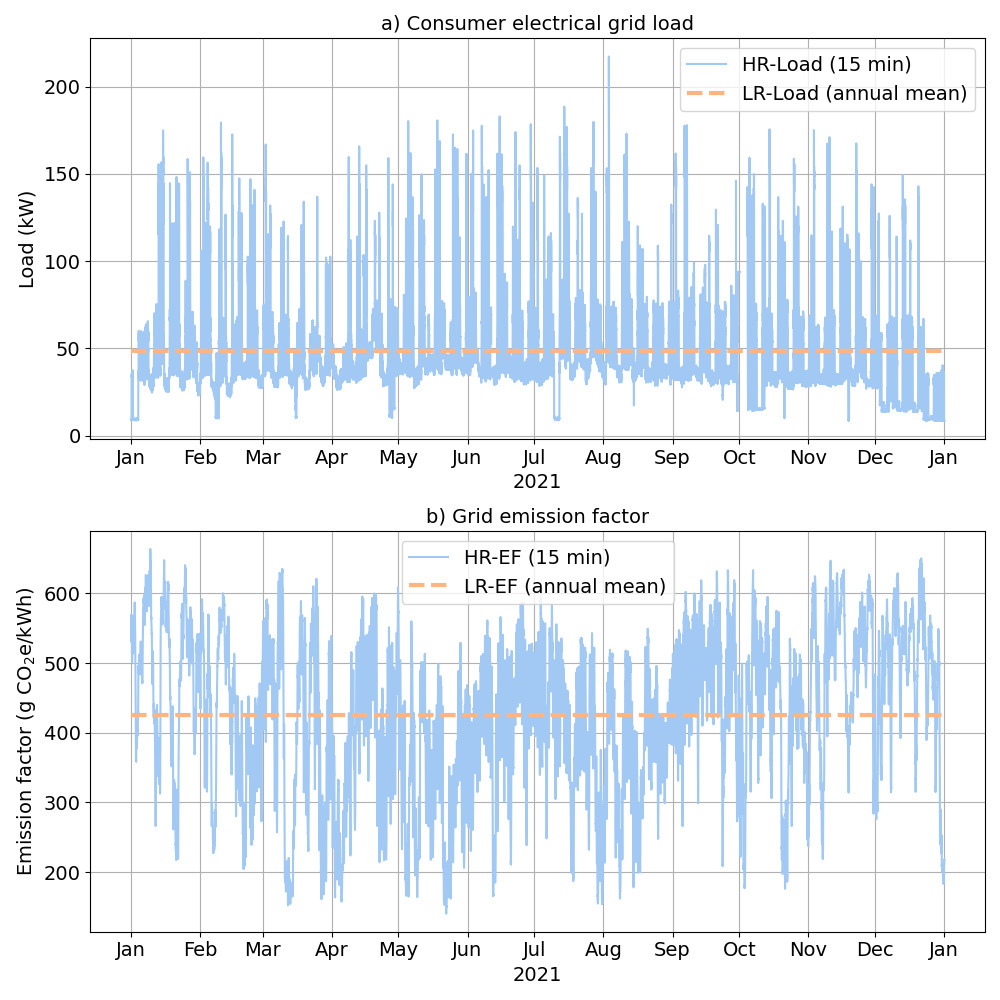}
    \caption{a) Electrical grid load profile of the Battery Lab Factory in Braunschweig (BLB) and b) corresponding grid EF for Germany in 2021. Both high (15-minute) and low (annual) resolutions are presented (\textit{HR}, \textit{LR}). The recommended grid EF configuration (cf. Section \ref{sec:discussion_rec_calc}) is applied.}
    \label{fig:result_effect_temp_blb_setup}
\end{figure}

Figure \ref{fig:result_effect_temp_blb_setup} presents the grid load profile for an exemplary electricity consumer, the Battery Lab Factory (BLB) in Braunschweig, Germany (for details on the BLB, see e.g. \cite{Turetskyy.2019, Drachenfels.2021, Wessel.2021}). The figure also displays the corresponding grid EF for Germany during the same time frame, in both high and low temporal resolutions. The configuration chosen for the grid EF is the one recommended in Section \ref{sec:discussion_rec_calc}.

The grid load profile reveals typical daily and weekly patterns, with a base electrical load ranging from 10 to 40 kW\(_{el}\). Notably, a drop in demand is observed around the holiday season at the end of December. The mean load hovers around 50 kW\(_{el}\), while the grid EF shows significant fluctuations, averaging between 430-440 g CO\(_2\)e/kWh.

Equations \ref{eqn:blb_temp_1} and \ref{eqn:blb_temp_2} detail the computational steps for determining total emissions at both resolutions.

\begin{equation}
    \label{eqn:blb_temp_1}
    Em_{LR} = \Delta T \cdot Load_{LR} \cdot EF_{LR} = \sum_t (\Delta t \cdot Load_{HR,t}) \cdot \sum_t EF_{HR,t}
\end{equation}

\begin{equation}
    \label{eqn:blb_temp_2}
    Em_{HR} = \sum_t (\Delta t \cdot Load_{HR,t} \cdot EF_{HR,t})
\end{equation}

Here, \( Em \) represents the total emissions, \( Load \) denotes the electrical load, \( EF \) is the grid emission factor, \( t \) is the time variable, \( \Delta T \) is one year and \( \Delta t \) is 15 minutes. The subscripts \( LR \) and \( HR \) refer to low and high resolutions, respectively. In this case, using a higher temporal resolution lowers the total emissions from 184.2 to 177.2 t CO\(_2\)e, a relative reduction of 3.8\%.

\section{Discussion}
\label{sec:discussion}

This section begins with a validation of the results (Section \ref{sec:discussion_val}), followed by an outline of recommendations grounded in this study's outcomes (Section \ref{sec:discussion_rec}). The ESI\dag contains sections that reflect on the limitations of this investigation and suggest avenues for future investigations.

\subsection{Validation}
\label{sec:discussion_val}

To benchmark our results and methodology, we compare them with both prior academic investigations (Section \ref{sec:discussion_val_lit}) and official grid EF figures (Section \ref{sec:discussion_val_off}).

\subsubsection{Benchmarking Against Academic Research.}
\label{sec:discussion_val_lit}

\begin{figure}[hbt]
\centering
    \includegraphics[keepaspectratio, width=\linewidth]{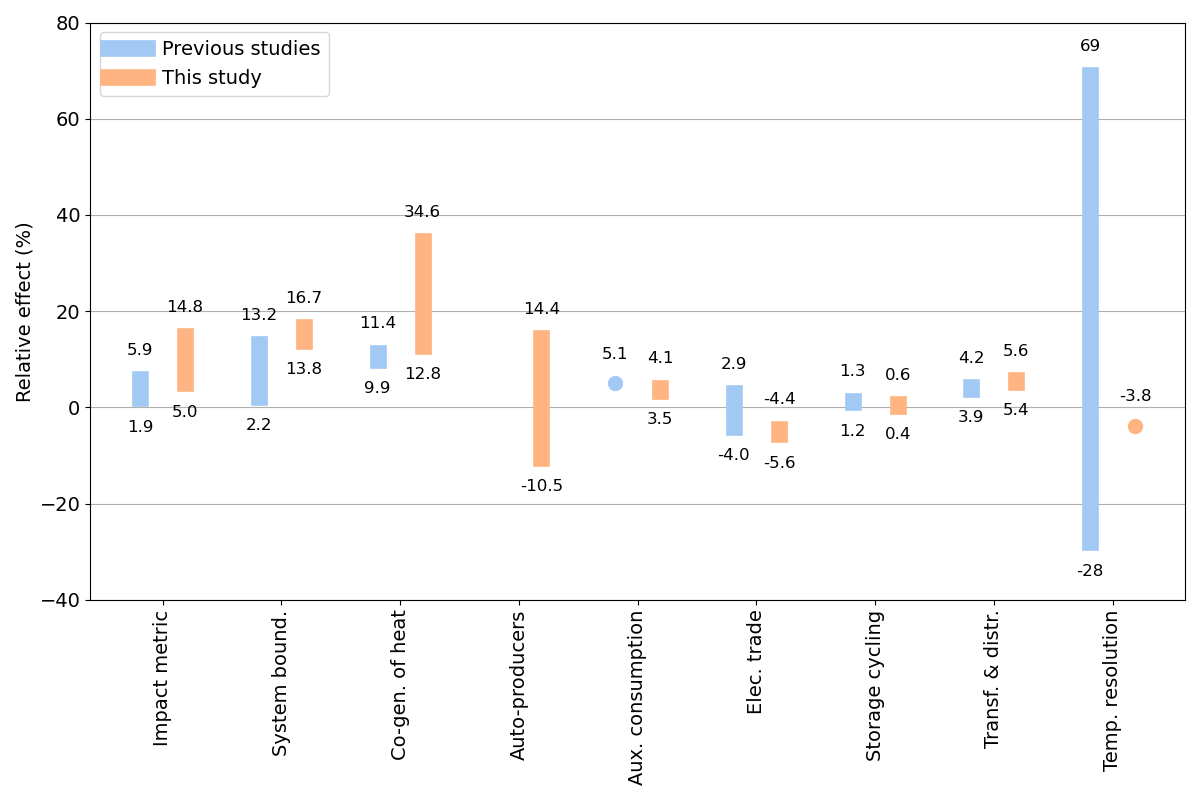}
    \caption{Benchmarking of our findings against prior research (data on prior research from Table \ref{tbl:lit_aspects_effect}). The plot delineates the effect range for each methodological aspect, defined as the relative difference in grid EF arising from different choices within each aspect. Note that the reference range (blue bar) for \textit{Temporal resolution} is the only one that does not refer to values for Germany, but other countries.}
    \label{fig:disc_val_lit}
\end{figure}

We revisit Table \ref{tbl:lit_aspects_effect} to contrast its summary of prior research with our own findings, as visualized in Figure \ref{fig:disc_val_lit}. The graph captures the range of relative differences in grid EF that result from varying choices within methodological aspects. It underscores that the alignment between our results and prior research varies across aspects.

For \textit{Impact metric}, our findings indicate a larger effect than previous studies. However, when only comparing \textit{\ch{CO2}} and \textit{GWP100} (for \textit{GWP20}, the effect has not been previously quantified), the effect is limited to 5.0-5.9\%---well in line with previous results.

For \textit{System boundaries}, our results skew towards the high end of previous findings. This may be explained by our choice of primary energy emission factors (\textit{UBA)}, for which the upstream emissions make up a relatively large share of the life-cycle emissions compared to other sources.

Emission allocation with respect to \textit{Co-generation of heat appears} to have a much larger effect in this study than in previous research articles. However, the upper end (34.6\% divergence) can be explained by comparing extreme allocation methods (all emissions allocated to heat only (\textit{TH100}) vs. to electricity output only (\textit{EL100})), a comparison not found in previous studies. When comparing only the \textit{EN} and the \textit{EL100} allocation method (as it was done in the only reference study for CHP allocation methods \cite{Soimakallio.2012}), the relative differences between the two methods for this study (10.7\%-12.7\%) are comparable to those from the previous study (9.9-11.4\%).

For \textit{Auto-producers}, with up to 14.4\%, the effect appears to be quite large (no previous studies have quantified this effect). However, the larger effects occur only when either only emissions or only electricity from auto-producers are considered, but not both. The difference between considering neither emissions nor electricity from auto-producers and considering both emissions and electricity from auto-producers is less than 1\%.

The results for \textit{Auxiliary consumption} are close to those of previous studies and are based on well-documented data on gross and net electricity production.

The effect size for \textit{Electricity trade} in this study is similar to that documented in other studies. However, not all other studies come to the conclusion that trading reduces Germany's grid EF. The direction of the effect depends on the trade deficit, and on the grid EF of Germany compared to its neighbors' grid EF. A detailed analysis of the effect of electricity trade can be found in the ESI\dag.

The effect of \textit{Storage cycling} is relatively small for the case of Germany (0.4-0.6\%), and does not differ greatly from previous findings (1.2-1.3\%)

\textit{Transformation \& distribution (T\&D)} losses, approximately in line with previous results, have a notable effect on the grid EF (5.4-5.6\% in our study vs. 3.9-4.2\% in previous ones).

The effect of changing the \textit{Temporal resolution} cannot be directly compared to other studies, since no previous study quantified the effect for Germany. The largest relative effects of +69\% and +36\% were observed for countries with a relatively low overall grid EF (Switzerland and France, see ESI\dag). In these countries, 
a small absolute effect results in a relatively large relative effect, due to the low baseline. For the UK, with a baseline grid EF closer to that of Germany, \citeauthor{Mehlig.2022} observe a relative effect of +4.2\%. In absolute terms, this is close to the relative effect observed in our study (-3.8\%; cf. Section 
 \ref{sec:results_ind_temp}).

\subsubsection{Validation Against Official Sources.}
\label{sec:discussion_val_off}

\begin{figure}[hbt]
\centering
    \includegraphics[keepaspectratio, width=\linewidth]{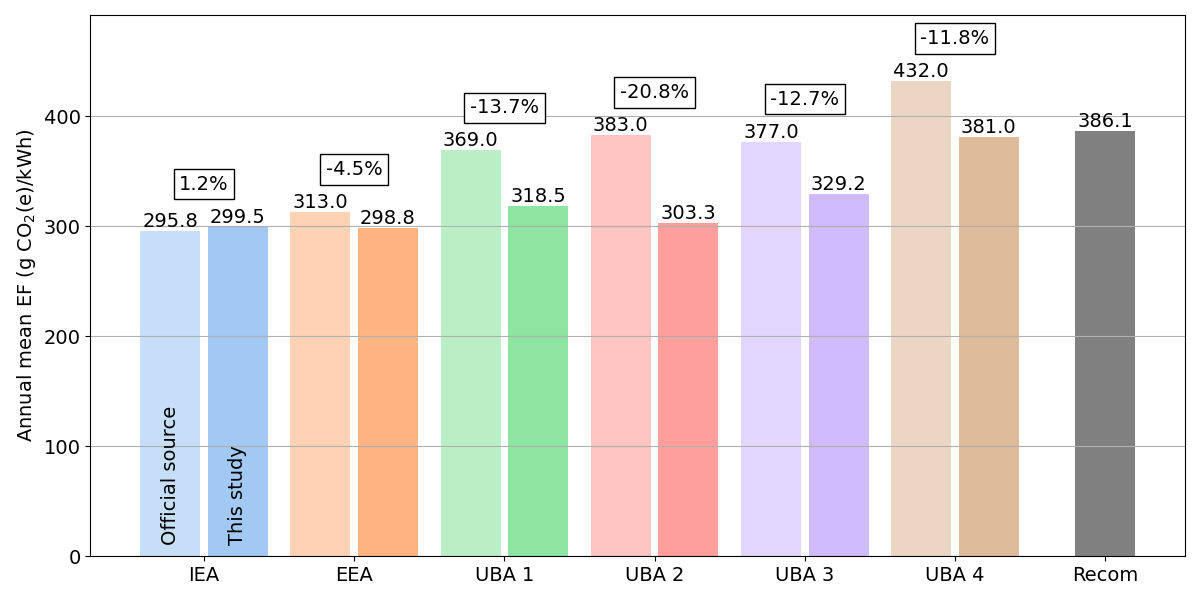}
    \caption{Methodological validation of this study against official grid EF data. The figure contrasts the grid EF figures from IEA, EEA, UBA, (cf. Figure \ref{fig:motivation}) against the ones generated in this study, all for Germany in 2020. The bars labeled \textit{This study} (darker shade) are calculated using the methodology from this study and the methodological choices from the respective documentations \cite{IEA.2021, EEA.2023, UBA.2023}. Data labels indicate the annual mean grid EF atop each bar and the relative difference between the official and our calculated figures between bars. The far-right bar, labeled \textit{Recom}, shows the annual mean grid EF based on our study's recommended configuration of methodological choices (cf. Section \ref{sec:discussion_rec_calc}). The methodological aspects defining the configurations \textit{UBA 1}-\textit{UBA 4} are provided in the caption of Figure \ref{fig:motivation}.}
    \label{fig:disc_val_off}
\end{figure}

To examine the credibility of our methodology, we scrutinize how it stacks up against the reported figures from the IEA, EEA, and UBA (cf. Figure \ref{fig:motivation}). Informed by the documented methodologies of these institutions \cite{IEA.2021, EEA.2023, UBA.2023}, we recreate their grid EF calculations for Germany for the year 2020, presented in Figure \ref{fig:disc_val_off}.

Our results align closely with the IEA's grid EF, deviating 1.2\%. For the EEA's value, the divergence is larger, with a 4.5\% difference. The gap widens considerably with the UBA's figures, with the difference ranging from 11.8\% to 20.8\%. Aspects that may explain this divergence include differences in the characterization factors (CF) used: the UBA relies on CF from the 5th IPCC Assessment Report (AR), while this study applies CF from the more recent 6th IPCC AR. Furthermore, the different data sources used may have an influence. The UBA applies a top-down approach, relying on national emission and energy statistics, while this study pursues a bottom-up approach, multiplying energy flows with production-type specific EFs. As illustrated by \citeauthor{Unnewehr.2022}, these two approaches can yield different results \cite{Unnewehr.2022}.

Finally, the UBA takes a different approach to electricity trade: an UBA grid EF that takes trade into account is larger than one that does not, while the opposite is true for this study. This effect can be observed when comparing the values for \textit{UBA 1} (w/o trade) and \textit{UBA 2} (with trade), and explains why the difference between this study and the official value is largest for \textit{UBA 2}. Following the UBA logic, a country exporting more electricity than it imports (like Germany in 2020) has a higher grid EF after accounting for trade, while the opposite is true for this study. In addition, our study takes into account the grid EF both of the importing and of the exporting nation, while the UBA only considers the EF of the exporting nation (Germany).

\subsection{Steering the Course: Recommendations}
\label{sec:discussion_rec}

In light of the insights gathered throughout our investigation, we articulate a series of recommendations. These not only aim to guide the mechanics of grid EF calculation (Section \ref{sec:discussion_rec_calc}) thereby addressing RQ3, but also touch upon broader considerations we believe are crucial in the context of calculating grid EF for Scope 2 emission accounting (Section \ref{sec:discussion_rec_stand}).

\subsubsection{Recommended Grid EF Configuration.}
\label{sec:discussion_rec_calc}

With nine key methodological aspects uncovered and discussed in this study, we seek to recommend a set of choices for calculating Scope 2 emissions. This set is grounded in five guiding principles borrowed from the GHG Protocol Scope 2 Guidance \cite{WRI.2015}: relevance, completeness, consistency, transparency, and accuracy. We find that the choices summarized in Table \ref{tbl:recom_choices} best represents these principles.

\begin{table}[htb]
\small
    \caption{Recommended set of methodological choices for calculating a grid emission factor to be used in Scope 2 emission accounting}
    \label{tbl:recom_choices}
    \begin{tabular*}{0.48\textwidth}{@{\extracolsep{\fill}}ll}
        \hline
        Aspect & Recommended choice \\
        \hline
        Impact metric & GWP100 \\
        System boundaries & LC \\
        Co-generation of heat & EX \\
        Auto-producers & MAPonly \\
        Auxiliary consumption & with \\
        Electricity trade & with \\
        Storage cycling & with \\
        Transformation \& distribution & with\\
        Temporal resolution & high (15 min) \\
        \hline
    \end{tabular*}
    
\end{table}

By including all losses and transformations that occur between electricity production and consumption (\textit{Auxiliary consumption}, \textit{Electricity trade}, \textit{Storage cycling} and \textit{T\&D losses}), the recommended configuration considers the consumer perspective relevant for Scope 2 accounting, meeting the relevance, completeness, and consistency criteria. The impact metric \textit{GWP100} is more complete than \textit{\ch{CO2}}, as it considers multiple
GHG, and is consistent with most other studies, which typically use \textit{GWP100} over \textit{GWP20}. Similarly, life-cycle (\textit{LC}) system boundaries are more complete than operational (\textit{OP}) boundaries \cite{dixit2014calculating}. Emission allocation by exergy (\textit{EX}) reflects the usefulness of the heat and electricity output energy flows better than all other allocation methods, thus meeting the relevance and accuracy criteria. Excluding generators not feeding into the grid (\textit{MAPonly}) from the grid EF calculation appears to be the most consistent and accurate way of handling auto-producers among all the choices available. Including auto-producers (which do not feed electricity into the grid) in the calculation of a grid EF would be logically inconsistent. Finally, a higher temporal resolution (\textit{15 minutes}) certainly yields more accurate result than a lower one (e.g. \textit{one year}). For a nuanced justification of why we believe this set of choices best embodies the five guiding principles, the reader is directed to the ESI\dag.

However, the necessary data may not be available for all regions to calculate a grid EF with the recommended configuration. This study only demonstrates that the data is available, and the computation is viable for the case of Germany. For regions where some input data are lacking, compromises may be required. For example, should no data on the share of auto-producers in a region exist, then they may be included in the calculation of a grid EF against better knowledge. Figure \ref{fig:disc_val_lit} can provide orientation on how much neglecting a specific aspect may potentially influence the resulting grid EF.

\subsubsection{Recommendations for Standardization and Harmonization.}
\label{sec:discussion_rec_stand}

The area of grid EF calculation for Scope 2 emission accounting would benefit from further standardization and harmonization. Below are specific recommendations to address this need, based on results and insights from this study.

\paragraph{Standardize Data Categories.}
Harmonizing the categories for production types between Eurostat and ENTSO-E is advisable. The current disparity in categorization presented challenges in our study and may affect the accuracy of the results.

\paragraph{Provide Detailed Methodologies.}
Institutions such as the IEA, EEA, and UBA that publish grid EF should also offer comprehensive methodology descriptions. While some existing methodologies are accessible \cite{EEA.2023, UBA.2021, UBA.2023}, they occasionally lack detail on essential aspects. Greater transparency and comparability in documentation is recommended (e.g. with regard to the methodological aspects discussed in this study).

\paragraph{Open Data Accessibility.}
The availability of data is crucial for advancing both scientific research and climate change mitigation efforts. In the case of this study, data availability posed certain challenges. For instance, the IEA offers grid emission factors for global application but restricts access behind a paywall. Similarly, while ENTSO-E provides free data access upon account creation, the licensing terms limit its further dissemination by researchers. Such restrictions can impede the progress of science and the broader climate agenda. Therefore, we advocate for more open licensing arrangements and the removal of paywalls for such vital data.

\paragraph{Align Methodological Approaches.}
A common methodology for calculating grid EF should be considered by institutions that publish these figures. Such standardization would provide clear benefits for various stakeholders, ranging from power plant operators to electricity consumers. If multiple grid EF figures are to be published, clarity on which metric is appropriate for Scope 2 emission accounting is essential.

\paragraph{Disclose Source of EF in Reports.}
It is advisable for organizations reporting their Scope 2 emissions to include the grid EF value and source used in their calculations. This information is often missing from current sustainability reports, making it challenging to validate and compare emissions data.

\paragraph{Incorporate Guidelines into Existing Protocols.}
The GHG Protocol and other institutions publishing guidelines on Scope 2 emission accounting could include specific recommendations on grid EF calculation in their Scope 2 Guidelines. This could encompass the nine methodological aspects identified in this study. The currently ongoing review process for the Scope 2 Guidelines may serve as an appropriate context for such an inclusion.

\section{Conclusion}
\label{sec:conclusion}

This study started with a practical question in mind: how can one accurately account for a company's Scope 2 emissions? Through the course of this research, we have shed light on the methodological aspects and choices involved in calculating grid emission factors, a critical component in Scope 2 accounting. 

We identified nine key methodological aspects (e.g. impact metric, temporal resolution) that significantly influence the outcome of a grid emission factor. For each of these aspects, we explored various choices (e.g. \ch{CO2}, GWP100) and quantified their impacts, some of which alter the emission factor by more than 10\%. Building upon these findings, we proposed a set of recommended choices grounded in the principles of relevance, completeness, consistency, transparency, and accuracy. These recommendations are aimed at providing a more standardized approach for calculating Scope 2 emissions.

Standardized emission calculations not only benefit corporate GHG accounting, but also other areas where electricity-related emissions are relevant. Energy systems at various scales are increasingly optimized for low emissions \cite{de2019city}, as is electric vehicle charging \cite{holdway2010indirect} and hydrogen production \cite{millinger2021electrofuels}. All these applications require a transparent and consistent calculation procedure to determine the resulting emissions.

Moreover, the study underscores the need for further standardization and harmonization in the domain of corporate GHG accounting and reporting. Various stakeholders, including practitioners, researchers, and data providers, can contribute to these standardization efforts. 

In a move toward greater transparency and academic rigor, this study makes all its data and calculations openly available in the ESI\dag. We invite the scholarly community and interested parties to review, reuse, and build upon this foundation, further contributing to the robustness and comparability in the field of Scope 2 emissions accounting.

\section*{Author Contributions}
Conceptualization: MS, FC; Data Curation: MS; Formal Analysis: MS, FC; Funding Acquisition: CH; Investigation: MS, FC; Methodology: MS, FC; Project Administration: MS; Resources: FC, CH; Software: MS; Supervision: FC, CH; Validation: MS; Visualization: MS; Writing – Original Draft: MS; Writing – Review \& Editing: MS, FC, CH.

\section*{Conflicts of interest}
There are no conflicts to declare.

\section*{Acknowledgements}
The authors thank Benoît Gschwind for his help with the mapping of production type categories.

The research that led to the results presented in this paper was funded by the Federal Ministry for Economic Affairs and Climate Action (BMWK) as part of the research project `flexess' under Grant No. 03EI4005A.



\balance


\bibliography{rsc} 
\bibliographystyle{rsc} 

\end{document}


\pagestyle{fancy}
\thispagestyle{plain}
\fancypagestyle{plain}{
\renewcommand{\headrulewidth}{0pt}
}

\makeFNbottom
\makeatletter
\renewcommand\LARGE{\@setfontsize\LARGE{15pt}{17}}
\renewcommand\Large{\@setfontsize\Large{12pt}{14}}
\renewcommand\large{\@setfontsize\large{10pt}{12}}
\renewcommand\footnotesize{\@setfontsize\footnotesize{7pt}{10}}
\makeatother

\renewcommand{\thefootnote}{\fnsymbol{footnote}}
\renewcommand\footnoterule{\vspace*{1pt}%
\color{cream}\hrule width 3.5in height 0.4pt \color{black}\vspace*{5pt}} 
\setcounter{secnumdepth}{5}

\makeatletter 
\renewcommand\@biblabel[1]{#1}            
\renewcommand\@makefntext[1]%
{\noindent\makebox[0pt][r]{\@thefnmark\,}#1}
\makeatother 
\renewcommand{\figurename}{\small{Fig.}~}
\sectionfont{\sffamily\Large}
\subsectionfont{\normalsize}
\subsubsectionfont{\bf}
\setstretch{1.125} 
\setlength{\skip\footins}{0.8cm}
\setlength{\footnotesep}{0.25cm}
\setlength{\jot}{10pt}
\titlespacing*{\section}{0pt}{4pt}{4pt}
\titlespacing*{\subsection}{0pt}{15pt}{1pt}

\fancyfoot{}
\fancyfoot[LO,RE]{\vspace{-7.1pt}\includegraphics[height=9pt]{head_foot/LF}}
\fancyfoot[CO]{\vspace{-7.1pt}\hspace{13.2cm}\includegraphics{head_foot/RF}}
\fancyfoot[CE]{\vspace{-7.2pt}\hspace{-14.2cm}\includegraphics{head_foot/RF}}
\fancyfoot[RO]{\footnotesize{\sffamily{1--\pageref{LastPage} ~\textbar  \hspace{2pt}\thepage}}}
\fancyfoot[LE]{\footnotesize{\sffamily{\thepage~\textbar\hspace{3.45cm} 1--\pageref{LastPage}}}}
\fancyhead{}
\renewcommand{\headrulewidth}{0pt} 
\renewcommand{\footrulewidth}{0pt}
\setlength{\arrayrulewidth}{1pt}
\setlength{\columnsep}{6.5mm}
\setlength\bibsep{1pt}

\makeatletter 
\newlength{\figrulesep} 
\setlength{\figrulesep}{0.5\textfloatsep} 

\newcommand{\topfigrule}{\vspace*{-1pt}%
\noindent{\color{cream}\rule[-\figrulesep]{\columnwidth}{1.5pt}} }

\newcommand{\botfigrule}{\vspace*{-2pt}%
\noindent{\color{cream}\rule[\figrulesep]{\columnwidth}{1.5pt}} }

\newcommand{\dblfigrule}{\vspace*{-1pt}%
\noindent{\color{cream}\rule[-\figrulesep]{\textwidth}{1.5pt}} }

\makeatother

\twocolumn[
  \begin{@twocolumnfalse}
{\includegraphics[height=30pt]{head_foot/journal_name}\hfill\raisebox{0pt}[0pt][0pt]{\includegraphics[height=55pt]{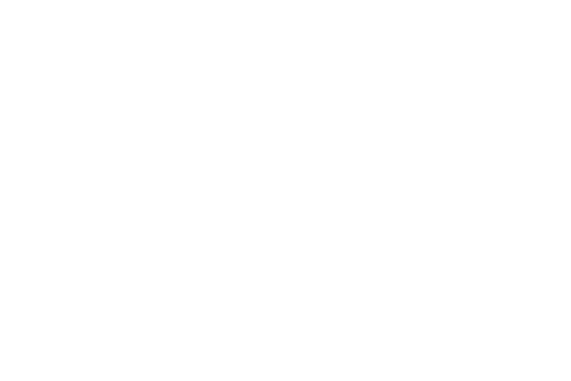}}\\[1ex]
\includegraphics[width=18.5cm]{head_foot/header_bar}}\par
\vspace{1em}
\sffamily
\begin{tabular}{m{4.5cm} p{13.5cm} }

\includegraphics{head_foot/DOI} & \noindent\LARGE{\textbf{Towards Standardized Grid Emission Factors: Methodological Insights and Best Practices (Electronic Supplementary Information)$^\dag$}} \\
\vspace{0.3cm} & \vspace{0.3cm} \\

 & \noindent\large{Malte Schäfer,$^{\ast}$\textit{$^{a}$} Felipe Cerdas,\textit{$^{a}$} and Christoph Herrmann\textit{$^{a}$}} \\

\includegraphics{head_foot/dates} 

\end{tabular}

 \end{@twocolumnfalse} \vspace{0.6cm}

  ]

\renewcommand*\rmdefault{bch}\normalfont\upshape
\rmfamily
\section*{}
\vspace{-1cm}


\footnotetext{\textit{$^{a}$~Institute of Machine Tools and Production Technology (IWF), Technische Universität Braunschweig, 38106 Braunschweig, Germany. Tel: +49 (0)531 391-7650; E-mail: malte.schaefer@tu-braunschweig.de}}

\footnotetext{\dag~Electronic Supplementary Information (ESI) available online. Code: \url{https://doi.org/10.24355/dbbs.084-202309131139-0}, Data: \url{https://doi.org/10.24355/dbbs.084-202309111514-0} }



\section{Theoretical Background \label{sec:SI-theory}}

\begin{figure*}[htbp]
\centering
    \includegraphics[keepaspectratio, width=\textwidth]{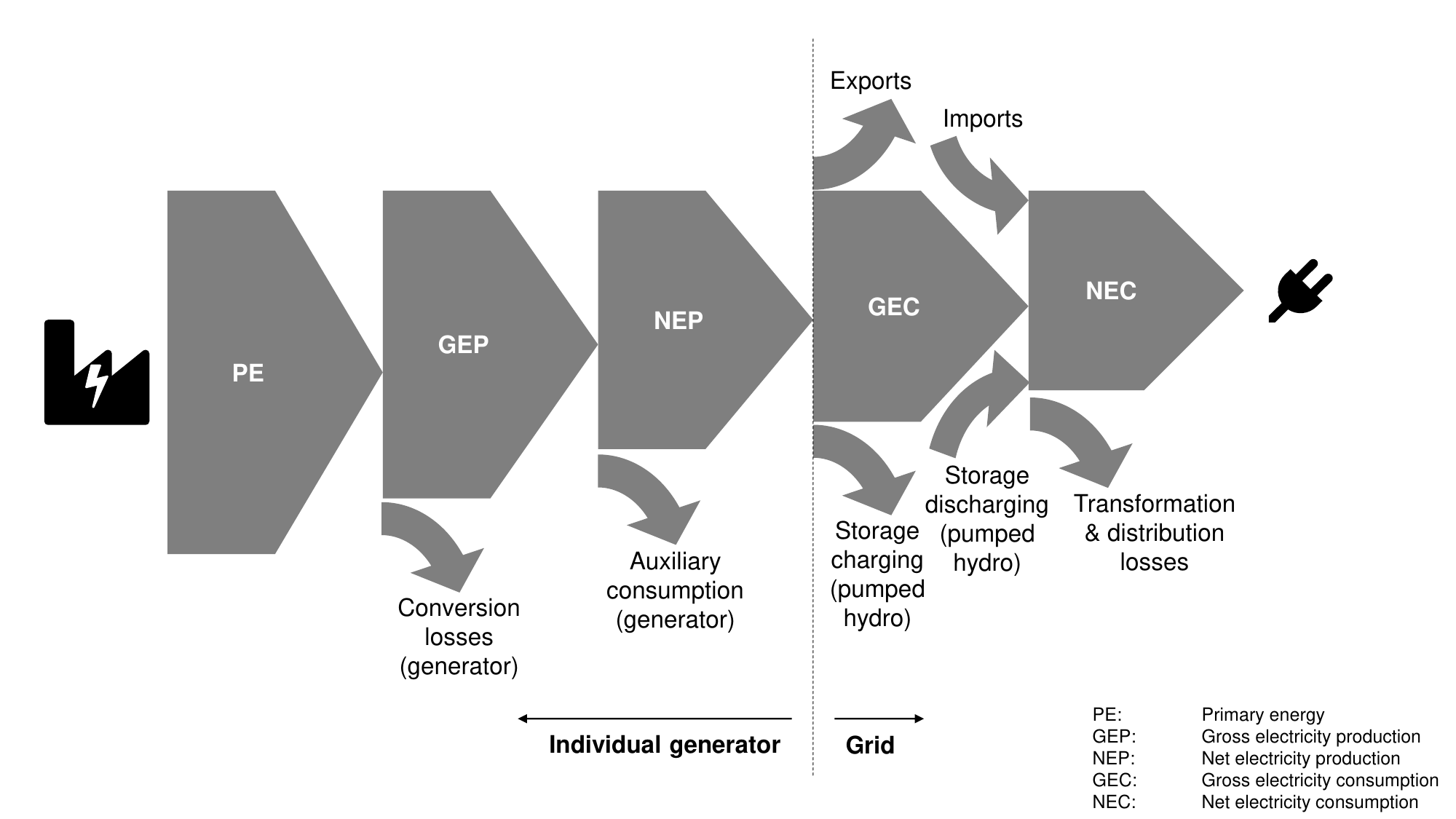}
    \caption{Energy conversion steps from the power plant to the plug, inspired by ENTSO-E \cite{ENTSOE.2016}}
    \label{fig:planttoplug}
\end{figure*}

Grid EF can be calculated at different steps along the conversion chain from the power plant to the plug. The following illustration (Figure \ref{fig:planttoplug}) provides an overview of these conversion steps.

Primary energy (PE) in the form of e.g., wind, solar radiation or lignite is converted in generators/power plants into gross electricity production (GEP) and waste heat (conversion losses). After subtracting the electrical energy these generators need for e.g., powering its own pumps (auxiliary consumption), the result is net electricity production (NEP). This is the amount of electrical energy fed into the grid. Gross electricity consumption (GEC) is NEP minus electricity exports to, and plus imports from, neighboring regions (e.g., countries, bidding zones); and minus the losses occurring from cycling grid storage (e.g., pumped hydro). Finally, after subtracting the grid losses due to transformation and distribution, the result is net electricity consumption (NEC).

Going from left to right, the EF referenced to each of these stages necessarily grows larger (with the exception of electricity trade - this may have the opposite effect under certain circumstances). The total amount of emissions (numerator) remains the same, while the amount of electrical energy (denominator) decreases.

\section{Extended State of Research}
\label{sec:literature_SI}

This section expands on the literature review covered in the main article.

\subsection{Scope and Retrieval Process}
\label{sec:literature_scope}
The review of the literature is aimed at identifying any studies that meet the following criteria regarding the studies' content:
\begin{itemize}
  \setlength\itemsep{0em}  
  \item ELECTRICITY FOCUS: Discusses electricity-related emissions (not emissions related to other forms of energy, e.g., heat)
  \item CONSUMER FOCUS: Focuses on emissions primarily from a consumer perspective (not from the producer perspective, i.e. power plant or grid operators)
  \item CLIMATE FOCUS: Focuses on GHG emissions and/or climate change impact (extending the scope to other types of emissions/impact is acceptable if climate change impact/GHG emissions are included)
  \item METRIC MATCH: Assesses emissions on the basis of an indicator which relates emissions to the amount of electricity produced, i.e. an EF (and discusses not only e.g., total emissions)
  \item TRANSPARENCY: Transparently documents most or all (i.e. more than half of the) methodological choices made in calculating emission factors
  \item ACCOUNTING PERSPECTIVE: Focuses on average (not marginal) EF
  \item GRID SCALE: Assesses emissions within interconnected electricity systems and markets of significant scale, typically national grids (excluding e.g., off-grid, micro-grid or island grid cases)
  \item REAL SETTING: Assesses emissions of real, existing electricity systems and markets, using real world data and realistic assumptions (excluding fictional grid setups)
  \item RETROSPECTIVITY: Assesses past emissions (not projections of possible future emissions)
\end{itemize}

Additionally, the references have to meet these formal requirements to be included in the review:
\begin{itemize}
    \setlength\itemsep{0em}
    \item Only peer-reviewed journal articles
    \item Primary research (no review articles)
    \item Must be cited at least once
    \item At least one citation from authors other than the study authors
\end{itemize}

In the literature retrieval process, we used various combinations of relevant key-words (e.g., “emission factor”, “electricity”, “average”, “grid”) in literature search engines and software tools, primarily Google Scholar and ResearchRabbit. From initially discovered references, we then conducted an upstream- and downstream-search, i.e. we reviewed the citing and citing literature for relevant articles. We also looked through articles of the same author to uncover further relevant studies.

\subsection{Detailed Literature Review of Methodological Aspects}
\label{sec:literature_results_detail}
The following paragraphs briefly describe each methodological aspect reviewed in the literature review of the main article. Furthermore, they summarize the findings of studies where the impacts of each aspect are quantified. An emphasis lies on studies that quantify the effect for Germany, as this is the focus of our own investigation.

Where impacts are quantified, we compare grid EF for two (or more) choices regarding a methodological aspect. We list the resulting grid EF for each choice, and calculate the absolute (\textit{Abs.dif.}) and relative difference (\textit{Rel.dif. (\%)}) between them using the equations $Abs.dif. = (EF_{choice 2}-EF_{choice 1})$ and $Rel.dif. (\%) = Abs.dif./EF_{choice 1}$.

\subsubsection{Impact Metric} describes the metric used to assess the global warming/climate change impact per kWh of electricity. In order to fully define an impact metric, one needs to specify the impact assessment model used. An impact assessment model shall contain information on which substances it includes, the characterization factors used to evaluate their impact, and the time period over which their impact is assessed.

Most studies use GWP as a metric to calculate climate impact, and only a few \cite{Jiusto.2006, Soimakallio.2012, Louis.2014, Tamayao.2015, Qu.2017, Qu.2018, Donti.2019, Braeuer.2020, Mehlig.2022} rely on just \ch{CO2}. In two studies, \ch{CO2} and GWP (\ch{CO2}e) appear to be mixed together and used interchangeably \cite{Schwabeneder.2019, Chalendar.2021}. Some studies assess environmental impact besides climate change, either on the basis of the amount of substances (e.g., \ch{CH4}) emitted, or relying on impact assessment methodologies used in LCA \cite{Maurice.2014, Roux.2016, Kono.2017, Collinge.2018, Milovanoff.2018, Vuarnoz.2018, BeloinSaintPierre.2019, Chalendar.2019, Donti.2019, Walzberg.2019, Braeuer.2020, Mehlig.2022}. Most studies use the IPCC impact assessment models to calculate the resulting climate impact, while a few rely on other impact assessment models, namely CML \cite{MunneCollado.2019,Papageorgiou.2020}, EF3.0 \cite{Peters.2022}, GREET \cite{Colett.2016}, IMPACT2002+ \cite{Maurice.2014, Milovanoff.2018, Walzberg.2019}, ReCiPe \cite{Messagie.2014} and TRACI \cite{Collinge.2018}. These "other" impact models, however, employ the IPCC models as well \cite{PRe.2020}, so it is safe to say that the IPCC models dominate impact assessment. Not all studies explicitly state which impact assessment model they use \cite{KopsakangasSavolainen.2017, Mills.2017, Baumann.2019, Rupp.2019, Schram.2019, Schwabeneder.2019, Tranberg.2019, Worner.2019, Chalendar.2021, Fleschutz.2021}. To highlight the impact of short-lived GHG such as methane, one study \cite{Pereira.2020} assesses both the impact over 20 years (GWP20) and over 100 years (GWP100), with the GWP20 values being 15 to 20\% higher than the GWP100 values. In the same study, the authors conduct a probabilistic estimate of the amount of emissions emitted per source and for the characterization factors of the substances emitted using probability density functions, to account for the inherent uncertainty surrounding these parameters.

Of all reviewed studies, only one assesses both \ch{CO2} and \ch{CO2}e emissions \cite{Worner.2019}. The authors do not specify which impact assessment model (characterization factors) they use to calculate \ch{CO2}e emissions. In the following Table \ref{tbl:metric}, we assume these to represent GWP100, as it is the most common indicator used in impact assessment models. \citeauthor{Pereira.2020} \cite{Pereira.2020} also provide different impact metrics, consisting of a comparison of GWP20 and GWP100 as well as a probabilistic estimate of different characterization factors for non-\ch{CO2} substances. Unfortunately, the detailed data is locked behind a paywall, and thus is not included in the quantitative analysis in Table \ref{tbl:metric}.

\begin{table*}[ht!]
\small
    \caption{Impact metric effect in the literature, based on data from \citeauthor{Worner.2019} \cite{Worner.2019} Values in EF-columns are in g \ch{CO2}(e)/kWh. All values are for Germany, 2017}
    \label{tbl:metric}
    \begin{tabular*}{\textwidth}{@{\extracolsep{\fill}}llllll}
        \hline
        Perspective & Sys. bound. & EF$_{\ch{CO2}}$ & EF$_{GWP100}$ & Abs.dif. & Rel.dif. (\%) \\
        \hline
        Production & OP & 439 & 448 & +9 & +2.1 \\
        Production & LC & 479 & 507 & +28 & +3.8 \\
        Consumption & OP & 515 & 525 & +10 & +1.9 \\
        Consumption & LC & 561 & 594 & +33 & +5.9 \\
        \hline
    \end{tabular*}

    OP: Operational;
    LC: Life cycle
    
\end{table*}

For the case of Germany, the study reports a +1.9 to +5.9\% relative difference (+9 to +33 g/kWh) between \ch{CO2} and GWP100 EF. The relative and absolute differences are larger for EF that include multiple life cycle stages (LC) than those that only cover the operational stage (OP).

\subsubsection{System Boundaries} describe the life cycle stages included in calculating the EF of a generator or a set of generators. In general, one can distinguish between the plant life cycle and the fuel life cycle. For the purpose of this review, we only distinguish between operational (OP) and life cycle (LC) system boundaries. As per the definition used in this study, LC boundaries differ from OP boundaries in that they cover at least one additional life cycle stage, i.e. an upstream and/or downstream stage from the fuel and/or plant life cycle.

Several studies limit their assessment to operational emissions \cite{Jiusto.2006, Li.2013, Mills.2017, Qu.2017, Khan.2018, Chalendar.2019, Donti.2019, Braeuer.2020, Chalendar.2021, Fleschutz.2021, Mehlig.2022}. The other studies consider life cycle emissions to a different extent, including e.g., fuel upstream emissions \cite{Soimakallio.2012, Louis.2014, Tamayao.2015, Colett.2016, Ji.2016, Qu.2018, Baumgartner.2019, BeloinSaintPierre.2019}, fuel upstream emissions and the plant life cycle \cite{Maurice.2014, Roux.2016, Kono.2017, KopsakangasSavolainen.2017, Clauss.2018, Collinge.2018, Rupp.2019, Schram.2019, Walzberg.2019, Papageorgiou.2020, Pereira.2020, Peters.2022} or the complete fuel and plant life cycle \cite{Messagie.2014, Milovanoff.2018, Vuarnoz.2018, Baumann.2019, Clauss.2019, MunneCollado.2019, Neirotti.2020, Noussan.2020, Scarlat.2022}. One study distinguishes between operational (combustion) emissions, upstream and downstream emissions, which are again split up into fuel extraction and power plant construction (upstream) and T\&D as well as pumping losses (downstream), respectively \cite{Blizniukova.2023}.

For two studies \cite{Stoll.2014, Fiorini.2018}, it is unclear which life cycle stages are included. These studies draw the generator-specific EF from the 2021 IPCC Special Report on Renewable Energy (SRREN) \cite{IPCC.2012}, which provides aggregate EF values based on a review of LCA studies. The qualifying criterion regarding system boundaries in the IPCC SRREN report states that two or more life cycle stages must be covered, so the EF reported by \citeauthor{Stoll.2014} and \citeauthor{Fiorini.2018} are considered LC EF. \citeauthor{Nilsson.2017} combine the aggregate values from the IPCC SRREN report with values from the IEA (which covers operational and fuel upstream emissions). \citeauthor{Schwabeneder.2019} appears to mix system boundaries, reporting operating emissions only for renewables and nuclear, and adding fuel upstream emissions for all other generation technologies. \citeauthor{Tranberg.2019} reports two separate EF, one including the operational stage and the fuel upstream stage, and the other adding the plant upstream stage to it. Similarly, \citeauthor{Worner.2019} provides an EF based on only operational emissions, and one which additionally includes both fuel and plant upstream emissions.

Studies which calculate EF with different system boundaries, keeping all other aspects constant, are required to quantify the effect that using different system boundaries has on the results. Only four of the studies we review do so \cite{Tranberg.2019, Worner.2019, Noussan.2020, Scarlat.2022}. Unfortunately, two of those studies only provide plots without the underlying data that could be used for a comparison \cite{Noussan.2020, Scarlat.2022}. Data from the remaining two studies \cite{Tranberg.2019, Worner.2019} is documented in Table \ref{tbl:boundaries} to quantify the system boundary effect. From the study by \citeauthor{Tranberg.2019}, only the countries with the largest and smallest relative effect, and Germany (as it is the same country assessed by \citeauthor{Worner.2019}) are selected. The system boundaries are documented for the lower and higher bound of the EF range, which describes the EF resulting from applying those system boundaries.

\begin{table}[ht!]
\small
    \caption{System boundary effect in the literature. Values in EF-columns are in g \ch{CO2}(e)/kWh}
    \label{tbl:boundaries}
    \begin{tabular*}{0.48\textwidth}{@{\extracolsep{\fill}}llllll}
        \hline
        Study & Region & EF$_{OP}$ & EF$_{LC}$ & Abs.dif. & Rel.dif. (\%) \\
        \hline
        \cite{Tranberg.2019} & DE\textsuperscript{\emph{a}} & 643 & 657 & +14 & +2.2 \\
        \cite{Tranberg.2019} & EU28\textsuperscript{\emph{a}} & 440 & 446 & +6 & +1.4 \\
        \cite{Tranberg.2019} & PL\textsuperscript{\emph{a}} & 1030 & 1030 & 0 & 0 \\
        \cite{Tranberg.2019} & SE\textsuperscript{\emph{a}} & 39.3 & 42.3 & +3.0 & +7.7 \\
        \cite{Worner.2019} & DE\textsuperscript{\emph{b}} & 439 & 479 & +40 & +9.1 \\
        \cite{Worner.2019} & DE\textsuperscript{\emph{c}} & 448 & 507 & +59 & +13.2 \\
        \cite{Worner.2019} & DE\textsuperscript{\emph{d}} & 515 & 561 & +46 & +8.9 \\
        \cite{Worner.2019} & DE\textsuperscript{\emph{e}} & 525 & 594 & +69 & +11.6 \\
        \cite{Blizniukova.2023} & DE\textsuperscript{\emph{f}} & 354 & 377 & +23 & +6.5 \\
        \hline
    \end{tabular*}
    
    OP: Operational;
    LC: Life cycle
    
    \textsuperscript{\emph{a}}Low voltage mix;
    \textsuperscript{\emph{b}}Production-based \& \ch{CO2};
    \textsuperscript{\emph{c}}Production-based \& GWP100;
    \textsuperscript{\emph{d}}Consumption-based \& \ch{CO2};
    \textsuperscript{\emph{e}}Consumption-based \& GWP100;
    \textsuperscript{\emph{f}}Germany, 2020, EF$_{up}$ and EF$_{Total}$;
\end{table}

At the low end, \citeauthor{Tranberg.2019} find no difference between OP emissions and emissions considering more LC stages for the case of Poland. This is not entirely plausible, since the coal-heavy generation mix of Poland should generate at least some upstream plant and feedstock emissions. Perhaps this fact can be explained by a combination of using average European data instead of country-specific data and numerical (rounding) errors. At the high end, \citeauthor{Worner.2019} finds a relative difference of up to 13.2\% between OP and LC emissions for Germany. Across studies and calculation methods, the relative difference between OP and LC EF for Germany is in the range of 2.2 to 13.2\%.

\subsubsection{Co-generation of Heat} describes the co-production of electricity and other products, usually heat and/or steam, within the same plant. Since these co-products provide a value to their users, one can argue that the emissions from these co-generation plants should be allocated to both their electricity and their heat (steam) output. Different principles exist to decide on how exactly this allocation is to be implemented, based e.g., on economic value, energy or exergy content of the outputs \cite{Soimakallio.2012}. Alternatively, using a substitution approach, one can calculate the emissions created when using alternative heat (or steam) sources, and subtract these emissions from the co-generating plant's emissions to receive the emissions for producing electricity only \cite{Scarlat.2022}. The share of generators that participate in co-generation within a set of generators, the share of primary energy converted to electricity vs. other products, and the method used to allocate generator emissions to the various outputs all have an impact on the resulting EF of the electricity produced.

Few studies mention co-generation at all \cite{Soimakallio.2012, Louis.2014, KopsakangasSavolainen.2017, Milovanoff.2018, Vuarnoz.2018, Baumann.2019, Baumgartner.2019, MunneCollado.2019, Worner.2019, Scarlat.2022, Unnewehr.2022, Blizniukova.2023}. \citeauthor{Soimakallio.2012} provide the only study with a detailed assessment of the impact that allocation methods have on the resulting EF. \citeauthor{Louis.2014} distinguish between co-generation of heat and power (CHP) for district heating and for industrial use, and when generation is driven by demand for electrical power and when by demand for heat. \citeauthor{KopsakangasSavolainen.2017} list CHP plants as a separate type of generator technology, but do not disclose how they arrive at their estimate of an electricity EF for CHP plants. \citeauthor{Vuarnoz.2018} use allocation by exergetic content, as do \citeauthor{MunneCollado.2019}, for those countries where data is available (BG, DE, NL). \citeauthor{Baumann.2019} consider co-generation of heat in the plant dispatch model they use to calculate EF, but do not specify how exactly. Similarly, \citeauthor{Baumgartner.2019} appear to provide details on how they consider CHP plants, but the details are available only in the supplementary material that is locked behind a paywall. \citeauthor{Worner.2019} allocate all emissions from CHP to electricity, knowing that this may lead to an overestimation of electricity-related emissions. \citeauthor{Braeuer.2020} supposedly to do the same, as they state that some outliers in their data from the high end may be explained by the fact that CHP plant emissions are allocated to electricity only. \citeauthor{Scarlat.2022} consider CHP plant emissions using a substitution approach (as is used by the IEA). They calculate the amount of emissions that would have been generated if the heat from CHP plants had been produced in heat-only plants with an efficiency of 85-90\%, and subtract these emissions from the CHP plant emissions. \citeauthor{Unnewehr.2022} allocate emissions based on free allowances of emission certificates for heat generation under the European Emission Trading Scheme (ETS). \citeauthor{Blizniukova.2023} employ the IEA method ("fixed-heat-efficiency approach").

Multiple studies probably implicitly consider CHP plants via the data that they use (e.g., the unit processes for CHP units in the Ecoinvent LCA database \cite{Wernet.2016} without explicitly stating whether and how they allocate emissions \cite{Maurice.2014, Messagie.2014, Tamayao.2015, Colett.2016, Kono.2017, KopsakangasSavolainen.2017, Clauss.2019, Collinge.2018, Milovanoff.2018, Baumgartner.2019, BeloinSaintPierre.2019, Chalendar.2019, Tranberg.2019, Walzberg.2019, Noussan.2020, Papageorgiou.2020}.

To assess the impact that co-generation allocation methods have on the resulting EF, we use data by \citeauthor{Soimakallio.2012}, as it is the only study in our review quantifying this effect. We select seven countries from the list of OECD countries with various shares of electricity generation originating in CHP plants ("CHP share") and absolute levels of grid EF. We document the range of EF for each of these countries when using each of the two different allocation methods applied in the study, allocation based on energy content of the co-products (EN) and all emissions allocated to the electrical energy ("Motivation electricity" / EL100: 100\% of emissions allocated to electricity). All values are from the latest year documented in the study (2008) and use production-based EF only, to avoid confounding effects from electricity trading. The results are documented in Table \ref{tbl:cogen}.

\begin{table*}[ht!]
\small
    \caption{Co-generation effect in the literature, based on data from \citeauthor{Soimakallio.2012} \cite{Soimakallio.2012} (year 2008). Values in EF columns are in g \ch{CO2}/kWh}
    \label{tbl:cogen}
    \begin{tabular*}{\textwidth}{@{\extracolsep{\fill}}llllll}
        \hline
        Country & CHP share (\%) & EF$_{EN}$ & EF$_{EL100}$ & Abs.dif. & Rel.dif. (\%) \\
        \hline
        DE\textsuperscript{\emph{a}} & 13 & 547 & 601 & +54 & +9.9 \\
        DE\textsuperscript{\emph{b}} & 13 & 525 & 585 & +60 & +11.4 \\
        DK\textsuperscript{\emph{a}} & 81 & 351 & 663 & +312 & +88.9\\
        FI\textsuperscript{\emph{a}} & 36 & 185 & 316 & +131 & +70.8 \\
        MX\textsuperscript{\emph{a}} & 0 & 566 & 566 & 0 & 0 \\
        NO\textsuperscript{\emph{a}} & 0 & 3 & 4 & +1 & +33 \\
        PL\textsuperscript{\emph{a}} & 98 & 902 & 1229 & +327 & +36.3 \\
        SE\textsuperscript{\emph{a}} & 10 & 15 & 53 & +33 & +253 \\
        \hline
    \end{tabular*}
    
    \textsuperscript{\emph{a}}Production-based;
    \textsuperscript{\emph{b}}Consumption-based;
\end{table*}

For Denmark, with a relatively high CHP share of 81\%, the EF estimate almost doubles when using a different allocation method. Finland's EF estimate exhibits a very similar behavior, yet from a lower baseline. Poland, with a much higher CHP share of 98\%, but compared to Denmark and Finland also a much higher absolute EF to begin with, the allocation method only shifts the estimate by about one third. The biggest influence of the allocation method (in relative terms) can be observed with Sweden, even though its CHP share is only 10\%---mostly due to the baseline effect, with Sweden having a very low grid EF of 15 (53) g \ch{CO2}/kWh. As expected, for countries with a low CHP share (Mexico and Norway), the CHP allocation method has little impact on the grid EF. For Germany, with a CHP share of 13\%, the CHP allocation method can change the resulting grid EF by 9.9-11.4\% (production-/consumption-based).

\subsubsection{Auto-producers} are generators which do not feed into the electrical grid, but instead exclusively supply one consumer (e.g., an industrial facility) with electricity. Depending on whether the underlying dataset contains auto-producer generation and/or emission data, they may be included in grid EF calculations.

Only five studies address the issue of auto-producers \cite{Clauss.2018, Clauss.2019, Colett.2016, Unnewehr.2022, Blizniukova.2023}. In their first study, \citeauthor{Clauss.2018} compare two scenarios, one in which auto-producers are included in the dataset, and one in which they are excluded \cite{Clauss.2018}. They find that for the case of Norway, removing auto-producers reduces the resulting EF noticeably in three out of five bidding zones. In their second study, they only consider scenarios without auto-producers \cite{Clauss.2019}. \citeauthor{Colett.2016} consider the emissions from on-site generators in their aluminum smelting case study. \citeauthor{Unnewehr.2022} estimate the emissions from auto-producers based on the emissions profiles of main-activity producers. \citeauthor{Blizniukova.2023} include both auto-producers and main-activity-producers in their assessment. Unfortunately, none of these study quantifies the effect of removing auto-producers from the dataset on the resulting grid EF.

\subsubsection{Auxiliary Consumption} describes the amount of electricity used by generators for supporting their own operations (e.g., to power pumps). By subtracting auxiliary consumption from gross electricity production (GEP), one receives the net electricity production (NEP).

Nine studies in total address auxiliary consumption \cite{Soimakallio.2012, Tamayao.2015, Mills.2017, MunneCollado.2019, Rupp.2019, Worner.2019, Scarlat.2022, Unnewehr.2022, Blizniukova.2023}, using a simple calculation step consisting of subtracting the auxiliary consumption from GEP. One of these studies quantifies the effect that this subtraction has for multiple (primarily European) countries \cite{Scarlat.2022}. The findings for a selection of these countries are summarized in Table \ref{tbl:aux}.

\begin{table}[ht!]
\small
    \caption{Auxiliary consumption effect in the literature, based on data from \citeauthor{Scarlat.2022} \cite{Scarlat.2022} Values in EF-columns are in g \ch{CO2}e/kWh}
    \label{tbl:aux}
    \begin{tabular*}{0.48\textwidth}{@{\extracolsep{\fill}}lllll}
        \hline
        Region & EF$_{GEP}$ & EF$_{NEP}$ & Abs.dif. & Rel.dif. (\%) \\
        \hline
        DE & 390 & 410 & +20 & +5.1 \\
        EE & 571 & 659 & +88 & +15.4 \\
        SE & 33 & 33 & 0 & 0 \\
        EU27 & 296 & 310 & +14 & +4.7 \\
        \hline
    \end{tabular*}
    
    GEP: Gross electricity production;
    NEP: Net electricity production
    
\end{table}

The results show the for the EU27 states, auxiliary consumption increases the grid EF on average by 4.7\%. For a country like Estonia, with a comparatively high share of auxiliary consumption, the increase may be as high as 15.4\%, while the opposite is true of Sweden (no change). For Germany, the influence of the auxiliary consumption (5.1\%) is close to that of the EU27-average (in relative terms).

\subsubsection{Electricity Trading} is the exchange of electricity between different regions (e.g., countries, bidding zones), typically in return for money. It separates the net electricity production (NEP) within a grid from the gross electricity consumption (GEC) within that same grid. Note that, unlike depicted in Figure \ref{fig:planttoplug}, losses from storage cycling are not included at this point. They are covered in a paragraph further down.

The approaches that take into account electricity trading fall into two categories: 1) simple, first order trading (SFOT) approaches \cite{Jiusto.2006, Soimakallio.2012, Louis.2014, Stoll.2014, Tamayao.2015, Colett.2016, Roux.2016, Nilsson.2017, Khan.2018, Milovanoff.2018, Vuarnoz.2018, Baumann.2019, Baumgartner.2019, BeloinSaintPierre.2019, Schram.2019, Walzberg.2019, Worner.2019, Pereira.2020, Mehlig.2021, Peters.2022, Scarlat.2022} and 2) approaches based on multi-regional input-output (MRIO) models \cite{Li.2013, Ji.2016, Qu.2017, Clauss.2018, Qu.2018, Chalendar.2019, Clauss.2019, Tranberg.2019, Papageorgiou.2020, Chalendar.2021}. SFOT approaches only consider direct neighbors of the region of interest, the amount of electricity traded with these neighbors, and the EF of these neighbor regions. MRIO approaches (sometimes also referred to as "network-based" or "flow-tracing" approaches) rely on networks and graph theory, where regions are nodes with a specific generation and load, and the connections between regions are edges with specific, bidirectional flows. The MRIO networks include more regions than just the direct neighbors of the region of interest, and therefore-unlike SFOT models-consider higher-order effects as well (e.g., transit flows from region A via region B to region C). Like SFOT models, MRIO models may be based on data with different temporal resolution levels.

Table \ref{tbl:trading} contains data on all studies that provide both EF$_{NEP}$ and EF$_{GEC}$, thus allowing for an analysis that electricity trade has on the resulting grid EF. The table lists the regions covered by these studies, the method (SFOT or MRIO) used to calculate the EF, the production- (EF$_{NEP}$) and consumption-based EF (EF$_{GEC}$), and the absolute and relative difference between the two. For each study, we list minimum, mean and maximum values.

\begin{table*}[ht!]
\small
    \caption{Electricity trading effect in the literature. Listed are the minimum (min), mean and maximum (max) values for the production-based emission factor (EF$_{NEP}$-does not include trade), and the consumption-based emission factor (EF$_{GEC}$-does include trade). Further listed are the absolute (Abs.dif.) and relative (Rel.dif. (\%)) differences between these two values}
    \label{tbl:trading}
    \begin{tabular*}{\textwidth}{@{\extracolsep{\fill}}lcllllll}
        \hline \hline
        Study & Regions & Method & Value & EF$_{NEP}$ & EF$_{GEC}$ & Abs.dif. & Rel.dif. (\%) \\
        \hline \hline
        \multirow{3}{*}{\cite{Jiusto.2006}\textsuperscript{\emph{,a}}} & 51 & \multirow{3}{*}{SFOT} & min & 0 & 143 & -104 & -38.5 \\
         & US & & mean & 169 & 191 & +21.5 & +209 \\
         & States & & max & 291 & 250 & +174 & +8233 \\
         \hline
         \multirow{3}{*}{\cite{Soimakallio.2012}\textsuperscript{\emph{,b}}} & 25 & \multirow{3}{*}{SFOT} & min & 3 & 6 & -109 & -31.1 \\
         & OECD & & mean & 408 & 419 & +10.5 & +54.6 \\
         & Countries & & max & 902 & 880 & +167 & +1113 \\
         \hline
         \multirow{3}{*}{\cite{Ji.2016}\textsuperscript{\emph{,c}}} & 6 & \multirow{3}{*}{MRIO} & min & 398 & 408 & -123 & -12.4 \\
         & CN & & mean & 810 & 795 & -14.7 & -1.2\\
         & Regions & & max & 1197 & 1171 & +36 & +4.4 \\
         \hline
         \multirow{3}{*}{\cite{Qu.2017}\textsuperscript{\emph{,d}}} & 30 & \multirow{3}{*}{MRIO} & min & 219 & 265 & -162 & -12.4 \\
         & CN & & mean & 674 & 672 & -1.3 & +1.2 \\
         & Provinces & & max & 947 & 947 & +229 & +41.4 \\
         \hline
         \multirow{3}{*}{\cite{Qu.2018}\textsuperscript{\emph{,d}}} & 137 & \multirow{3}{*}{MRIO} & min & 0 & 0 & -790 & -49.8 \\
         & World & & mean & 452 & 451 & -1.5 & +48.7 \\
         & Countries & & max & 1587 & 1169 & +391 & +4340 \\
         \hline
         \multirow{3}{*}{\cite{Chalendar.2019}} & 66 & \multirow{3}{*}{MRIO} & min & 0 & 0 & -208 & -36.0 \\
         & US & & mean & 389 & 392 & +3.0 & +2088 \\
         & BA & & max & 1034 & 1003 & +409 & +\numprint{681029} \\
         \hline
         \multirow{3}{*}{\cite{Clauss.2019}} & 9 & \multirow{3}{*}{MRIO} & min & 7 & 9 & -145 & -56.0 \\
         & Scandinav. & & mean & 115 & 83 & -31.3 & +22.2 \\
         & BZ & & max & 461 & 316 & +9.0 & +114 \\
         \hline
         \multirow{3}{*}{\cite{Tranberg.2019}} & 27 & \multirow{3}{*}{MRIO} & min & 11 & 16 & -171 & -38.7 \\
         & European & & mean & 459 & 455 & -3.7 & +6.9 \\
         & Countries & & max & 994 & 947 & +161 & +82.4 \\
         \hline
         \multirow{3}{*}{\cite{Scarlat.2022}\textsuperscript{\emph{,e}}} & 42 & \multirow{3}{*}{MRIO} & min & 19 & 24 & -187 & -28.4 \\
         & European & & mean & 376 & 426 & +50.2 & +28.9 \\
         & Countries & & max & 1101 & 1101 & +236 & +253 \\
        \hline \hline
    \end{tabular*}

    SFOT: simple, first order trading approach;
    MRIO: multi-regional input-output approach;
    BA: balancing area;
    BZ: bidding zone;

    \textsuperscript{\emph{a}}in MtC/GWh;
    \textsuperscript{\emph{b}}for year 2008;
    \textsuperscript{\emph{c}}boundary 3;
    \textsuperscript{\emph{d}}network method;
    \textsuperscript{\emph{e}}net electricity production;
\end{table*}

The table illustrates that in every single study, trade has a moderating effect on the grid EF, i.e. the minimum values are higher and the maximum values are lower after accounting for electricity trade. In some cases, the maximum differences (both absolute and relative) between EF$_{NEP}$ and EF$_{GEC}$ can be very high (e.g., for \citeauthor{Chalendar.2019} \cite{Chalendar.2019}). This typically occurs when regions have little electricity generation compared to the amount of electricity traded with neighboring regions, and when the domestic grid EF (EF$_{NEP}$) differs a lot from the grid EF of its neighbors. In these instances, these outliers may have an impact not just on the maximum, but also on the mean grid EF (as is the case e.g., for \citeauthor{Chalendar.2019} \cite{Chalendar.2019}).

For Germany, which is of special interest for this study, the influence of trade on the grid EF is moderate. Most authors find that trade reduces the grid EF, including \citeauthor{Soimakallio.2012} (-22 absolute / -4.0\% relative difference), \citeauthor{Qu.2018} (-20 / -4.3\%) and \citeauthor{Tranberg.2019} (-10 / -1.9\%). On the contrary, \citeauthor{Scarlat.2022} detect the trade increases the German grid EF (+12 / +2.9\%). This may be due to different time periods covered by the studies, or due to other methodological differences.

\subsubsection{Storage Cycling} losses are the difference between the amount of electrical energy used for charging and the amount generated from discharging a grid-connected storage (e.g., a pumped hydro plant). Storage cycling losses complement electricity trading, in the sense that both together comprise the step from net electricity production (NEP) to gross electricity consumption (GEC).

A total of 17 studies consider the losses due to grid storage cycling \cite{Soimakallio.2012, Messagie.2014, Roux.2016, Kono.2017, Clauss.2018, Collinge.2018, Milovanoff.2018, Baumgartner.2019, BeloinSaintPierre.2019, Clauss.2019, MunneCollado.2019, Tranberg.2019, Worner.2019, Peters.2022, Scarlat.2022, Unnewehr.2022, Blizniukova.2023}. For some studies, it is not clear which efficiencies (losses), EF or share of electricity they assume for pumped hydro \cite{Soimakallio.2012, BeloinSaintPierre.2019, MunneCollado.2019, Worner.2019, Scarlat.2022, Unnewehr.2022}. Two studies indicate only the share of electricity contributed by pumped hydro \cite{Messagie.2014, Collinge.2018}, but not the losses or EF they assume. Two other studies assume an efficiency of 70\% for pumped hydro \cite{Roux.2016, Peters.2022}. \citeauthor{Baumgartner.2019} list efficiencies and losses, but they are in the supplementary information that is hidden behind a paywall. Six studies list the EF used for electricity coming from pumped hydro plants \cite{Kono.2017, Clauss.2018, Clauss.2019, Milovanoff.2018, Tranberg.2019, Blizniukova.2023}.

Since it is not always clear how these EF for pumped hydro are used in calculating grid EF, it is worth noting that if done incorrectly, double-counting may occur. The electricity used in pumping mode (storage charging) has already been accounted for in gross/net electricity production. Only the difference between the electricity used for storage charging and the electricity generated from storage discharging (pumping losses) should be accounted for, not the total amount of electricity discharged. Depending on how EF for pumped hydro (e.g., from the Ecoinvent database) factor into grid EF calculations, the resulting grid EF may be too high. For example, \citeauthor{Kono.2017} uses an EF for pumped hydro in Germany of 951.52 g \ch{CO2}e/kWh, which is higher than the grid-average. If this EF is used for pumped hydro just like the EF for all other sources (e.g., biomass, wind, coal), without considering double-counting, the resulting grid EF is too high.

Only one study quantifies the effect of pumped hydro losses, listed in table \ref{tbl:pump}. It shows that for Germany (the only country assessed in the study), the losses make up consistently 1.2-1.3 \% of the total EF.

\begin{table}[ht!]
\small
    \caption{Effect of pumped hydro storage losses in the literature, from \citeauthor{Blizniukova.2023} \cite{Blizniukova.2023}. The values refer to EF$_{Total}$ (EF$_{wP}$) and EF$_{Total}$ - $\Delta$EF$_{P}$ (EF$_{woP}$) for Germany in the Supplementary Information of the study}
    \label{tbl:pump}
    \begin{tabular*}{0.48\textwidth}{@{\extracolsep{\fill}}lllll}
        \hline
        Year & EF$_{woP}$ & EF$_{wP}$ & Abs.dif. & Rel.dif. (\%) \\
        \hline
        2017 & 493 & 499 & +6 & +1.2 \\
        2018 & 475 & 481 & +6 & +1.3 \\
        2019 & 412 & 417 & +5 & +1.2 \\
        2020 & 372 & 377 & +5 & +1.3 \\
        \hline
    \end{tabular*}
\end{table}

\subsubsection{Transformation \& Distribution} describe the conversion of electricity from one voltage level to another, and the transmission from producers to consumers. These steps are accompanied by dissipation of heat to the environment, i.e. losses. Following the logic illustrated in Figure \ref{fig:planttoplug}, T\&D losses separate gross electricity consumption (GEC) from net electricity consumption (NEC). In our review, we assume that study authors who do not mention T\&D losses in their study do not consider them.

T\&D losses are typically calculated using a fixed value that can be derived e.g., from the difference between generation and load (possibly including trade). Most studies considering grid losses fall into this category \cite{Jiusto.2006, Soimakallio.2012, Tamayao.2015, Kono.2017, Mills.2017, Baumann.2019, Baumgartner.2019, Rupp.2019, Worner.2019, Mehlig.2022, Peters.2022, Blizniukova.2023}. Some studies differentiate between losses in different regions (e.g., countries) \cite{Jiusto.2006, Li.2013, Colett.2016, Milovanoff.2018, Tranberg.2019, Fleschutz.2021} or at different grid voltage levels \cite{Collinge.2018, Papageorgiou.2020, Scarlat.2022}.

Table \ref{tbl:grid_losses} summarizes the findings from those studies that quantify T\&D losses. The losses are identical to the relative difference between EF$_{GEC}$ and EF$_{NEC}$.

\begin{table}[ht!]
\small
    \caption{Effect of transformation and distribution (T\&D) losses in the literature}
    \label{tbl:grid_losses}
    \begin{tabular*}{0.48\textwidth}{@{\extracolsep{\fill}}lll}
        \hline
        Study & Region & T\&D losses (\%) \\
        \hline
        \citeauthor{Mills.2017} \citeyear{Mills.2017} & AU & 10 \\
        \citeauthor{Li.2013} \citeyear{Li.2013} & CN & 2 \\
        \citeauthor{Kono.2017} \citeyear{Kono.2017} & DE & 4 \\
        \citeauthor{Baumgartner.2019} \citeyear{Baumgartner.2019} & DE & 3.9 \\
        \citeauthor{Rupp.2019} \citeyear{Rupp.2019} & DE & 4.15 \\
        \citeauthor{Blizniukova.2023} \citeyear{Blizniukova.2023} & DE & 4.4-4.9 \\
        \citeauthor{Milovanoff.2018} \citeyear{Milovanoff.2018} & EU & 1.9-2.9\textsuperscript{\emph{T}} \\
        \citeauthor{Fleschutz.2021} \citeyear{Fleschutz.2021} & EU & 3.8-15.3 \\
        \citeauthor{Roux.2016} \citeyear{Roux.2016} & FR & 3\textsuperscript{\emph{T}} / 6\textsuperscript{\emph{D}} \\
        \citeauthor{Papageorgiou.2020} \citeyear{Papageorgiou.2020} & SE & 1.97\textsuperscript{\emph{HV}} / 0.3\textsuperscript{\emph{MV}} / 3.12\textsuperscript{\emph{LV}} \\
        \citeauthor{Mehlig.2022} \citeyear{Mehlig.2022} & UK & 7.5 \\
        \citeauthor{Jiusto.2006} \citeyear{Jiusto.2006} & US & 8.5-10.5  \\
        \hline
    \end{tabular*}
    
    \textsuperscript{\emph{T}}Transformation;
    \textsuperscript{\emph{D}}Distribution;
    
    \textsuperscript{\emph{HV}}High voltage;
    \textsuperscript{\emph{MV}}Medium voltage;
    \textsuperscript{\emph{LV}}Low voltage.
\end{table}

A review of the studies that quantify the contribution of grid losses to the EF show that these losses lie in the range of 1.9 to 15.3\% (see Table \ref{tbl:grid_losses}). The two\% losses for the Chinese grid stated by \citeauthor{Li.2013} appear relatively low, and probably only include transformation losses (without distribution losses), even though the study does not state this explicitly. In the European cross-country study by \citeauthor{Fleschutz.2021}, Finland features the lowest grid losses (3.8\%), and Serbia the highest (15.3\%). Studies that distinguish losses by voltage levels or between transmission and distribution grid losses show that low voltage/distribution grid losses tend to be higher than high voltage/transmission grid losses \cite{Roux.2016, Papageorgiou.2020}. However, this assessment rests on sparse data. T\&D losses in Germany are in the range of 3.9-4.9\% \cite{Fleschutz.2021, Kono.2017, Baumgartner.2019, Rupp.2019, Blizniukova.2023}, and 2.1\% for losses in the high-voltage grid only \cite{Milovanoff.2018}.

\subsubsection{Temporal Resolution} describes the temporal reference frame used to calculate a temporally averaged EF. The length of the reference frame depends primarily on the user needs and the temporal resolution of the available data needed for EF calculation. Most of the studies included in our review calculate hourly EF, while some rely on a coarser resolution of one year \cite{Jiusto.2006, Soimakallio.2012, Li.2013, Colett.2016, Ji.2016, Qu.2017, Qu.2018, Scarlat.2022}, and others on finer resolutions of up to 30 minutes \cite{Mills.2017, Khan.2018, Mehlig.2021} or even 15 minutes \cite{Rupp.2019, Worner.2019, Blizniukova.2023}. Some studies that include multiple countries in their analysis harmonize data by choosing the lowest temporal resolution available for all countries of interest \cite{Milovanoff.2018, Noussan.2020, Fleschutz.2021}.

Several studies contain EF at multiple temporal resolution levels. They allow for an assessment of the impact that the choice of temporal resolution has on the resulting EF, as documented in Table \ref{tbl:temp_res}. All but one study compare yearly (y) and hourly (h) resolution levels, while \citeauthor{Kono.2017} additionally provides monthly (m) EF. We limit our comparative assessment in this study to the coarsest (yearly, EF$_y$) and finest (hourly, EF$_h$) spatial resolution level.

Table \ref{tbl:temp_res} provides an overview of all studies that quantify the effect that changing the temporal resolution of the grid EF has on the resulting emissions. It lists the effect that changing the temporal resolution from annual to hourly has on the electricity consumer's electricity-related GHG emissions. Besides the temporal resolution, all else remains constant (e.g., no change in the consumer load profile). The relative difference is calculated by dividing the difference between the emission results at an hourly resolution and the results at an annual resolution by the results and an annual resolution. The table contains all study results from the literature review that allow for such a quantiative comparison.

\begin{table}[ht!]
\small
    \caption{Temporal resolution effect in the literature. The relative difference describes the change in GHG emissions when applying an hourly instead of an annual grid EF.}
    \label{tbl:temp_res}
    \begin{tabular*}{0.48\textwidth}{@{\extracolsep{\fill}}lll}
        \hline
        Study & Region & Rel.dif. (\%) \\
        \hline
        \cite{Roux.2016} & France & +36 \\
        \cite{KopsakangasSavolainen.2017} & Finland & -5.7 \\
        \cite{Milovanoff.2018} & Spain & -28 \\
        \cite{Milovanoff.2018} & France & -2.6 \\
        \cite{Vuarnoz.2018} & Austria & +6.2 \\
        \cite{Vuarnoz.2018} & France & -4.9 \\
        \cite{Vuarnoz.2018} & Switzerland & +5.0 \\
        \cite{BeloinSaintPierre.2019} & Switzerland & +69 \\
        \cite{Walzberg.2019} & Canada, Ontario & +2.8 \\
        \cite{Mehlig.2022} & United Kingdom & +4.2 \\
        \cite{Peters.2022} & Spain & -7.6 \\
        \hline
    \end{tabular*}
\end{table}

The largest relative difference can be observed for the study by \citeauthor{BeloinSaintPierre.2019}, who find that the temporal resolution effect can increase emissions by 69\% for the case of Switzerland. The smallest effect is observed by \citeauthor{Milovanoff.2018} for the case of France, with an emission reduction of 2.6\%. No study quantifies the effect for the case of Germany.

\subsubsection{Other Methodological Aspects} for calculating grid EF besides those mentioned above exist as well. Most importantly, the choice of spatial \cite{Colett.2016, Chalendar.2019, Chalendar.2021, Tamayao.2015, Qu.2017} and technological resolution \cite{Fleschutz.2021, Braeuer.2020, Chalendar.2019, Chalendar.2021, Unnewehr.2022} stand out.

The spatial resolution describes the geographical reference frame chosen for calculating a grid EF. The most typical spatial resolution is that of a country, but it is also possible to calculate a grid EF for a state, a province, a region, a bidding zone, a balancing area, a grid region or a continent. \citeauthor{Colett.2016} discuss this choice in detail, including the (dis)advantages for each reference frame, and develop their own approach that spans several spatial resolution levels ("Nested approach"). Due to the grid topology in most places with interconnected grids, there is no "right" choice for choosing a spatial resolution level. Usually, the spatial resolution level is determined by the data availability-most data is available at country level. Due to this fact, and since data on how the spatial resolution influences grid EF, we do not pursue this aspect any further in our study.

The technological resolution describes the level of differentiation between individual generators (power plants). The most common technological resolution is one where all generators are grouped together by the type of primary energy input (e.g., wind, coal, biomass). Some further distinctions are often made, e.g., between offshore and onshore wind, or between hard coal and lignite. However, at this "generator type" level, no distinction is made between individual generators. Therefore, e.g., all hard coal fueled power plants and offshore wind turbines are lumped together into a homogeneous mass. A notably higher level of technological resolution is the "generator" level. At this level, each generator has individual technical aspects that influence the final result. These aspects may include different efficiencies due to age or different emissions due to the fuel type used (example: different chemical compositions of lignite used in East and West German lignite power plants \cite{Huber.2021}). We do not include an investigation on the technological resolution in our study, mostly because the data needed for such an assessment is sparse, and because the effort is relatively high.

Besides the spatial and technological resolution, there are several other methodological niche aspects not covered by our review. These aspects are either covered only in very few studies, their impact on emission accounting results is deemed negligible, the data availability to consider these aspects is insufficient, or a combination of these circumstances applies. Some of these aspects include the ramping (up/down) of generators \cite{Baumgartner.2019, Braeuer.2020}, the age of generators \cite{Clauss.2019}, the operational restrictions of power plants (e.g., minimum load \cite{Milovanoff.2018}) and the uncertainty regarding the characterization factors of the different greenhouse gases \cite{Pereira.2020}.

\subsection{Extended Summary of the Literature Review}
\label{sec:literature_results_summary_appendix}

Table \ref{tbl:lit_aspects_summary} summarizes which methodological aspects are considered in previous studies.

\begin{table*}[ht!]
\small
    \caption{Summary of methodological aspects in primary research articles. The count indicates how many studies addressed a specific aspect or which choice is made with respect to that aspect. A checkmark (\checkmark) indicates that an aspect is addressed, a dash (-) that it is not}
    \label{tbl:lit_aspects_summary}
    \begin{tabular*}{\textwidth}{@{\extracolsep{\fill}}llllllllll}
        \hline
        Metric & Sys. bound. & Co-gen. & AP & Aux. & Trade & Stor. & T\&D & Temp.res. \\
        \hline
        14 \ch{CO2} & 13 OP & 3 EL100 & 5 \checkmark & 9 \checkmark & 31 \checkmark & 17 \checkmark & 24 \checkmark & 8 y \\
        34 GWP100 & 37 LC & 1 EN & 43 - & 39 - & 17 - & 31 - & 24 - & 34 h \\
        1 GWP20 & & 2 IEA & & & & & & 6 <h \\
        & & 2 EX & & & & & & \\
        & & 6 other & & & & & & \\
        & & 35 - & & & & & & \\
        \hline
    \end{tabular*}

    AP: auto-producers;
    T\&D: transformation \& distribution losses;
    GWP: global warming potential;
    OP: operational;
    LC: life cycle;
    EL100: all emissions allocated to electricity;
    EN: allocation by energy content;
    IEA: allocation method used by the International Energy Agency;
    EX: allocation by exergy;
    
\end{table*}

\section{Extended Methodology}
\label{sec:methodology_SI}

This section expands on the methodology and data covered in the methodology section of the main article.

\subsection{Extended Methodology: Input Data}
\label{sec:methodology_SI_data}

\subsubsection{ENTSO-E Data}
\label{sec:methodology_input_entsoe}
ENTSO-E provides data on the \textit{Aggregated Generation per Type (AGPT)}, i.e. the net electricity production per individual energy carrier/generation technology \cite{ENTSOE.2014} at a temporal resolution of up to 15 minutes. At the same temporal resolution, ENTSO-E also documents \textit{Physical Flows (PF)}, i.e. the flow of electrical energy between bidding zones/countries \cite{ENTSOE.2014}.

\paragraph{Aggregated Generation per Type (AGPT).} The data downloaded from the ENTSO-E Transparency Portal (e.g., via FTP client) is available at a monthly resolution. The datasets for the four years of interest (2019-2022) are merged together into one combined dataframe. A filter is applied to the dataframe to limit it to the data and locations of interest (Germany and its neighboring countries). The data is then interpolated to 15 minute time steps, to take into account e.g., data gaps. The method chosen for the interpolation is "forward fill", i.e. a missing entry will be filled with the last available entry for that location and production type (type of generator). A plausibility check ensures that the total sum of the numerical data did not change too much with the interpolation step.

\paragraph{Physical Flows (PF).} The data manipulation steps for PF are identical to those for AGPT. Only the flows from and to Germany are considered.

\subsubsection{Eurostat Data}
\label{sec:methodology_input_eurostat}
Eurostat provides annual data on primary energy (PE) input, gross electricity production (GEP) and net electricity production (NEP), electricity imports and exports, inputs and outputs from pumped hydro storage and distribution losses (from T\&D). We assume Eurostat to be more accurate than ENTSO-E data, due to reasons listed e.g., in studies by \citeauthor{Hirth.2018} and \citeauthor{Worner.2019} \cite{Hirth.2018, Worner.2019}. In some cases, we therefore used it to normalize the annual sums of high resolution data from ENTSO-E.

All data from Eurostat is freely available on the Eurostat website. We use the {\tt xlsx} data format, but they can also be downloaded as {\tt csv} or {\tt tsv} files.

\paragraph{Primary Energy Demand (PE).} The dataset contains the PE demand for main-activity producers (MAP) and auto-producers (AP). These can be separated into electricity-only (EL)  generators and combined heat and power (CHP) units. The data is filtered for years (2019-2022), location (Germany) and production types of interest (those that are potentially relevant for Germany). \textit{NaN} values are replaced with zeros. If for any combination of year and production type a PE value does not exist, it is added as a zero value.

\paragraph{Gross Electricity Production (GEP).} The dataset contains the GEP for main-activity producers (MAP) and auto-producers (AP). These can be separated into electricity-only (EL)  generators and combined heat and power (CHP) units. Furthermore, it contains data on gross heat production (GHP) for CHP units (both MAP and AP). The data manipulation steps for GEP (\& GHP) are identical to those for PE.

\paragraph{Net Electricity Production (NEP).} The dataset contains the NEP for main-activity producers (MAP) and auto-producers (AP). These can be separated into electricity-only (EL)  generators and combined heat and power (CHP) units. The data manipulation steps for NEP are mostly identical to those for PE and GEP. The dataset differs primarily in that it uses different (fewer) categories for production types for NEP than for both PE and GEP. Therefore, these categories need to be matched.

\paragraph{Imports and Exports.} The dataset contains the amount of electricity traded within Europe. The data is filtered for the years (2019-2022) of interest, and for imports and exports to and from Germany.

\paragraph{Pumped Hydro (Storage Cycling).} The dataset for the electricity input into pumped hydro storage (charging) comes from Eurostat. It contains data for both pure and mixed plants. Both pure and mixed plants' input are added together into one single column. The other data manipulation steps for are identical to the other Eurostat datasets (filter etc.).

The dataset for the electricity output from pumped hydro storage (discharging) also comes from Eurostat, and is included in the NEP dataset (pumped hydro being a production type within that dataset). The date for pumped hydro is separated from the rest of the NEP data and manipulated similarly to the pumped hydro input.

\paragraph{Distribution Losses (T\&D).} The dataset on the distribution losses due to transformation and distribution (T\&D) contain all losses in Europe. The data manipulation steps for NEP (\& GHP) are identical to the other Eurostat datasets (filter etc.).

\subsubsection{UBA Data}
\label{sec:methodology_input_uba}
The German Umweltbundesamt (UBA)-the German Federal Environmental Agency-provides emission factors per individual production type (EF$_{PE}$), referenced to its primary energy content \cite{UBA.2021}. In addition, it publishes reference efficiency values for electricity and heat generation (eta\_ref), which are used for emission allocation in CHP units.

\paragraph{Emission Factor per Production Type.} The data is extracted from the report "Emissionsbilanz erneuerbarer Energieträger" (2020) by the Umweltbundesamt \cite{UBA.2021}. It contains \ch{CO2}, \ch{CH4} and \ch{N2O} upstream and operational EF per production type (e.g., biomass, wind onshore etc.), referenced to the PE input. They are mapped to the production type categories used by Eurostat. For those categories for which the UBA reference does not provide an EF, the EF of similar energy carriers are used. These assumptions are documented in (ref Excel UBA comments).

\paragraph{Reference Efficiencies} The UBA provides reference efficiencies of 0.4 for electricity production and 0.8 for heat production, which are used in the UBA allocation method ("Finnish Method") for allocating emissions to electricity and heat in CHP units. These values are used in the calculation steps on co-generation of heat for the UBA allocation method \cite{UBA.2021}.

\subsubsection{IPCC Data}
\label{sec:methodology_input_ipcc}
The Sixth IPCC Assessment Report (AR6) \cite{IPCC.2021} provides characterization factors (CF) used to calculate the global warming potential (GWP) of various greenhouse gases. In our assessment, we include \ch{CO2}, \ch{CH4} and \ch{N2O}. The IPCC AR6 CF are listed in Table \ref{tbl:method_data_cf}.

\begin{table}[ht!]
\small
    \caption{Characterization factors for different impact metrics, from the IPCC AR6 \cite{IPCC.2021}}
    \label{tbl:method_data_cf}
    \begin{tabular*}{0.48\textwidth}{@{\extracolsep{\fill}}llll}
        \hline
        \multirow{2}{*}{Substance} & \multicolumn{3}{c}{Impact metric}  \\
        & \ch{CO2} & GWP20 & GWP100 \\
        \hline
        \ch{CO2} & 1 & 1 & 1 \\
        \ch{CH4} & 0 & 81.2 & 27.9 \\
        \ch{N2O} & 0 & 273 & 273 \\
        \hline
    \end{tabular*}
\end{table}

\subsubsection{Further Notes on Input Data}
\label{sec:methodology_input_other}
All the input data can be found in \path{1_data\1_raw} in the folder structure for the code and data accompanying this article, organized by data source (e.g., Eurostat).

\subsection{Extended Methodology: Mapping}
\label{sec:methodology_SI_mapping}

\subsubsection{NEP to GEP Categories}
\label{sec:methodology_map_nep_gep}
The NEP dataset distinguishes between fewer production type categories than the GEP (and the PE) dataset. We want to use the more detailed GEP set of categories. To do so, we match the data as listed in Table \ref{tbl:method_map_nep_gep}.

\begin{table*}[ht!]
\small
    \caption{Mapping Eurostat NEP data to GEP categories}
    \label{tbl:method_map_nep_gep}
    \begin{tabular*}{\textwidth}{@{\extracolsep{\fill}}ll}
        \hline \hline
        NEP category & GEP category  \\
        \hline \hline
        Hydro + Tide, wave, ocean & Hydro \\
        \hline
        Geothermal & Geothermal \\
        \hline
        Wind & Wind \\
        \hline
        \multirow{2}{*}{Solar} & Solar photovoltaic \\
        & Solar thermal \\
        \hline
        Nuclear fuels and other fuels n.e.c. & Nuclear heat \\
        \hline
        \multirow{27}{*}{\vtop{\hbox{\strut Combustible fuels + Other fuels n.e.c. -}\hbox{\strut heat from chemical sources + Other fuels n.e.c.}}} & Anthracite \\
        & Coking coal \\
        & Other bituminous coal \\
        & Sub-bituminous coal \\
        & Lignite \\
        & Coke oven coke \\
        & Gas coke \\
        & Patent fuel \\
        & Brown coal briquettes \\
        & Coal tar \\
        & Manufactured gases \\
        & Peat and peat products \\
        & Oil shale and oil sands \\
        & Natural gas \\
        & {\vtop{\hbox{\strut Oil and petroleum products}\hbox{\strut (excluding biofuel portion)}}} \\
        & Primary solid biofuels \\
        & Charcoal \\
        & Pure biogasoline \\
        & Blended biogasoline \\
        & Pure biodiesels \\
        & Blended biodiesels \\
        & Pure bio jet kerosene \\
        & Blended bio jet kerosene \\
        & Other liquid biofuels \\
        & Biogases \\
        & Non-renewable waste \\
        & Renewable municipal waste \\
        \hline \hline
    \end{tabular*}
\end{table*}

The NEP values matched to the GEP categories are calculated using the logic from Equation \ref{eqn:method_map_nep}:

\begin{equation}
    \label{eqn:method_map_nep}
    NEP_{xj} = \frac{GEP_{xj}}{\sum_{i=1}^{n} GEP_{xi}} \cdot NEP_y
\end{equation}

Where NEP and GEP are the values for net and gross electricity production, respectively. y is the NEP category that encompasses all GEP categories x$_{1}$...x$_{n}$. n is the total number of GEP categories, and j is a specific GEP category of interest ($j \in [1...n]$).

Besides mapping NEP to GEP values, we also add values for net heat production (NHP). Eurostat does not provide these data, so we assume the NHP to be identical to GHP.

The calculation steps for mapping NEP to GEP categories are documented in section {\tt 1.2.2.1 Map NEP to PE/GEP categories} of the Jupyter Notebook attached to this study. The GEP and NEP data required for the calculations in Equation \ref{eqn:method_map_nep} are provided by Eurostat, as described in Section \ref{sec:methodology_input_eurostat}.

\subsubsection{Eurostat Data to ENTSO-E Categories}
\label{sec:methodology_map_eurostat_entsoe}

This section describes how Eurostat data is mapped to ENTSO-E production type categories.

\paragraph{Generation.} Eurostat provides data on PE, GEP, NEP, and ENTSO-E on NEP (at a higher temporal resolution than ENTSO-E - named \textit{Aggregated Generation per Production Type, AGPT}). These two entities, however, do not use the same categories when describing fuels/energy carriers. Eurostat uses the Standard International Energy Product Classification (SIEC) scheme, while ENTSO-E uses a proprietary type of classification. Table \ref{tbl:method_map_eurostat_entsoe} indicates how both categorization schemes are matched with one another.

\begin{table*}[ht!]
\small
    \caption{Mapping Eurostat data to ENTSO-E categories}
    \label{tbl:method_map_eurostat_entsoe}
    \begin{tabular*}{\textwidth}{@{\extracolsep{\fill}}lllll}
        \hline \hline
        Eurostat category & ENTSO-E category \\
        \hline
        Anthracite & \multirow{2}{*}{Fossil Hard coal} \\
        Other bituminous coal & \\
        \hline
        Coking coal & \multirow{9}{*}{Fossil Brown coal/Lignite} \\
        Sub-bituminous coal & \\
        Lignite & \\
        Coke oven coke & \\
        Gas coke & \\
        Patent fuel & \\
        Brown coal briquettes & \\
        Coal tar & \\
        Peat and peat products & \\
        \hline
        Manufactured gases & Fossil Coal-derived gas \\
        \hline
        Oil shale and oil sands & \multirow{2}{*}{Fossil Oil} \\
        \vtop{\hbox{\strut Oil and petroleum products}\hbox{\strut (excluding biofuel portion)}}& \\
        \hline
        \multirow{2}{*}{Natural gas} & Fossil Gas \\
        & Other \\
        \hline
        \multirow{2}{*}{Hydro} & Hydro Run-of-river and poundage \\
        & Hydro Water Reservoir \\
        \hline
        Geothermal & Geothermal \\
        \hline
        \multirow{2}{*}{Wind} & Wind Offshore \\
        & Wind Onshore \\
        \hline
        Solar photovoltaic & \multirow{2}{*}{Solar} \\
        Solar thermal & \\
        \hline
        Tide, wave, ocean & Marine \\
        \hline
        Primary solid biofuels & \multirow{4}{*}{Biomass} \\
        Biogases & \\
        Renewable municipal waste & \\
        Other liquid biofuels & \\
        \hline
        Charcoal & \multirow{7}{*}{Other renewable} \\
        Pure biogasoline & \\
        Blended biogasoline & \\
        Pure biodiesels & \\
        Blended biodiesels & \\
        Pure bio jet kerosene & \\
        Blended bio jet kerosene & \\
        \hline
        Non-renewable waste & Waste \\
        \hline
        Nuclear heat & Nuclear \\
        \hline \hline
    \end{tabular*}
\end{table*}

The matching of categories is done in the same manner as the matching of NEP and GEP categories, as described in Section \ref{sec:methodology_map_nep_gep}. The only difference is that (unlike for matching NEP and GEP categories) for Eurostat and ENTSO-E categories, two type of matching logics apply: one-to-many (e.g., "Natural gas" --> "Fossil Gas", "Other") and many-to-one (e.g., "Anthracite", "Other bituminous coal" --> "Fossil Hard coal"). One-to-many matching is done in the same manner as described in Equation \ref{eqn:method_map_nep}. Many-to-one matching is even simpler, as the values for two categories that are combined into one are simply added together.

Note that the Eurostat values include only electricity from main-activity producers (MAP), as it is assumed that the same is true for ENTSO-E data. The mapping logic depicted in Table \ref{tbl:method_map_eurostat_entsoe} is applied to PE, GEP and NEP, so that all of these datasets are organized according to the same categories of fuels/energy carriers.

The calculation steps for mapping Eurostat data to ENTSO-E categories are documented in section {\tt 1.2.2.2 Map NEP, GEP and PE to ENTSO-E categories} of the Jupyter Notebook attached to this study. The PE and GEP data required for the mapping steps described above are provided by Eurostat, as described in Section \ref{sec:methodology_input_eurostat}. The NEP data, mapped to GEP categories, comes from the previous mapping step described in Section \ref{sec:methodology_map_nep_gep}. The ENTSO-E categories are provided by ENTSO-E, as described in Section \ref{sec:methodology_input_entsoe}.

\paragraph{Imports and Exports.} The annual data for imports and exports provided by Eurostat does not match the annual sums of the high resolution data for physical flows (PF) by ENTSO-E, which tracks cross-border flows. Since we assume Eurostat data to be more accurate than ENTSO-E data, i.e. better representing the actual electricity trade between countries (bidding zones), we scale the ENTSO-E PF data to the Eurostat data on imports and exports.

We first sum up the ENTSO-E PF data by year. Then, we calculate the ratio of Eurostat annual import (export) flow to ENTSO-E annual PF (per year). Finally, we multiply this ratio with the high resolution PF data, to result in corrected PF data.

The calculation steps for scaling ENTSO-E import and export data to Eurostat values are documented in section {\tt 1.2.2.3 Map (scale) imports \& exports} of the Jupyter Notebook attached to this study. The annual data are provided by Eurostat, as described in Section \ref{sec:methodology_input_eurostat}. The hourly data are provided by ENTSO-E, as described in Section \ref{sec:methodology_input_entsoe}.

\subsubsection{UBA Emission Factors to ENTSO-E Categories}
\label{sec:methodology_map_uba}
The UBA emission factor input data is categorized by Eurostat categories. To map it to ENTSO-E categories, we use a similar method (and the same mapping logic) as described in Section \ref{sec:methodology_map_eurostat_entsoe}. For all many-to-one mappings, the mapping of UBA emission factors from Eurostat to ENTSO-E categories follows the following logic (see Equation \ref{eqn:method_map_uba_nto1}):

\begin{equation}
    \label{eqn:method_map_uba_nto1}
    EF_{PE,y} = \frac{\sum_{i=1}^{n} (PE_{xi} \cdot EF_{PE,xi})}{\sum_{i=1}^{n} PE_{xi}}
\end{equation}

Where EF$_{PE}$ is the emission factor referenced to primary energy for a certain production type, and PE is the primary energy input for a certain production type. y is the ENTSO-E category that encompasses all Eurostat categories x$_{1}$...x$_{n}$. n is the total number of Eurostat categories that make up the ENTSO-E category y.

For one-to-many mappings, the case is simpler: EF$_{PE}$ for the (many) ENTSO-E categories is identical to the (one) Eurostat category.

The calculation steps for mapping UBA emission factors to ENTSO-E categories are documented in section {\tt 1.3.2 Map (Eurostat --> ENTSO-E)} of the Jupyter Notebook attached to this study. The UBA emission factors following the Eurostat categorization scheme are provided by the UBA, as described in Section \ref{sec:methodology_input_uba}. The ENTSO-E categories are provided by ENTSO-E, as described in Section \ref{sec:methodology_input_entsoe}. The PE data following the Eurostat categorization scheme are provided by Eurostat, as described in Section \ref{sec:methodology_input_eurostat}.

\subsubsection{Other Data Mapping \& Manipulation Aspects}
\label{sec:methodology_map_other}
Some other relevant aspects of input data manipulation that come before the actual calculation step are listed here.

\paragraph{Missing UBA EF.}
The UBA does not provide EF for all Eurostat categories (production types). For some production types, the emission factor had to be estimated. Details are listed in the column labeled "reference, comment" in the input file \path{ef_pe.xlsx}.

\paragraph{Diverging CF.}
The UBA uses characterization factors (CF) from the Fifth IPCC Assessment Report (AR5) \cite{IPCC.2013}. In our own calculations, we use more recent CF from the Sixth Assessment report \cite{IPCC.2021}. Therefore, our calculated GWP100 values differ slightly from those calculated in the UBA report \cite{UBA.2021}.

\paragraph{Biomass EF.}
The UBA calculates individual EF for sub-production types of solid, liquid and gaseous biomass-based electricity production \cite{UBA.2021}. We summarize them into aggregated EF for solid, liquid and gaseous biomass-based electricity production, using CF from the IPCC AR6 \cite{IPCC.2021}. More details can be found in the input data file \path{ef_pe_biopower.xlsx}. The resulting aggregated EF are used in \path{ef_pe.xlsx}.

\subsection{Extended Methodology: Calculations}
\label{sec:methodology_SI_calculation}

\subsubsection{Generator Level}
\label{sec:methodology_calc_gen}
The calculations at the generator level address the methodological aspects \textit{Impact metric}, \textit{System boundaries}, \textit{Co-generation of heat}, \textit{Auxiliary consumption} and \textit{Auto-producers}.

\paragraph{Impact Metric.}
In a first step, the EF from UBA are transformed to reflect the chosen impact metric (\ch{CO2}, GWP20, GWP100). An EF that reflects a specific impact metric is calculated as follows (eq. \ref{eqn:ef_metric}):

\begin{equation}
    \label{eqn:ef_metric}
    \begin{split}
    EF_{m} = \\
    \sum_{i} (CF_{i,m} \cdot EF_{PE,i}) = \\
    CF_{CO_2,m} \cdot EF_{PE,CO_2} + CF_{CH_4,m} \cdot EF_{PE,CH_4} + CF_{N_2O,m} \cdot EF_{PE,N_2O}
    \end{split}
\end{equation}

Where $EF_{m}$ is the EF for the impact metric m, $CF_{i,m}$ is the characterization factor for a substance $i$ and impact metric $m$, and $EF_{PE,i}$ is the emission factor referenced to the primary energy content of the fuel/energy carrier for a substance $i$.

CF for substances other than the three listed above exist as well \cite{IPCC.2021}. However, they are not used, since the UBA reference lists EF only for \ch{CO2}, \ch{CH4} and \ch{N2O}. The characterization factors for the impact metrics \ch{CO2}, GWP20 and GWP100 are listed in Table \ref{tbl:method_data_cf}.

The impact metric calculation steps are documented in section {\tt 2.1 Impact metric} of the Jupyter Notebook attached to this study. The EF$_{PE}$ data that is used here is the result of the mapping process as described in Section \ref{sec:methodology_map_uba}. The CF data is provided by the IPCC, as described in Section \ref{sec:methodology_input_ipcc}.

\paragraph{System Boundaries.}
An EF can reflect the system boundaries of choice: operational (OP), upstream (UP) or life cycle (LC) emissions. They relate to one another as described in Equation \ref{eqn:ef_lc}

\begin{equation}
    \label{eqn:ef_lc}
    EF_{LC} = EF_{OP} + EF_{UP}
\end{equation}

Where $EF_{LC}$ is the life cycle EF, $EF_{OP}$ the operational EF and $EF_{OP}$ the upstream EF.

In section {\tt 2.2 System boundaries} of the Jupyter Notebook attached to this study the system boundaries calculation steps are documented. The EF data that is used here is the result of the previous calculation step on impact metrics.

\paragraph{Co-generation of Heat.}
For all generation units that produce not only electricity, but also heat (which is used e.g., for district heating), one has to decide how to allocate the emissions to each of these outputs. Multiple methods exist for allocating emissions in combined heat and power (CHP) units. The emission factor for a set of generators, consisting of both electricity-only (EL) and CHP units, is calculated as follows (see Equation \ref{eqn:ef_cogen}):

\begin{equation}
    \label{eqn:ef_cogen}
    EF_{GEP} =\frac{EF_{PE,EL} \cdot PE_{EL} + x \cdot EF_{PE,CHP} \cdot PE_{CHP}}{GEP_{EL} + GEP_{CHP} + y}
\end{equation}

Where $EF_{GEP}$ is the EF for a set of generators, consisting of both EL and CHP units, referenced to the gross electricity production (GEP) from that set of generators. $EF_{PE,EL}$ is the EF of the EL units within that set, referenced to the primary energy (PE) input, and $EF_{PE,CHP}$ is the EF of the CHP units within that set, referenced to the PE input. $PE_{EL}$ is the amount of PE going into EL units within that set, and $PE_{CHP}$ is the amount of PE going into CHP units within that set. $GEP_{EL}$ is the GEP from EL units, $GEP_{CHP}$ is the GEP from CHP units. GHP is gross heat production (from CHP units only). $x$ and $y$ are variables that depend on the allocation method, and are listed in Table \ref{tbl:cogen_allocation_xy}.

\begin{table}[ht!]
\small
    \caption{Variables $x$ and $y$, to be applied to Equation \ref{eqn:ef_cogen}, for different allocation methods}
    \label{tbl:cogen_allocation_xy}
    \begin{tabular*}{0.48\textwidth}{@{\extracolsep{\fill}}lll}
        \hline
        Method & $x$ & $y$ \\
        \hline
        EL100 & 1 & 0 \\
        TH100 & 0 & 0 \\
        EN & $\frac{GEP_{CHP}}{GEP_{CHP} + GHP}$ & 0 \\
        EX & 1 & $(1 - \frac{T_0}{T}) \cdot GHP$ \\
        IEA & $1 - \frac{GHP}{eta_{th,iea}}$ & 0 \\
        UBA & 1 & $\frac{eta_{el,uba}}{eta_{th,uba}} \cdot GHP$ \\
        \hline
    \end{tabular*}

    $eta_{th,iea}$: reference efficiency for heat producer, IEA method (= 0.9);
    
    $eta_{el,uba}$: reference efficiency for electricity producer, UBA method (= 0.4);

    $eta_{th,uba}$: reference efficiency for heat producer, UBA method (= 0.8);

    T$_0$: reference surrounding temperature (assumed to be 282 K);

    T: reference CHP output temperature (assumed to be 363 K);
    
\end{table}

The allocation calculation steps for co-generation of heat are documented in section {\tt 2.3 Co-generation of heat} of the Jupyter Notebook attached to this study. The EF$_{PE}$ data that is used here is the result of the previous calculation step on system boundaries. The PE and GEP (GHP) data that is used here is the result of the mapping process described in Section \ref{sec:methodology_map_eurostat_entsoe}. The reference efficiencies eta$_i$ are provided by the UBA \cite{UBA.2021} and the IEA \cite{IEA.2021}. T and T$_0$ are based on assumptions.

\paragraph{Auxiliary Consumption.}
The first part of the changes/losses along the path from gross electricity production (GEP) to net electricity consumption (NEC)-auxiliary consumption of generators-occurs at the generator-level, while the other losses/changes (electricity trading, storage cycling losses, T\&D losses) occur at the grid level. Due to this fact, they are addressed in a separate sub-section (\ref{sec:methodology_calc_grid}) of the methodology description.

An EF considering auxiliary consumption, and thus referenced to net electricity production (NEP), is calculated as described in Equation \ref{eqn:ef_nep}:

\begin{equation}
    \label{eqn:ef_nep}
    EF_{NEP} = \frac{GEP \cdot EF_{GEP}}{NEP}
\end{equation}

Where $EF_{NEP}$ is the EF referenced to NEP, $GEP$ is the GEP, $EF_{GEP}$ is the EF referenced to NEP, and $NEP$ is the NEP.

The calculation steps for considering auxiliary consumption are documented in section {\tt 2.4 Auxiliary consumption} of the Jupyter Notebook attached to this study. The EF$_{GEP}$ data that is used here is the result of the previous calculation step on co-generation of heat. The GEP and NEP data that is used here is the result of the mapping process described in Section \ref{sec:methodology_map_eurostat_entsoe}.

\paragraph{Auto-producers.}
Depending on the scope and goal of an assessment, one may want to include auto-producers (AP) into the calculation of a grid EF or exclude them. A grid EF for a grid with both main-activity producers (MAP, connected to the grid) and AP (not connected to the grid) is calculated as follows (see Equation \ref{eqn:ef_ap}):

\begin{equation}
    \label{eqn:ef_ap}
    EF_{GEP,a} = \frac{GEP_{MAP} \cdot EF_{GEP,MAP} + x \cdot GEP_{AP} \cdot EF_{GEP,AP}}{GEP_{MAP} + y \cdot GEP_{AP}}
\end{equation}

Where $EF_{GEP,a}$ is the EF for a set of generators consisting of both MAP and AP units, referenced to the GEP from that set of generators, for an auto-producer rule $a$ (see below). $EF_{GEP,MAP}$ is the EF of only the MAP within that set, referenced to the GEP. $EF_{GEP,AP}$ is the EF of only the AP within that set, referenced to the GEP. $GEP_{MAP}$ is the GEP by MAP, and $GEP_{AP}$ is the GEP by AP. $x$ and $y$ are variables that depend on the auto-producer rule, and are listed in Table \ref{tbl:map_ap_xy}.

\begin{table}[ht!]
\small
    \caption{Variables $x$ and $y$, to be applied to Equation \ref{eqn:ef_ap}, for different auto-producer rules}
    \label{tbl:map_ap_xy}
    \begin{tabular*}{0.48\textwidth}{@{\extracolsep{\fill}}lll}
        \hline
        Rule & $x$ & $y$ \\
        \hline
        MAPonly & 0 & 0 \\
        APem & 0 & 1 \\
        APen & 1 & 0 \\
        MAP\&AP & 1 & 1 \\
        \hline
    \end{tabular*}

    MAPonly: emissions and electricity from main-activity producers only,
    
    APem: emissions from all generators, electricity from main-activity producers only,
    
    APen: emissions from main-activity producers only, electricity from all generators,
    
    MAP\&AP: emissions and energy from all generators;
    
\end{table}

For calculating an EF referenced to NEP instead of GEP, all the instances of GEP in Equation \ref{eqn:ef_ap} have to be replaced with NEP.

The calculation steps for considering auto-producers are documented in section {\tt 2.5 Auto-producers} of the Jupyter Notebook attached to this study. The EF$_{GEP}$ data that is used here is the result of the previous calculation step on auxiliary consumption. The GEP data that is used here is the result of the mapping process described in Section \ref{sec:methodology_map_eurostat_entsoe}.

\subsubsection{Grid Level}
\label{sec:methodology_calc_grid}
This section describes the calculations at the grid level, which address the methodological aspects \textit{Temporal resolution}, \textit{Electricity trading}, \textit{Storage cycling} and\textit{ Transformation \& distribution}. From here on, the EF calculated refer to the grid mix, while the EF in Section \ref{sec:methodology_calc_gen} refer to individual production types.

\paragraph{Temporal Resolution.}
At the transition from generator to grid level, the decision is made which temporal resolution is required for the grid EF. The EF per production type is assumed to remain constant within a year, while the grid EF changes with the production shares of the individual production types. These shares change much more frequently. Therefore, the temporal resolution is introduced at this point, at the transition from production type to grid EF. Equation \ref{eqn:ef_t} describes how the temporal dimension is included in EF calculation:

\begin{equation}
    \label{eqn:ef_t}
    \begin{split}
    EF_{HR} =
    \frac{EM_{HR}}{GEP_{HR}} =
    \frac{\sum_i EM_{HR,i}}{\sum_i GEP_{HR,i}} = \\
    \frac{\sum_i (GEP_{HR,i} \cdot EF_{HR,i})}{\sum_i GEP_{HR,i}} = 
    \frac{\sum_i (GEP_{HR,i} \cdot EF_{LR,i})}{\sum_i GEP_{HR,i}} \\
    \end{split}
\end{equation}

Where $EF_{HR}$ is the high resolution grid EF, $EM_{HR}$ are the high resolution grid emissions, $GEP_{HR}$ is the high resolution GEP summed up across all production types $i$, $EM_{HR,i}$ are the high resolution emissions from production type $i$, $GEP_{HR,i}$ is the high resolution GEP from production type $i$, $EF_{HR,i}$ is the high resolution EF for production type $i$, and $EF_{LR,i}$ is the low resolution EF for production type $i$. As noted above, $EF_{LR,i}$ and $EF_{HR,i}$ are assumed to be identical.

For calculating an EF referenced to NEP instead of GEP, all the instances of GEP in Equation \ref{eqn:ef_t} have to be replaced with NEP.

In section {\tt 2.6 Temporal resolution} of the Jupyter Notebook attached to this study, the temporal resolution calculation steps are documented. The EF$_{LR,i}$ data that is used here is the result of the previous calculation step on auto-producers. The GEP data that is used here is a combination of the high resolution energy data provided by ENTSO-E (see Section \ref{sec:methodology_input_entsoe}) and the mapping process described in Section \ref{sec:methodology_map_eurostat_entsoe}.

\paragraph{Electricity Trading.}
An EF that considers electricity trading effectively starts out with a production-based EF, subtracts the carbon flows out of the region of interest, and adds the carbon flows into the same region. The carbon flow is calculated from the production-based grid EF for the region of origin multiplied with the amount of electricity transferred from the region of origin to the destination region. The result is a consumption-based EF, which can be calculated as described in Equation \ref{eqn:ef_trade}:

\begin{equation}
    \label{eqn:ef_trade}
    \begin{split}
    EF_{CONS} = 
    \frac{EM_{CONS}}{GEC} =
    \frac{EM_{DOM} + EM_{IMP} - EM_{EXP}}{GEP_{DOM} + IMP - EXP} = \\
    \frac{EF_{DOM} \cdot GEP_{DOM} + \sum_n (EF_{DOM,n} \cdot IMP_n) - EF_{DOM} \cdot \sum_n EXP_n}{GEP_{DOM} + \sum_n IMP_n - \sum_n EXP_n} \\
    \end{split}
\end{equation}

Where $EF_{CONS}$ is the consumption-based grid EF for a region of interest, $EM_{CONS}$ are the total, consumption-based emissions from the region of interest, $GEC$ is the gross electricity consumption within the region of interest, $EM_{DOM}$ are the domestic, production-based total emissions within the region of interest, $EM_{IMP}$ are the imported emissions into the region of interest, $EM_{EXP}$ are the exported emissions from the region of interest, $GEP_{DOM}$ is the domestic gross electricity production within the region of interest, $IMP$ is the amount of electricity imported into the region of interest, $EXP$ is the amount of electricity exported from the region of interest, $EF_{DOM}$ is the domestic, production-based grid EF for the region of interest, $EF_{DOM,n}$ is the production-based grid EF for neighbor region $n$, $IMP_n$ is the amount of electricity imported into the region of interest from neighbor region $n$ and $EXP_n$ is the amount of electricity exported to the neighbor region $n$ from the region of interest. $EF_{DOM,n}$ can be calculated just like $EF_{DOM}$, following the previous steps described in this methodology section-but replacing the GEP values with those of the neighbor region.

Note that this method only considers influences on the EF of a region from direct neighbors. Knock-on effects, such as imports/exports from/to second-order neighbors, are not taken into account (SFOT approach, see Section \ref{sec:literature_results_detail}.) Note also that the results differ depending on the temporal resolution of choice. If the variables in Equation \ref{eqn:ef_trade} are calculated at an hourly level, the resulting EF will be different from one calculated based on annual variables, as the temporal variability of the grid EFs, generation and imports/exports is "smoothed out". In our calculations, we consider trade at a high temporal resolution (15 minutes).

For calculating a consumption-based grid EF referenced to NEP instead of GEP, all the instances of GEP in Equation \ref{eqn:ef_trade} have to be replaced with NEP.

The electricity trade-related calculation steps are documented in section {\tt 2.7 Electricity trading} of the Jupyter Notebook attached to this study. The EF$_{DOM}$ data that is used here is the result of the previous calculation step on the temporal resolution. The GEP data that is used here is a combination of the high resolution generation data (AGPT) provided by ENTSO-E (see Section \ref{sec:methodology_input_entsoe}) and the mapping process described in Section \ref{sec:methodology_map_eurostat_entsoe}. The high resolution import and export data (EXP$_n$ and IMP$_n$) is a combination of the high resolution trade data (PF) provided by ENTSO-E (see Section \ref{sec:methodology_input_entsoe}) and the mapping process described in Section \ref{sec:methodology_map_eurostat_entsoe}.

\paragraph{Storage Cycling.}
To calculate an EF that reflects gross electricity consumption within a grid, the cycling losses of grid storage (pumped hydro) have to be considered. Equation \ref{eqn:ef_cyc} describes how this EF is calculated:

\begin{equation}
    \label{eqn:ef_cyc}
    EF_{grid,PH} = 
    \frac{GEP_{grid} + PH_{in}}{GEP_{grid} + PH_{out}} \cdot EF_{grid}
\end{equation}

Where $EF_{grid,PH}$ is the grid EF with consideration of pumped hydro storage, $GEP_{grid}$ is the GEP within that grid, $PH_{in}$ is the amount of electricity consumed by pumped hydro plants (charging), $PH_{out}$ is the amount of electricity produced by pumped hydro plants (discharging), and $EF_{grid}$ is the grid EF without considering pumped hydro storage.

Note that the results differ depending on the temporal resolution of choice. If the variables in Equation \ref{eqn:ef_cyc} are calculated at an hourly level, the resulting EF will be different from one calculated based on annual variables, as the temporal variability of the grid EFs, generation and imports/exports is "smoothed out". In our calculations, we consider pumped hydro storage cycling at a low temporal resolution (yearly).

The pumped hydro storage cycling calculation steps are documented in section {\tt 2.8 Storage cycling (pumped hydro)} of the Jupyter Notebook attached to this study. The EF$_{grid}$ data that is used here is the result of the previous calculation step on electricity trading. The GEP data that is used here is the result of the mapping process described in Section \ref{sec:methodology_map_eurostat_entsoe}. The data on pumped hydro input and output (PH$_{in}$ and PH$_{out}$) are provided by Eurostat, as described in Section \ref{sec:methodology_input_eurostat}.

\paragraph{Transformation \& Distribution.}
In a final calculation step, the EF is transformed to reflect the dissipation losses from transformation and distribution in the grid. It is calculated as described in Equation \ref{eqn:ef_td}:

\begin{equation}
    \label{eqn:ef_td}
    EF_{NEC} = 
    \frac{GEC}{NEC} \cdot EF_{GEC} = 
    \frac{GEC}{GEC - L_{TD}} \cdot EF_{GEC}
\end{equation}

Where $EF_{NEC}$ is the grid EF referenced to the net electricity consumption (NEC), $GEC$ is the gross electricity production, $NEC$ is the net electricity consumption, $EF_{GEC}$ is the grid EF referenced to the net electricity consumption (GEC, not including T\&D losses), and $L_{TD}$ are the relative transformation and distribution (T\&D) losses.

The T\&D calculation steps are documented in section {\tt 2.9 Grid losses (T\&D)} of the Jupyter Notebook attached to this study. The EF$_{GEC}$ and GEC data that is used here is the result of the previous calculation step on storage cycling. on distribution losses (L$_{TD}$) are provided by Eurostat, as described in Section \ref{sec:methodology_input_eurostat}.

\section{Extended Results}
\label{sec:results_SI}

This section expands on the results described in the results section of the main article.

\subsection{Extended Results: Data Sampling}
\label{sec:results_SI_sampling}

All violin plots in this study plot a sample of 10\% of the full dataset of \numprint{323149824} data points, i.e. \numprint{32314982} data points. The sample data is drawn randomly from across all rows (time steps) and columns (ways to calculate a grid EF). We plot only a sample of the data, and not the full dataset, to reduce computational burden. The full dataset takes up around 2.5 GB of storage space ({\tt pickle} file format). In some instances, we ran into limits with respect to working memory on a computer with 16 GB RAM when trying to plot the full dataset. To ensure that others can replicate our work, we decided to reduce the computational burden by relying on a sample of the full dataset for plotting.

To ensure that the sample dataset represents the full dataset well, we compare key statistical indicators for both, and calculate the relative difference. Comparing the mean, median and standard deviation of the full dataset and the sample dataset, the relative difference between them is 0.0045\%, 0.0053\% and 0.0047\%, respectively. We deem these differences to be small enough to be acceptable.

\subsection{Extended Results: Individual Effects}
\label{sec:results_SI_effects}

\subsubsection{System Boundaries}
\label{sec:results_ind_bound}

\begin{figure}[ht!]
\centering
    \includegraphics[keepaspectratio, width=\linewidth]{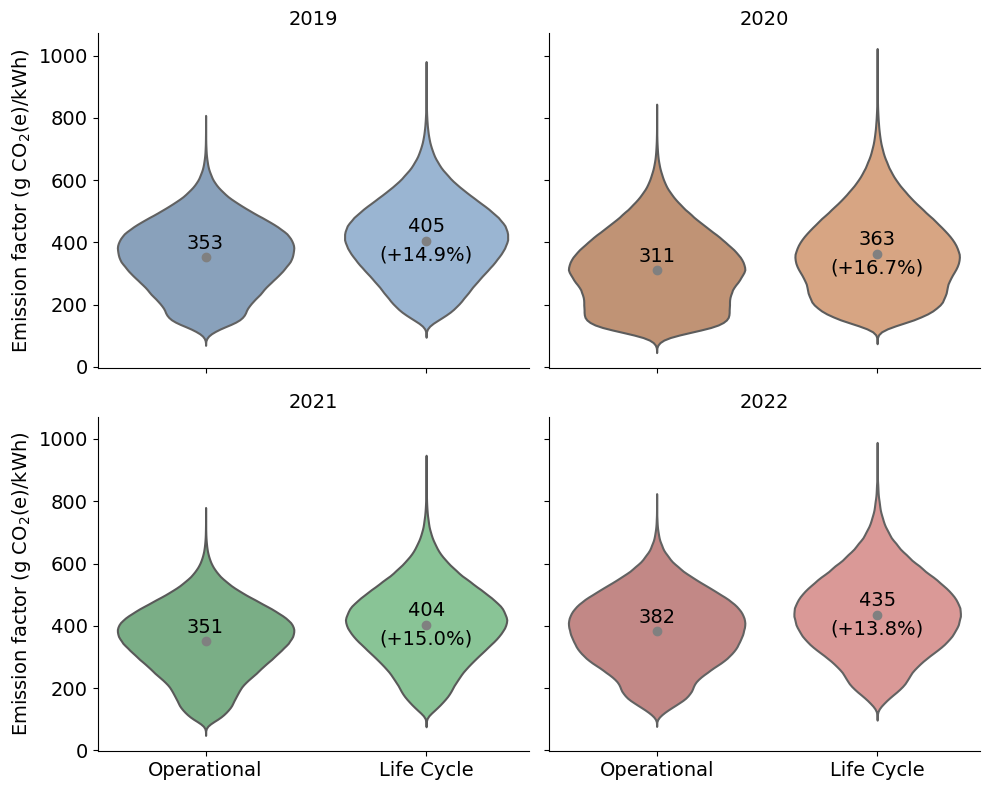}
    \caption{System boundary effect, disaggregated by year-the distribution of grid EF values for the system boundaries \textit{Operational} and \textit{Life Cycle} and the years 2019-2022. Labeled are the mean values for all data points, and the relative difference of the mean compared to the mean of the first impact metric (\textit{Operational}), for each year individually.}
    \label{fig:result_effect_boundaries_multi}
\end{figure}

In Figure \ref{fig:result_effect_boundaries_multi}, the effect of varying the system boundaries is plotted. The choices for the system boundaries are \textit{Operational (OP)} and \textit{Life Cycle (LC)}. The displayed values represent the mean for each choice, and the relative difference between the mean for a specific choice and the mean for the reference choice (in this case, \textit{OP}).

Figure \ref{fig:result_effect_boundaries_multi} indicates that when disaggregated by year, a \textit{LC}-EF is on average 13.8-16.7\% larger than a \textit{OP}-EF. This observation matches our expectations, since a \textit{LC}-EF covers more emission sources than a \textit{OP}-EF. The variation of the effect between years, which is smallest in 2022 (13.8\%) and largest in 2020 (16.7\%), may be explained by the baseline effect. While the absolute difference between the \textit{OP}-EF and the \textit{LC}-EF is larger in 2022 than in 2020 (53 vs. 52 g/kWh), the relative difference is smaller in 2022 than in 2020 due to the higher baseline value in 2022 compared to 2020 (382 vs. 311 g/kWh).

\subsubsection{Co-generation of Heat}
\label{sec:results_ind_chp}

\begin{figure}[ht!]
\centering
    \includegraphics[keepaspectratio, width=\linewidth]{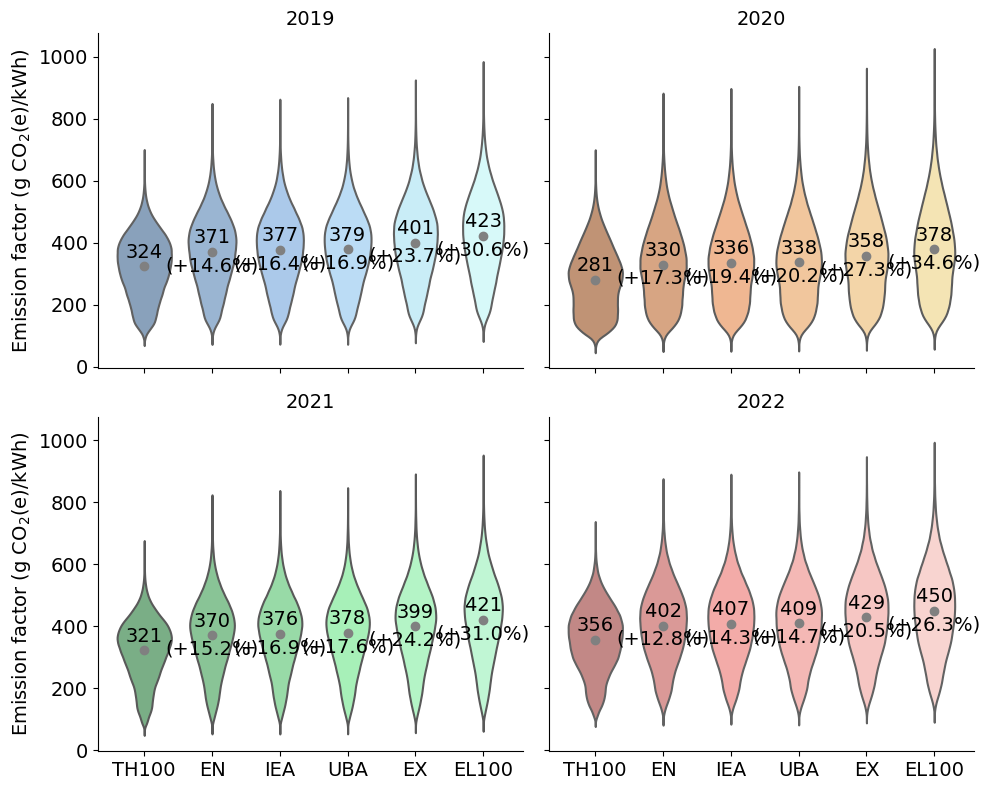}
    \caption{Co-generation of heat allocation effect, disaggregated by year-the distribution of grid EF values for the allocation methods \textit{TH100}, \textit{EN}, \textit{IEA}, \textit{UBA}, \textit{EX}, and \textit{EL100} and the years 2019-2022. Labeled are the mean values for all data points, and the relative difference of the mean compared to the mean of the first allocation method (\textit{TH100}), for each year individually.}
    \label{fig:result_effect_cogen_multi}
\end{figure}

In Figure \ref{fig:result_effect_cogen_multi}, the effect of varying the allocation method for CHP units is plotted. The choices for the allocation method are \textit{100\% to heat (TH100)}, \textit{By Energy Content (EN)}, \textit{IEA Method (IEA)}, \textit{UBA Method (UBA)}, \textit{By Exergy Content (EX)} and \textit{100\% to Electricity (EL100)}. The displayed values represent the mean for each choice, and the relative difference between the mean for a specific choice and the mean for the reference choice (in this case, \textit{TH100}).

Figure \ref{fig:result_effect_cogen_multi} indicates that when disaggregated by year, a \textit{EN}-EF is on average 12.8-17.3\% larger than a \textit{TH100}-EF, a \textit{IEA}-EF is on average 14.3-19.4\% larger than a \textit{TH100}-EF, a \textit{UBA}-EF is on average 14.7-20.2\% larger than a \textit{TH100}-EF, a \textit{EX}-EF is on average 20.5-27.3\% larger than a \textit{TH100}-EF, and a \textit{EL100}-EF is on average 26.3-34.6\% larger than a \textit{TH100}-EF.

The variation across years of the absolute values and the relative differences have already been discussed in the main article. The consistent ranking of the allocation methods, with \textit{TH100} yielding the lowest values, and \textit{EL100} yielding the highest values, match our expectations based on the internal logic of the allocation methods. \textit{TH100} is an extreme case which allocates all emissions from CHP plants to the heat and no emissions to the electricity that these plants produce. Naturally, this will yield relatively low electricity-EF for a fleet of generators that contains CHP plants. The opposite is true of the \textit{EL100} method: all CHP emissions are allocated to the electricity produced, and none to the heat. The \textit{EN} method yields relatively low electricity-EF, since it treats both heat and electricity as equal, irrespective of the fact that electricity is a more "valuable" form of energy, and typically has a lower conversion efficiency than heat when produced from a chemical energy carrier (e.g., coal). The \textit{IEA} and the \textit{UBA} method are both substitution methods that do consider the differences between conversion efficiencies. They take into account reference efficiencies for alternative methods of either producing heat ($\eta_{ref,th} = 0.8$ for UBA, $0.9$ for IEA) or electricity ($\eta_{ref,el} = 0.4$ for both UBA and IEA), and subtract these emissions for alternative heat / electricity production processes from the total CHP plant emissions. The small difference between the \textit{UBA} and the \textit{IEA} values can be explained by the different $\eta_{ref,th}$ values employed by these methods. The \textit{EX} method assesses the value of the two outputs of CHP plants, heat and electricity, based on the exergy of the output streams. The exergy of electricity is equal to one, while the exergy of heat depends on the both the temperature of the CHP heat output and the temperature of the environment. In this study, we assume the former to be 363 K (90°C) and the latter 282 K (9°C). The exergy of the heat output increases when the difference between the two values increases, which would mean a higher heat output temperature or a lower environmental temperature. An increase of the exergy of the heat output leads to decrease of the respective electricity-EF: the higher the "value" (=exergy) of an output, the larger the emission share assigned to it.

\subsubsection{Auxiliary Consumption}
\label{sec:results_ind_aux}

\begin{figure}[ht!]
\centering
    \includegraphics[keepaspectratio, width=\linewidth]{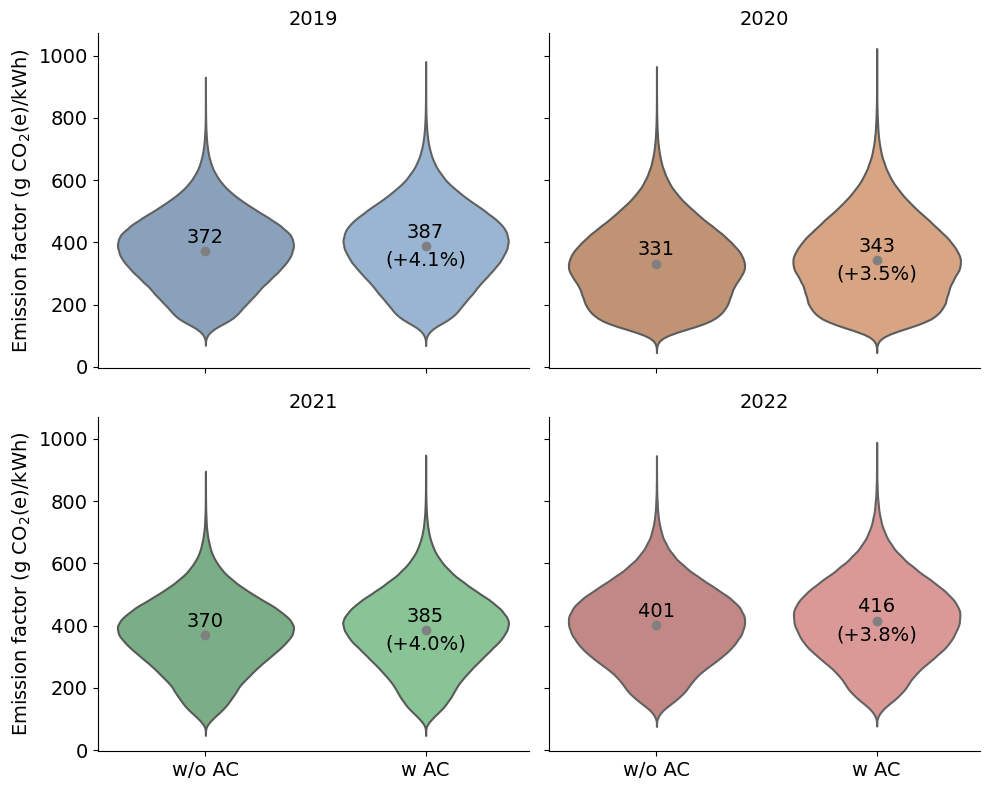}
    \caption{Auxiliary consumption effect, disaggregated by year-the distribution of grid EF values for EF without (\textit{w/o AC}) and with auxiliary consumption (\textit{w AC}), and the years 2019-2022. Labeled are the mean values for all data points, and the relative difference of the mean compared to the mean of the first EF (\textit{w/o AC}), for each year individually.}
    \label{fig:result_effect_aux_multi}
\end{figure}

In Figure \ref{fig:result_effect_aux_multi}, the effect of considering auxiliary consumption is plotted. The choices for considering auxiliary consumption are \textit{Without Auxiliary Consumption (w/o AC)} and \textit{With Auxiliary Consumption (w AC)}. The displayed values represent the mean for each choice, and the relative difference between the mean for a specific choice and the mean for the reference choice (in this case, \textit{w AC}).

Figure \ref{fig:result_effect_aux_multi} indicates that when disaggregated by year, an EF \textit{w AC} is on average 3.5-4.1\% larger than an EF \textit{w/o AC}. This effect meets our expectations---if more losses are included, then the resulting EF is higher.

\subsubsection{Auto-producers}
\label{sec:results_ind_ap}

\begin{figure}[ht!]
\centering
    \includegraphics[keepaspectratio, width=\linewidth]{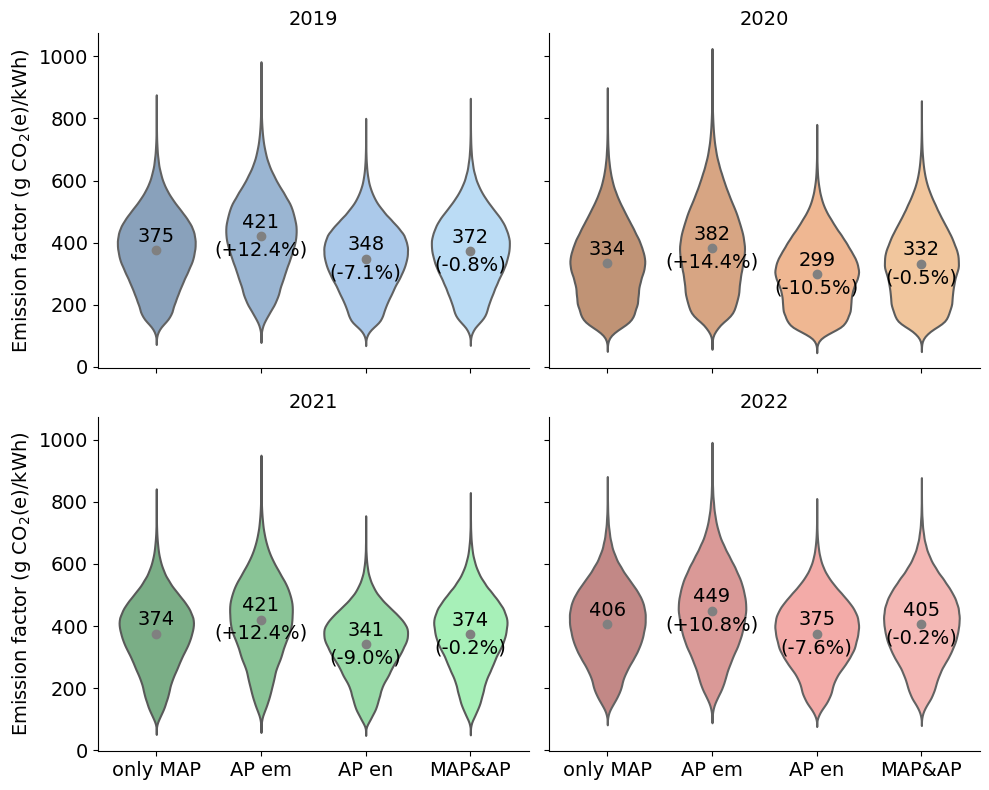}
    \caption{Auto-producer effect, disaggregated by year-the distribution of grid EF values for different auto-producer inclusion rules \textit{only MAP}, \textit{AP emissions}, \textit{AP energy}, \textit{MAP \& AP}, and the years 2019-2022. Labeled are the mean values for all data points, and the relative difference of the mean compared to the mean of the first rule (\textit{only MAP}), for each year individually.}
    \label{fig:result_effect_ap_multi}
\end{figure}

In Figure \ref{fig:result_effect_ap_multi}, the effect of considering auto-producers is plotted. The choices for considering auto-producers are \textit{Electricity \& Emissions from MAP only (only MAP)}, \textit{Electricity from MAP, Emissions from MAP \& AP (AP emissions)}, \textit{Electricity from MAP \& AP, Emissions from MAP (AP energy)} and \textit{Electricity \& Emissions from MAP \& AP (MAP\&AP)}. The displayed values represent the mean for each choice, and the relative difference between the mean for a specific choice and the mean for the reference choice (in this case, \textit{only MAP}).

Figure \ref{fig:result_effect_ap_multi} indicates that when disaggregated by year, an \textit{AP emissions}-EF is on average 10.8-14.4\% larger than a \textit{only MAP}-EF, an \textit{AP energy}-EF is on average 7.1-10.5\% smaller than a \textit{only MAP}-EF, and a \textit{MAP\&AP}-EF is on average 0.2-0.8\% smaller than a \textit{only MAP}-EF.

It appears worth noting that \textit{AP emissions} and \textit{AP energy} are cases of incorrect data matching. Normally, one would either consider both electricity and emissions from auto-producers (\textit{MAP\&AP}), or omit them (\textit{only MAP}). Considering only one of the two, electricity or emissions, may however happen in cases where datasets don't match. One might e.g., calculate a grid EF from an emission dataset which contains both MAP and AP, and an electricity dataset which contains only MAP. Since this case cannot be entirely ruled out in practice, we include them in our assessment.

\subsubsection{Electricity Trading}
\label{sec:results_ind_trade}

\begin{figure}[ht!]
\centering
    \includegraphics[keepaspectratio, width=\linewidth]{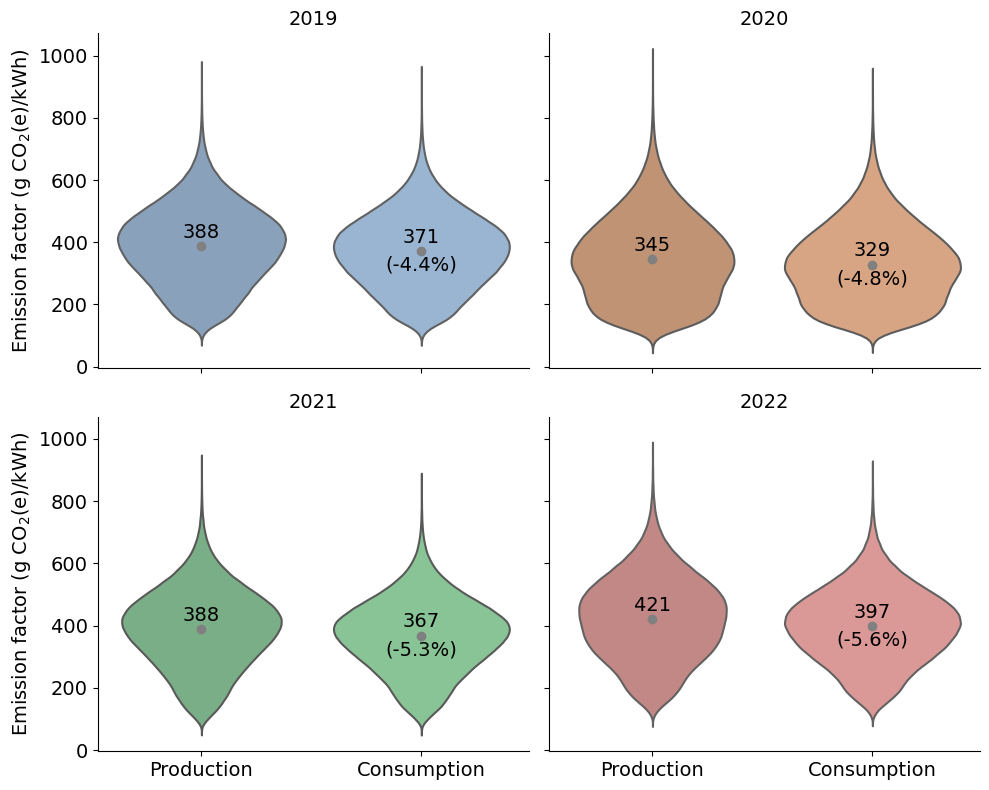}
    \caption{Electricity trade effect, disaggregated by year-the distribution of grid EF values for the perspectives \textit{Production} and \textit{Consumption}, and the years 2019-2022. Labeled are the mean values for all data points, and the relative difference of the mean compared to the mean of the first allocation perspective (\textit{Production}), for each year individually.}
    \label{fig:result_effect_trade_multi}
\end{figure}

In Figure \ref{fig:result_effect_trade_multi}, the effect of considering electricity trade is plotted. The choices for considering electricity trade are \textit{Without Trade (Production)} and \textit{With Trade (Consumption)}. The displayed values represent the mean for each choice, and the relative difference between the mean for a specific choice and the mean for the reference choice (in this case, \textit{Production}).

Figure \ref{fig:result_effect_trade_multi} indicates that when disaggregated by year, a \textit{Consumption}-EF is on average 4.4-5.6\% smaller than a \textit{Production}-EF.

\begin{figure}[!ht]
\centering
    \includegraphics[keepaspectratio, width=1.1\linewidth]{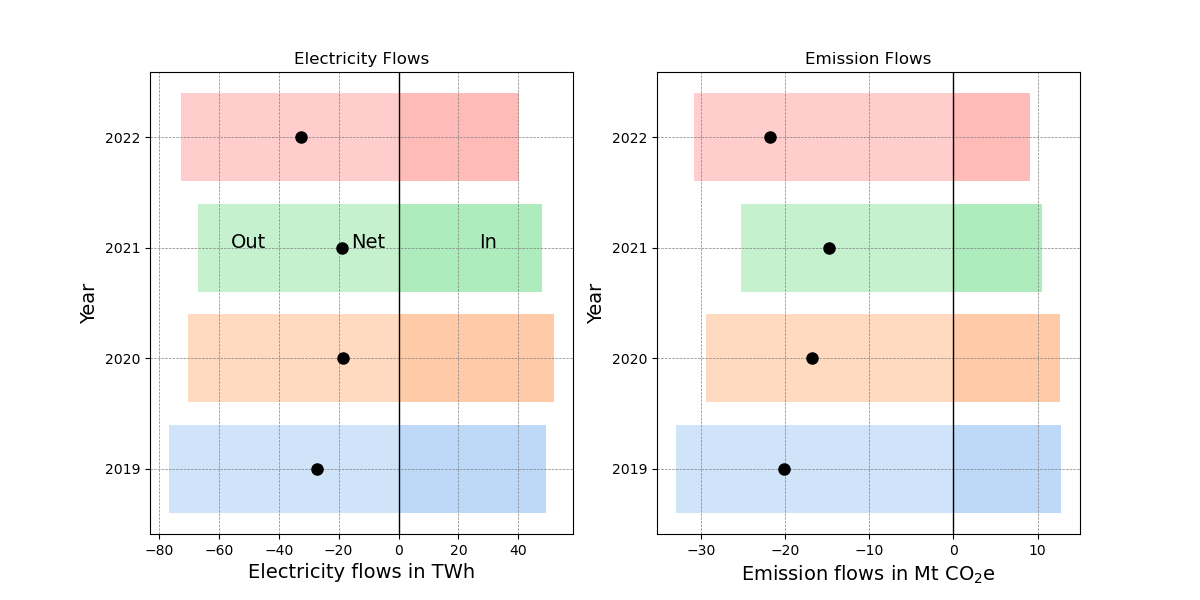}
    \caption{Trade balance.}
    \label{fig:result_effect_trade_balance}
\end{figure}

These results indicate that either Germany exports more electricity than it imports, that the exported electricity has a higher EF on average than the imported electricity, or both. To shed more light on this aspect, Figure \ref{fig:result_effect_trade_balance} depicts the net electricity (left) and emission (right) flow balance for the years 2019-2022.

Both the net electricity and net emission flow balance is negative for every year depicted. The ratio of outflows (bars to the left of the vertical line at zero) to inflows (bars to the right) is higher for the emission flows than for electricity flows, indicating that the exported electricity was on average more emission-intensive than the imported electricity.

\subsubsection{Storage Cycling}
\label{sec:results_ind_cyc}

\begin{figure}[ht!]
\centering
    \includegraphics[keepaspectratio, width=\linewidth]{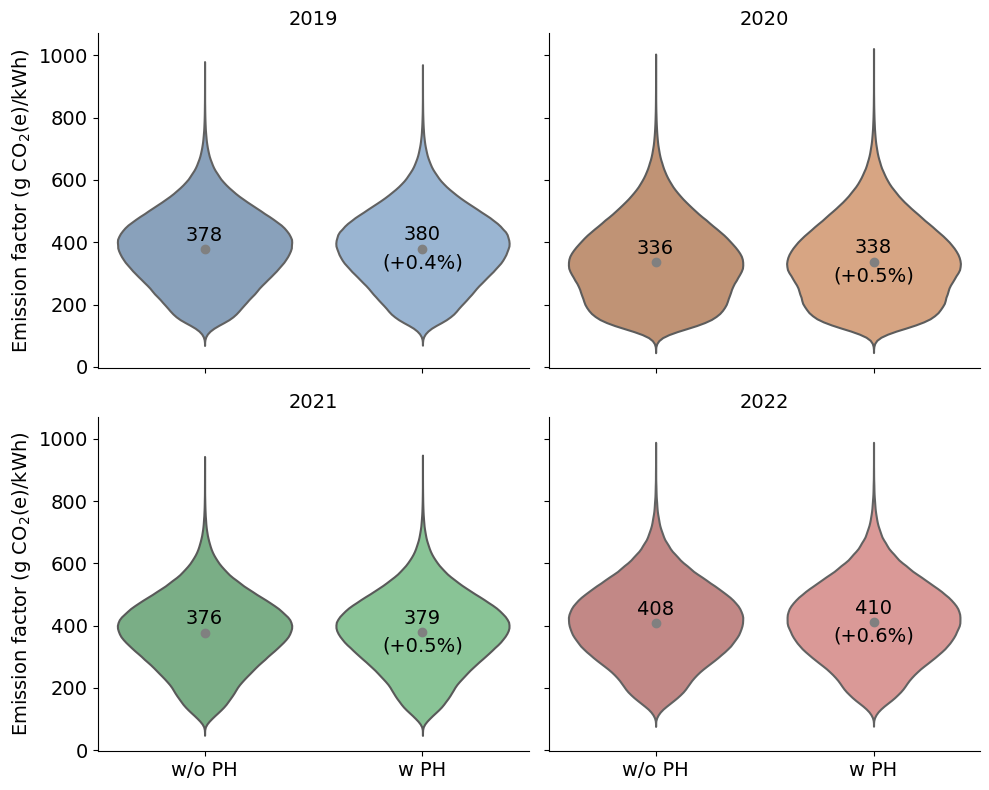}
    \caption{Storage cycling effect, disaggregated by year-the distribution of grid EF values for EF without (\textit{w/o PH}) and with pumped hydro storage cycling (\textit{w PH}), and the years 2019-2022. Labeled are the mean values for all data points, and the relative difference of the mean compared to the mean of the first EF (\textit{w/o PH}), for each year individually.}
    \label{fig:result_effect_ph_multi}
\end{figure}

In Figure \ref{fig:result_effect_ph_multi}, the effect of considering pumped hydro storage cycling is plotted. The choices for considering pumped hydro storage cycling are \textit{Without Pumped Hydro (w/o PH)} and \textit{With Pumped Hydro (w PH)}. The displayed values represent the mean for each choice, and the relative difference between the mean for a specific choice and the mean for the reference choice (in this case, \textit{w/o PH}).

Figure \ref{fig:result_effect_ph_multi} indicates that when disaggregated by year, an EF \textit{w PH} is on average 0.4-0.6\% larger than an EF \textit{w/o PH}. Just as with auxiliary consumption (cf. section \ref{sec:results_ind_aux}), this effect meets our expectations---if more losses are included, then the resulting EF is higher.

\subsubsection{Transformation \& Distribution}
\label{sec:results_ind_td}

\begin{figure}[ht!]
\centering
    \includegraphics[keepaspectratio, width=\linewidth]{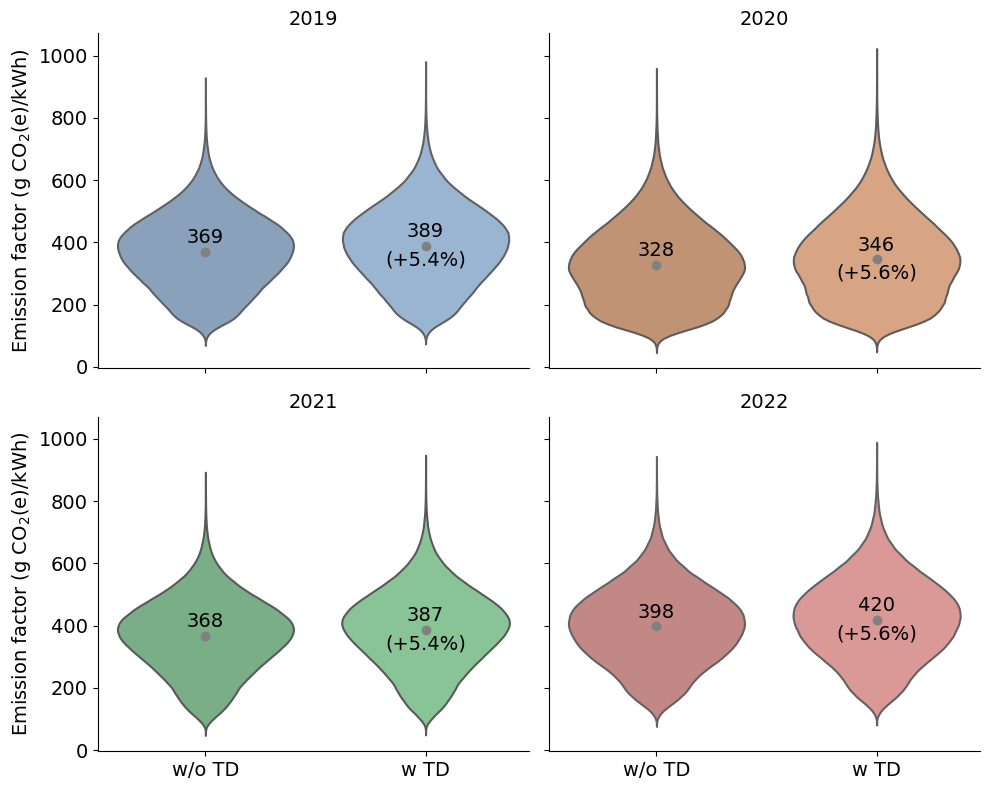}
    \caption{Transformation \& distribution (T\&D) effect, disaggregated by year-the distribution of grid EF values for EF without (\textit{w/o TD}) and with T\&D losses (\textit{w TD}), and the years 2019-2022. Labeled are the mean values for all data points, and the relative difference of the mean compared to the mean of the first EF (\textit{w/o TD}), for each year individually.}
    \label{fig:result_effect_td_multi}
\end{figure}

In Figure \ref{fig:result_effect_td_multi}, the effect of considering transformation \& distribution (T\&D) losses is plotted. The choices for considering T\&D losses are \textit{Without T\&D losses (w/o TD)} and \textit{With T\&D losses (w TD)}. The displayed values represent the mean for each choice, and the relative difference between the mean for a specific choice and the mean for the reference choice (in this case, \textit{w/o TD}).

Figure \ref{fig:result_effect_td_multi} indicates that when disaggregated by year, an EF \textit{w TD} is on average 5.4-5.6\% larger than an EF \textit{w/o TD}. Just as with auxiliary consumption (cf. section \ref{sec:results_ind_aux}) and storage cycling (cf. section \ref{sec:results_ind_cyc}), this effect meets our expectations---if more losses are included, then the resulting EF is higher.

\subsubsection{Temporal Resolution}
\label{sec:results_ind_tr}

\paragraph{Grid EF Temporal Trends.}

\begin{figure*}[ht!]
\centering

    \begin{subfigure}[b]{0.42\textwidth}
    \includegraphics[keepaspectratio, width=\textwidth]{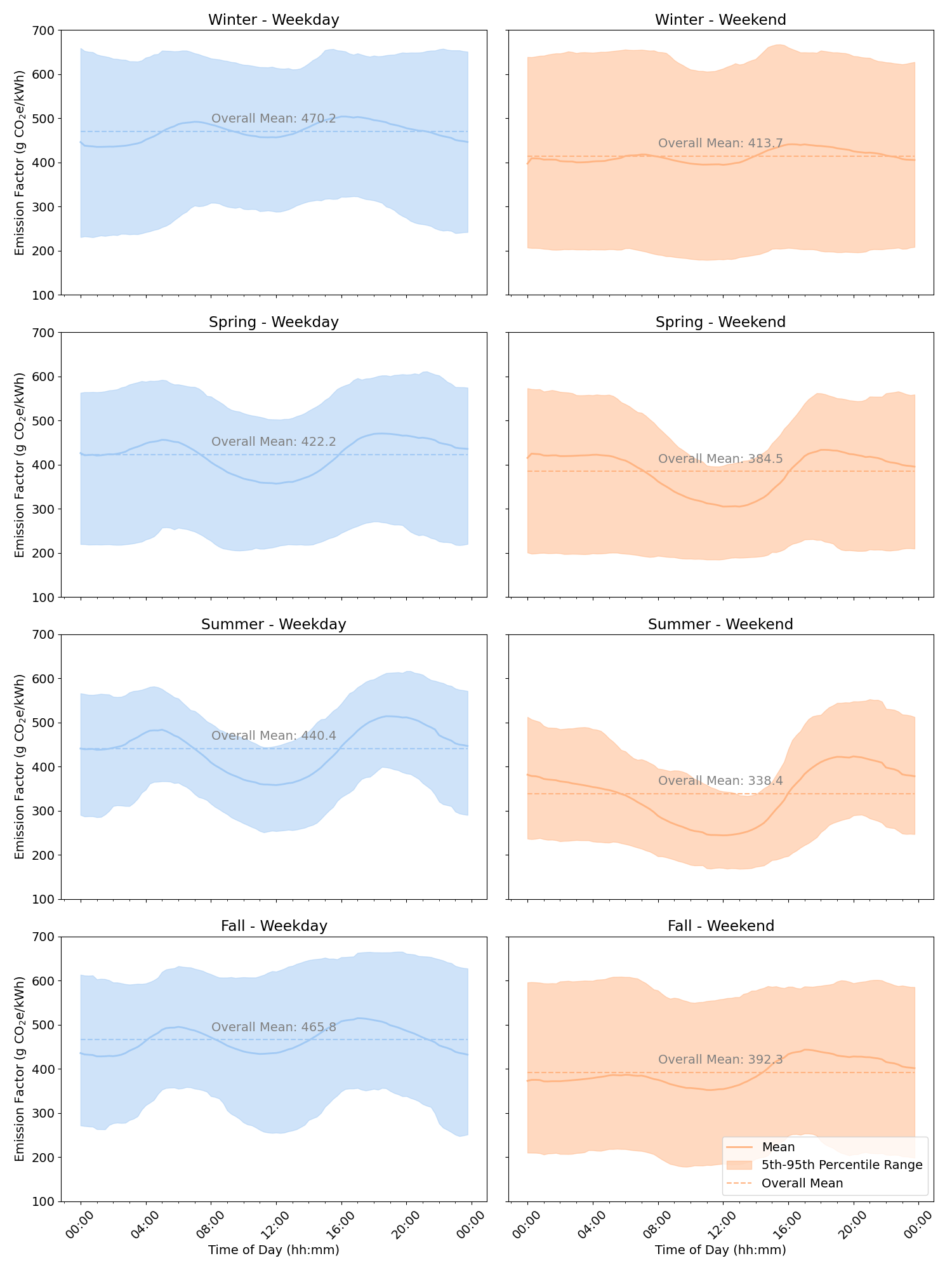}
    \caption{2019}
    \label{fig:gridef_daytype_season_2019}
    \end{subfigure}
    \hfill
    \begin{subfigure}[b]{0.42\textwidth}
    \includegraphics[keepaspectratio, width=\textwidth]{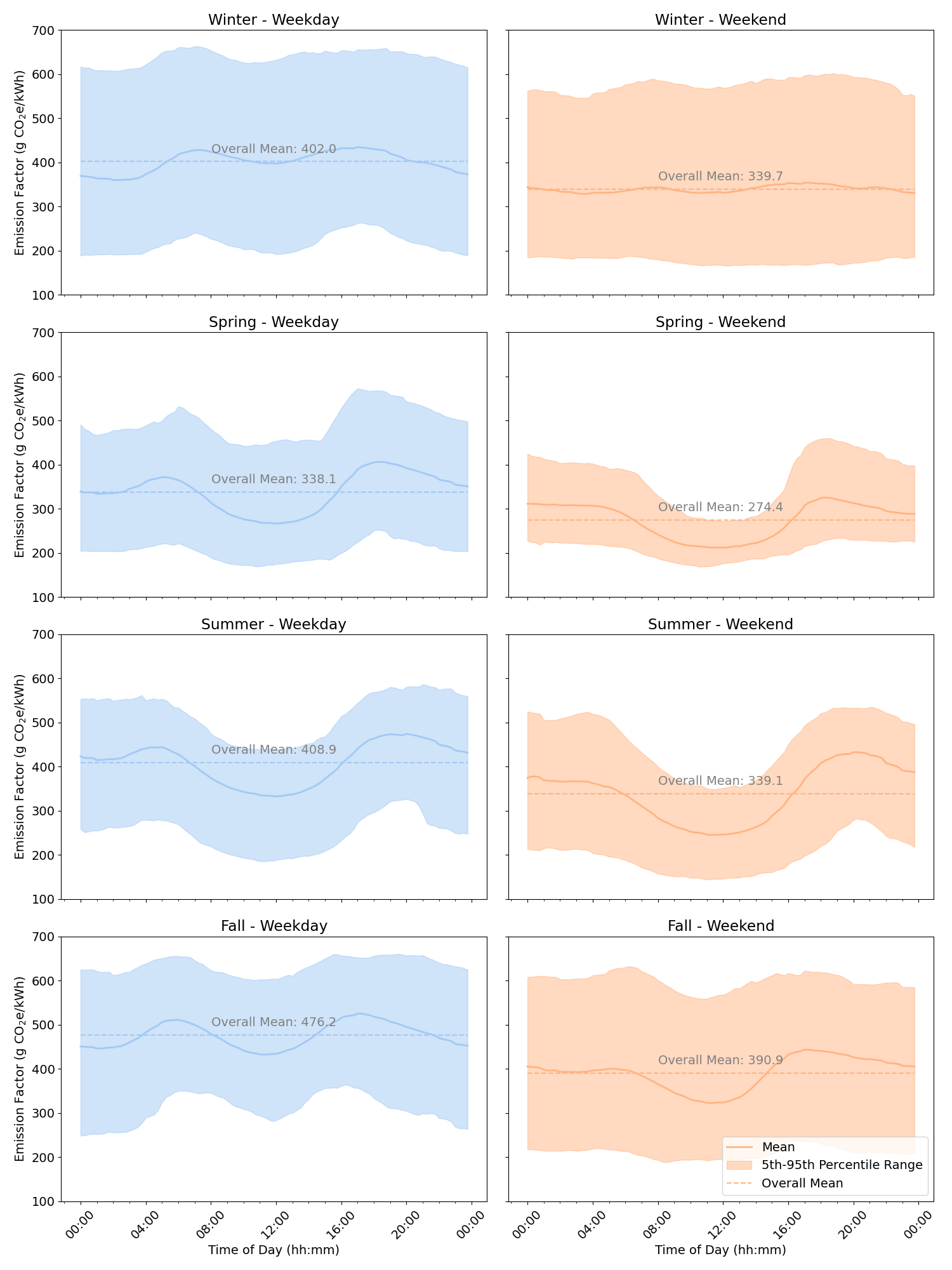}
    \caption{2020}
    \label{fig:gridef_daytype_season_2020}
    \end{subfigure}

    \begin{subfigure}[b]{0.42\textwidth}
    \includegraphics[keepaspectratio, width=\textwidth]{Figures/Results/Temporal_resolution/ef_correlation_weekday_season_tod_2021.png}
    \caption{2021}
    \label{fig:gridef_daytype_season_2021}
    \end{subfigure}
    \hfill
    \begin{subfigure}[b]{0.42\textwidth}
    \includegraphics[keepaspectratio, width=\textwidth]{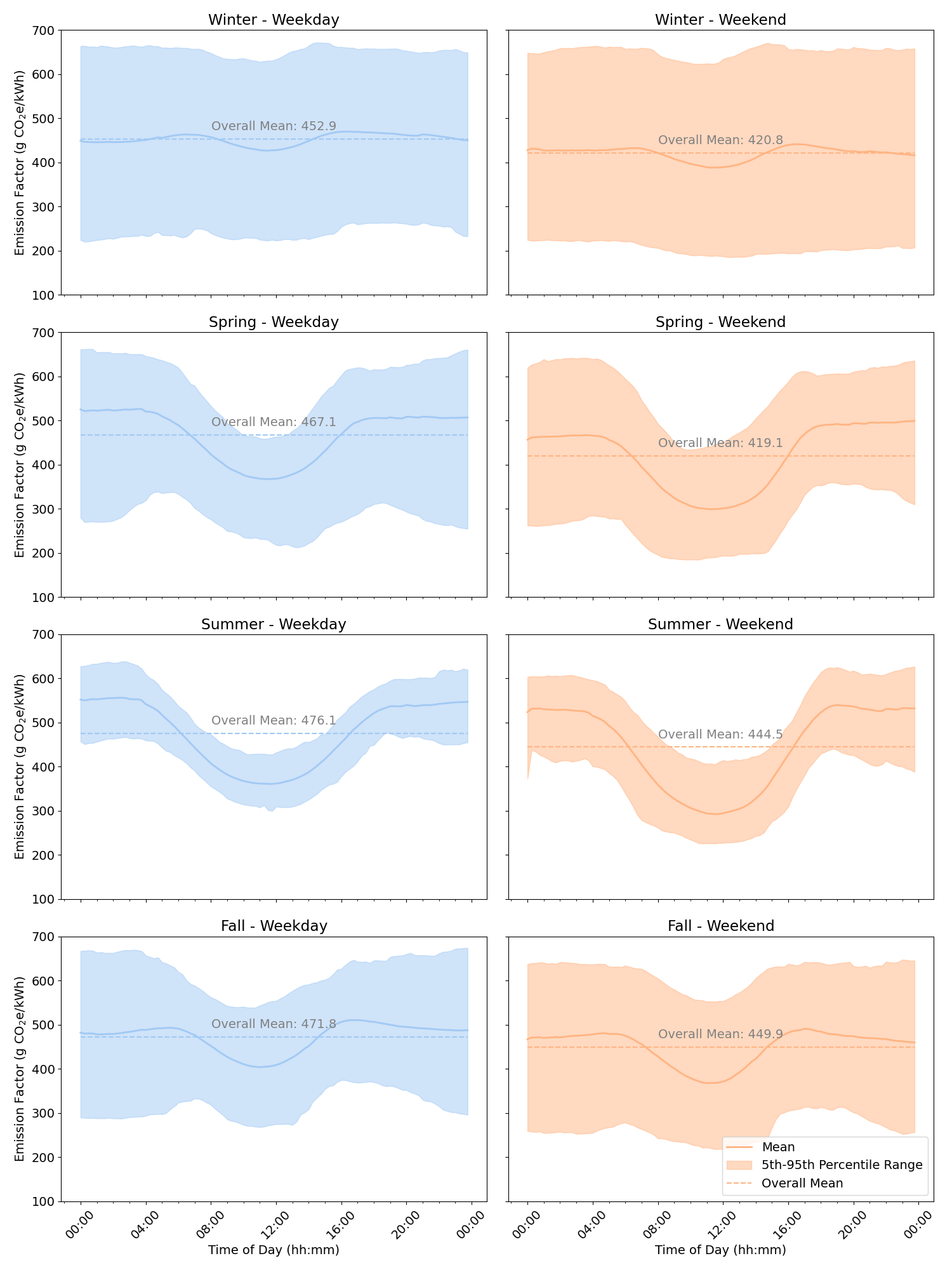}
    \caption{2022}
    \label{fig:gridef_daytype_season_2022}
    \end{subfigure}

\caption{Grid EF by day type, season, and time of day for the years 2019-2022. The solid line represents the mean value for specific time points (e.g., 12:00 h), day type (e.g., weekday), and season (e.g., Summer). The shaded area delineates the range between the 5th and 95th percentiles of the data, highlighting the distribution's variability and indicating where 90\% of the values lie for the given time and year. The dashed line represents the overall mean, i.e., the daily mean for a given day type and season.}
\label{fig:gridef_daytype_season}
\end{figure*}

Figure \ref{fig:gridef_daytype_season} depicts the grid EF by day type, season, and time of day for the years 2019-2022 (2021 is also covered in the main article).

Various possible explanations exist for the grid EF patterns described in the main article.

The relatively high grid EF in the morning and evening correlate with an overall high level of electricity production and consumption during those times. Besides hydro pumped storage plants, it is mainly fossil power plants (especially hard coal and natural gas) that increase their output to meet the increased electricity demand during those peak hours. Likewise, when demand recedes (e.g. during night), these fossil plants reduce electricity production, thus lowering the grid EF. The relatively low grid EF at noon benefits from the large share of solar PV production during those times.

Part of the reason that the `nightly dip' can barely be observed in 2022 may be twofold: 1) the Russian invasion of Ukraine during that year and the subsequent energy crisis that unfolded, and 2) the shutting down of nuclear reactors in Germany. Skyrocketing natural gas prices in Germany, which until then procured most of its gas from Russia, led to a change in the power plant merit order, so that natural gas plants ran less often, and hard coal plants ran more often. Since the latter are more emission intensive than the former, an electricity mix that relies more on hard coal at the expense of natural gas results in an increase of the grid EF. In addition, three German nuclear power plants went offline on December 31st, 2021. The lack of nuclear base load, and natural gas based generation being replaced by hard coal based generation, may explain the relatively high grid EF at night during 2022 compared to previous years.

The grid EF drop around noon that varies with the season is most likely primarily due to solar PV production. Lower demand on weekends (which include public holidays) may explain why the grid EF is lower than on weekdays. As long as there is no curtailment, generation from renewables is comparable for weekdays and weekends, meaning that less emission-intensive fossil generation is required to meet residual demand on weekends, resulting in a lower grid EF. Since the share of renewable generation is relatively low in the Winter compared to other seasons, this effect may be less pronounced then (or may even be non-existent, as for the year 2021 shown here). The narrower range between the 5th and the 95th percentile during the Summer may indicate that more solar PV generation reduces variability of the grid EF. This could be due to the remaining, non-solar generation exhibiting less variability with respect to its emission intensity. The more low-emission solar PV is included in the generation mix, the less residual generation (e.g., from natural gas or hard coal, both with notably different emission intensities) can impact the overall grid EF.

\begin{figure}[ht!]
\centering
    \includegraphics[keepaspectratio, width=\linewidth]{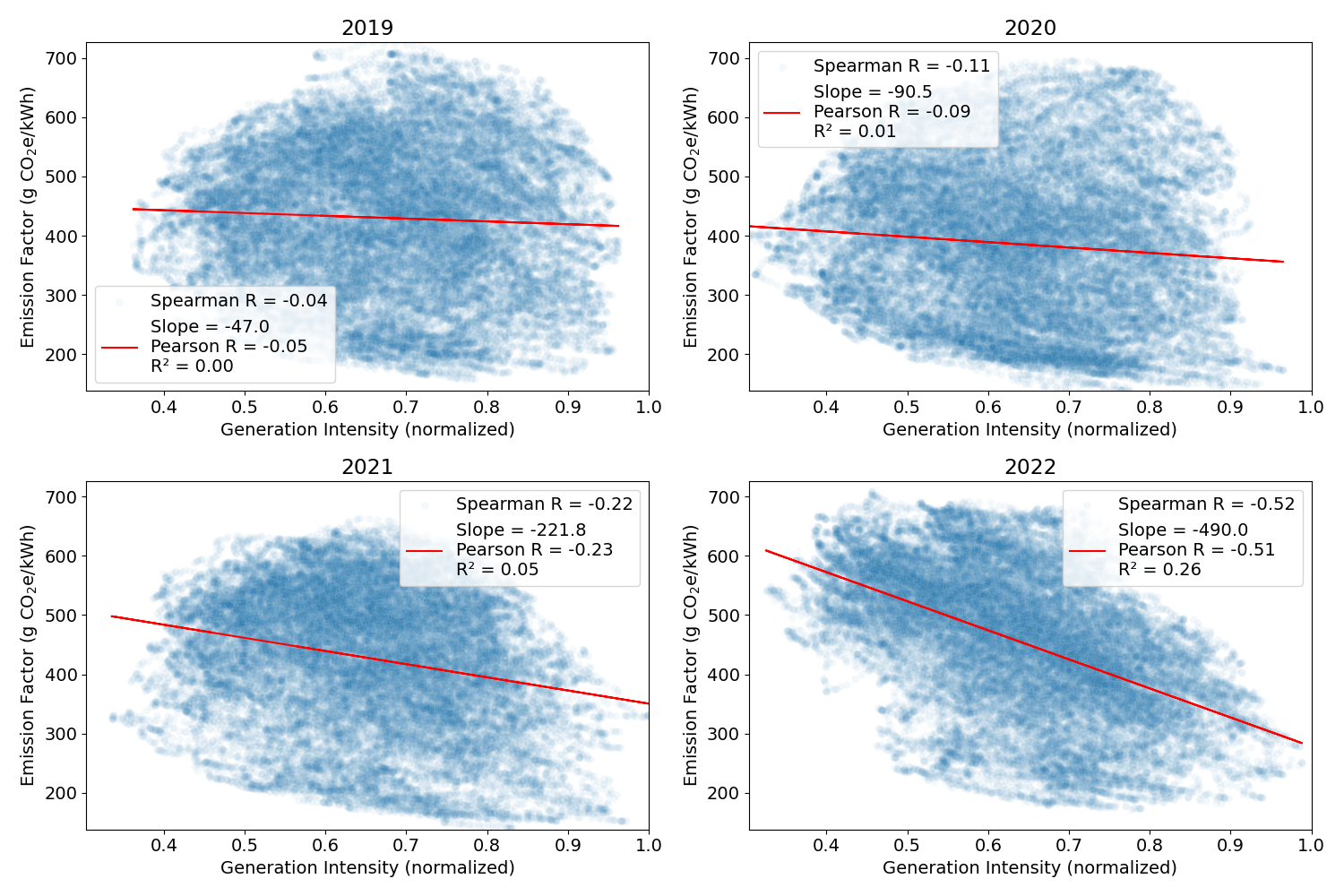}
    \caption{Correlation analysis for generation intensity and grid EF for the years 2019-2022. The Spearman R indicates the degree to which the data can be represented by a strictly monotonous function, while the Pearson R indicates the degree to which the data can be represented by a linear function (in both cases, 0 describes no representation at all, and -1 or 1 a perfect representation).}
    \label{fig:spearman_pearson_by_year}
\end{figure}

\paragraph{Correlation Analysis.}
The grid EF does not only correlate with the time of day, day type, season, and year, but also with overall electricity generation. Figure \ref{fig:spearman_pearson_by_year} depicts a correlation of generation intensity and grid EF for the years 2019-2022. Generation intensity describes overall generation (AGPT, cf. section \ref{sec:methodology_input_entsoe}) normalized to the maximum overall generation, resulting in values between 0.3 and 1. The grid EF is the recommended configuration discussed in the main article.

The plot indicates that with every year, the two variables become increasingly correlated. While the distribution is almost random in 2019, the Spearman R is -0.52 and the Pearson R -0.51 in 2022. The slope of the regression line also decreases continuously from year to year.

\begin{figure*}[hbt]
\centering

    \begin{subfigure}[b]{0.45\textwidth}
    \includegraphics[keepaspectratio, width=\textwidth]{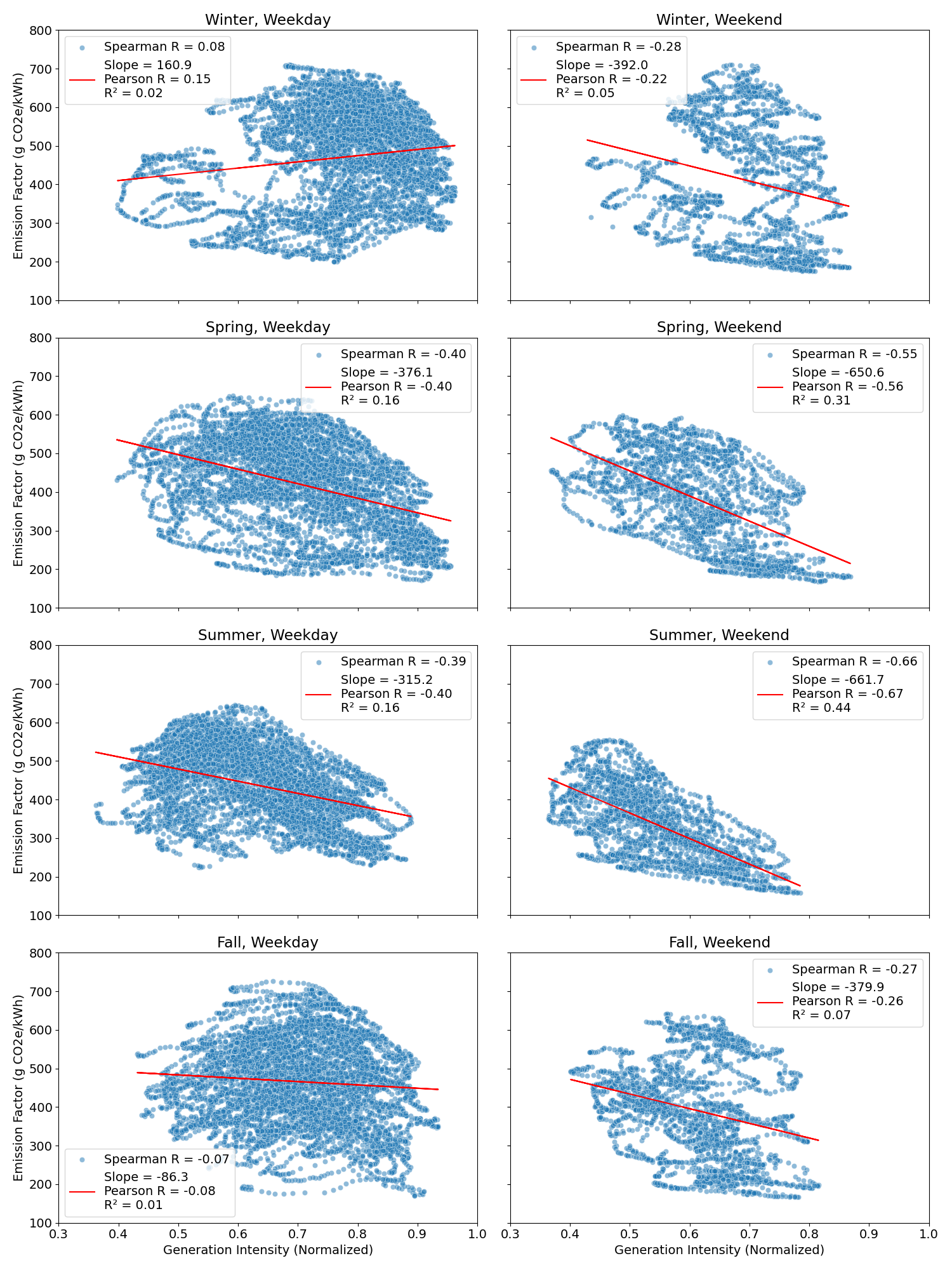}
    \caption{2019}
    \label{fig:spearman_daytype_season_2019}
    \end{subfigure}
    \begin{subfigure}[b]{0.45\textwidth}
    \includegraphics[keepaspectratio, width=\textwidth]{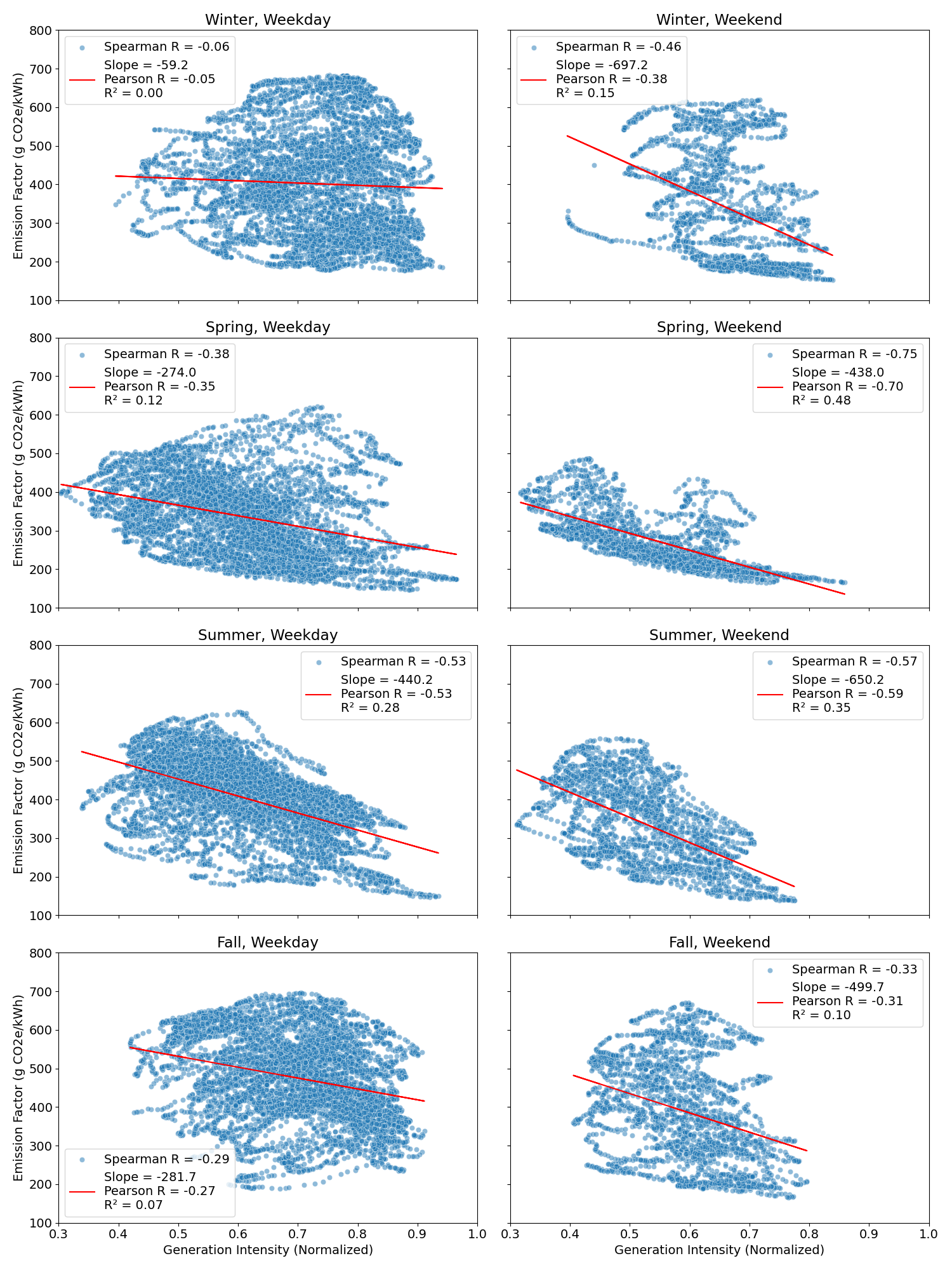}
    \caption{2020}
    \label{fig:spearman_daytype_season_2020}
    \end{subfigure}

    \begin{subfigure}[b]{0.45\textwidth}
    \includegraphics[keepaspectratio, width=\textwidth]{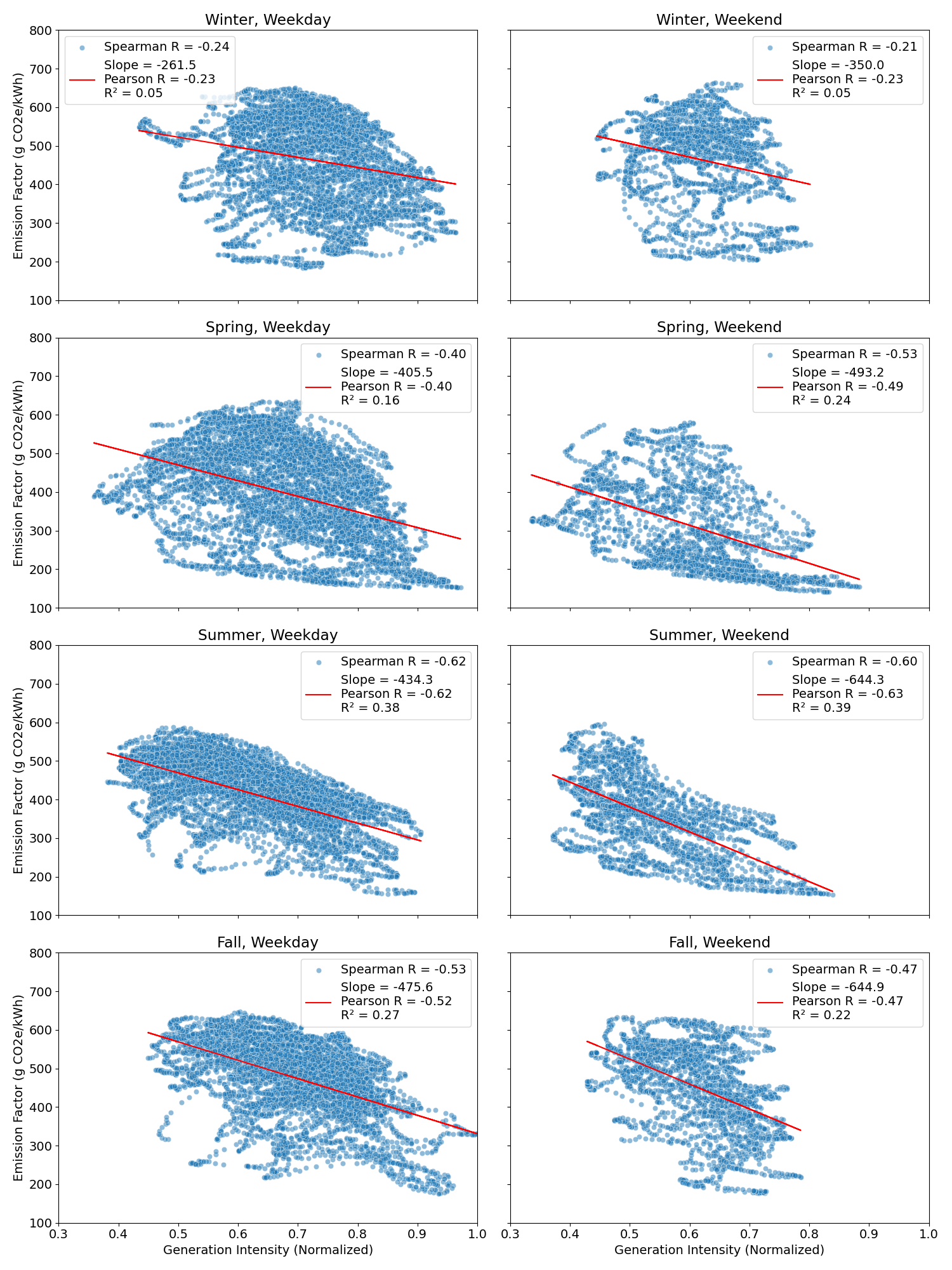}
    \caption{2021}
    \label{fig:spearman_daytype_season_2021}
    \end{subfigure}
    \begin{subfigure}[b]{0.45\textwidth}
    \includegraphics[keepaspectratio, width=\textwidth]{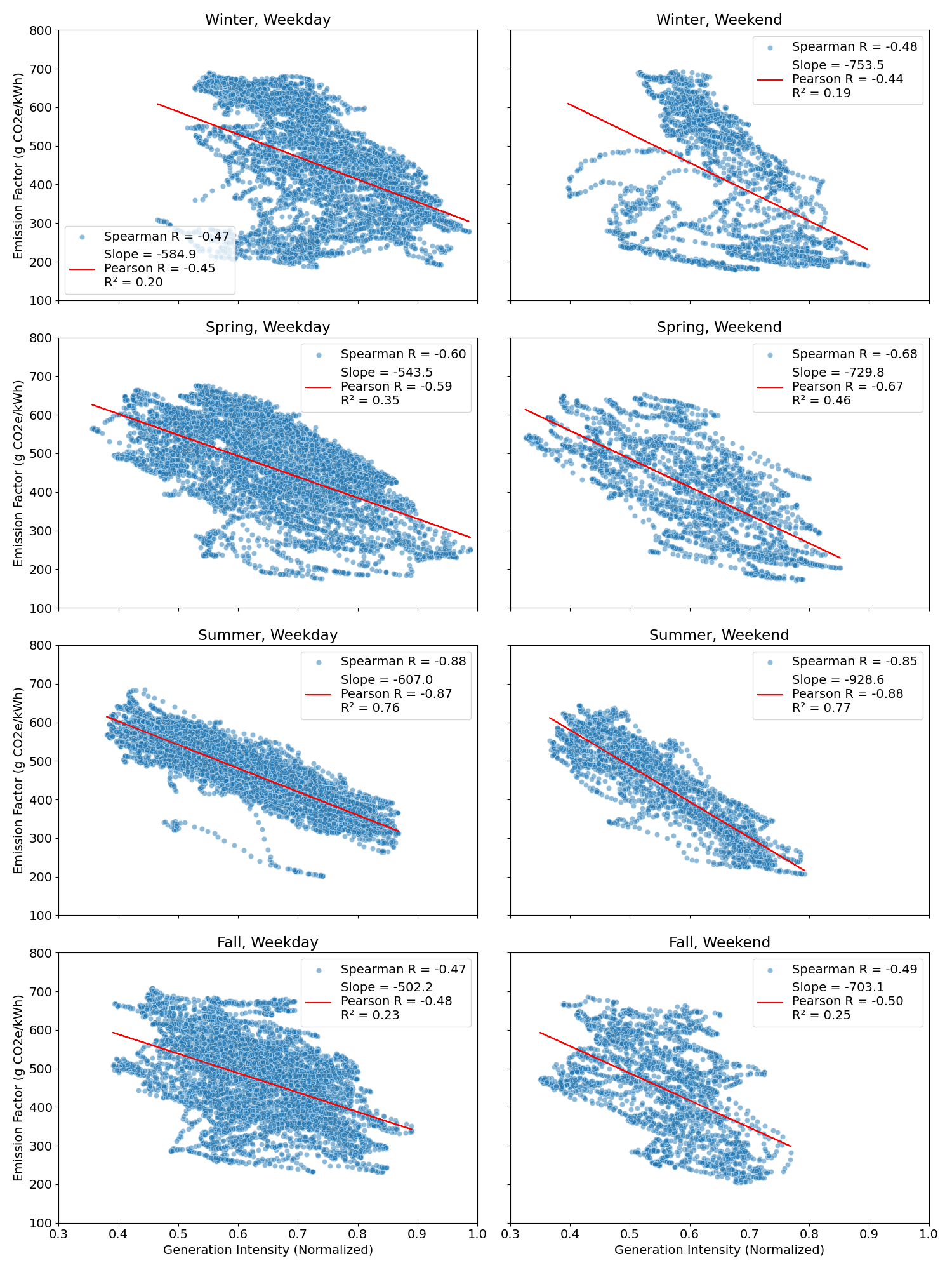}
    \caption{2022}
    \label{fig:spearman_daytype_season_2022}
    \end{subfigure}

\caption{Correlation analysis for generation intensity and grid EF by day type and season for years 2019-2022. The Spearman R indicates the degree to which the data can be represented by a strictly monotonous function, while the Pearson R indicates the degree to which the data can be represented by a linear function (in both cases, 0 describes no representation at all, and -1 or 1 a perfect representation).}
\label{fig:spearman_daytype_season}
\end{figure*}

For a more detailed analysis of temporal trends, figure \ref{fig:spearman_daytype_season} depicts the correlation of generation intensity and grid EF for the years 2019-2022 by day type (weekday, weekend) and season (Winter, Spring, Summer, Fall).

The plots illustrate that the two variables tend to correlate more strongly on weekends than on weekdays, and more strongly during Spring and Summer than during Fall and Winter. The difference between the Pearson R and Spearman R is relatively small across all time periods, indicating that a linear regression line (in red) represents the data similarly well as a strictly monotonous function (not plotted).

\begin{figure*}[ht!]
\centering

    \begin{subfigure}[b]{0.45\textwidth}
    \includegraphics[keepaspectratio, width=\textwidth]{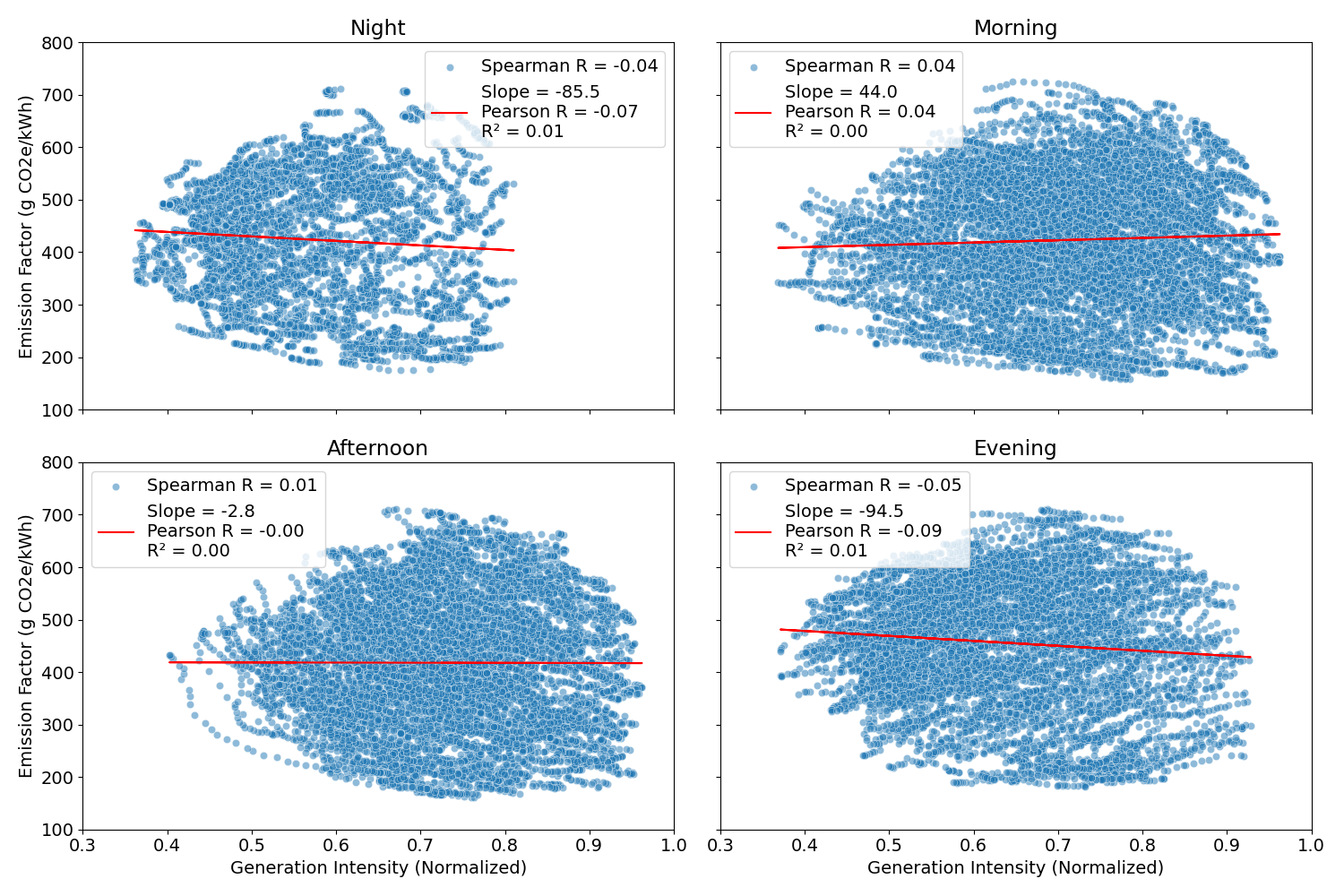}
    \caption{2019}
    \label{fig:spearman_tod_2019}
    \end{subfigure}
    \begin{subfigure}[b]{0.45\textwidth}
    \includegraphics[keepaspectratio, width=\textwidth]{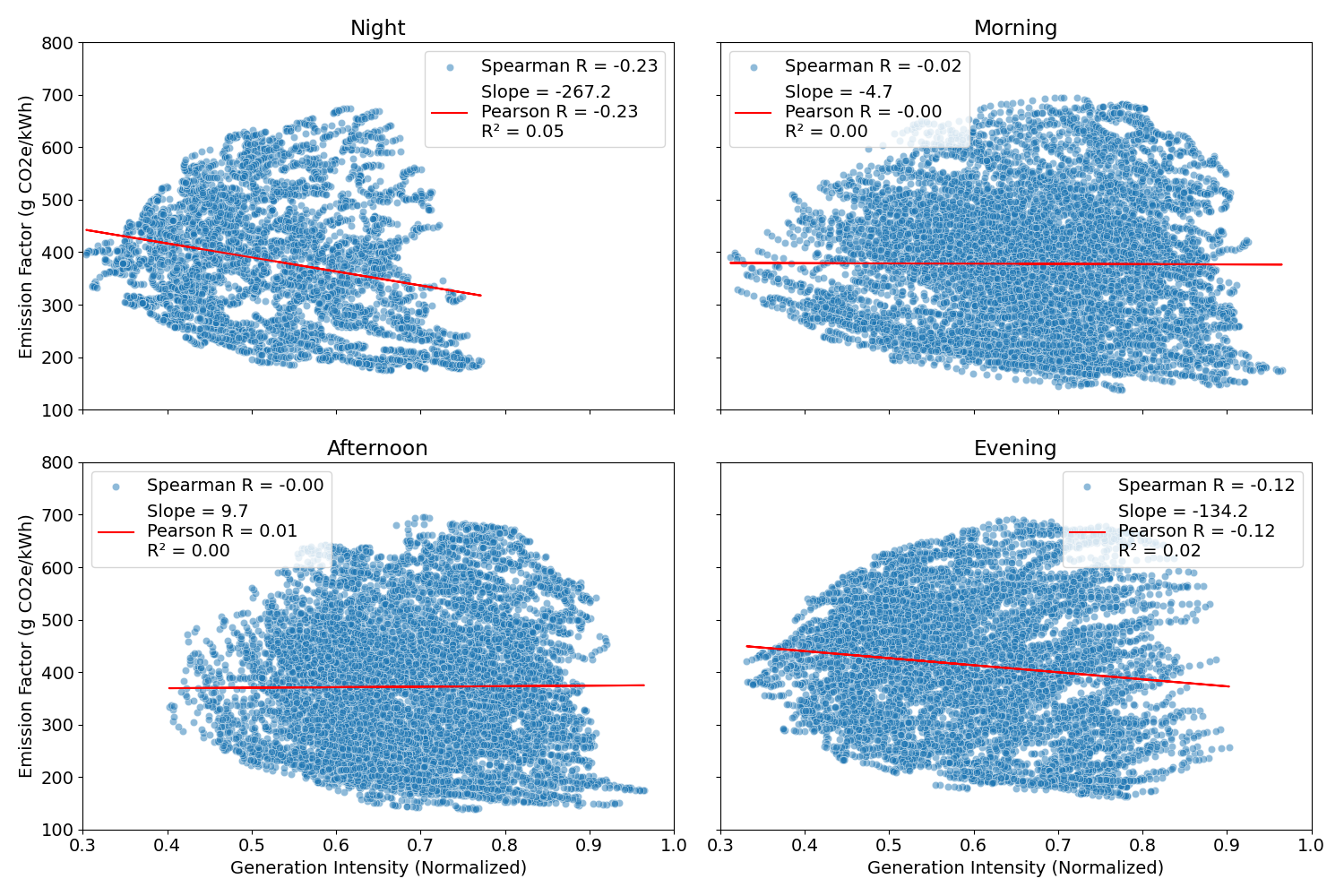}
    \caption{2020}
    \label{fig:spearman_tod_2020}
    \end{subfigure}

    \begin{subfigure}[b]{0.45\textwidth}
    \includegraphics[keepaspectratio, width=\textwidth]{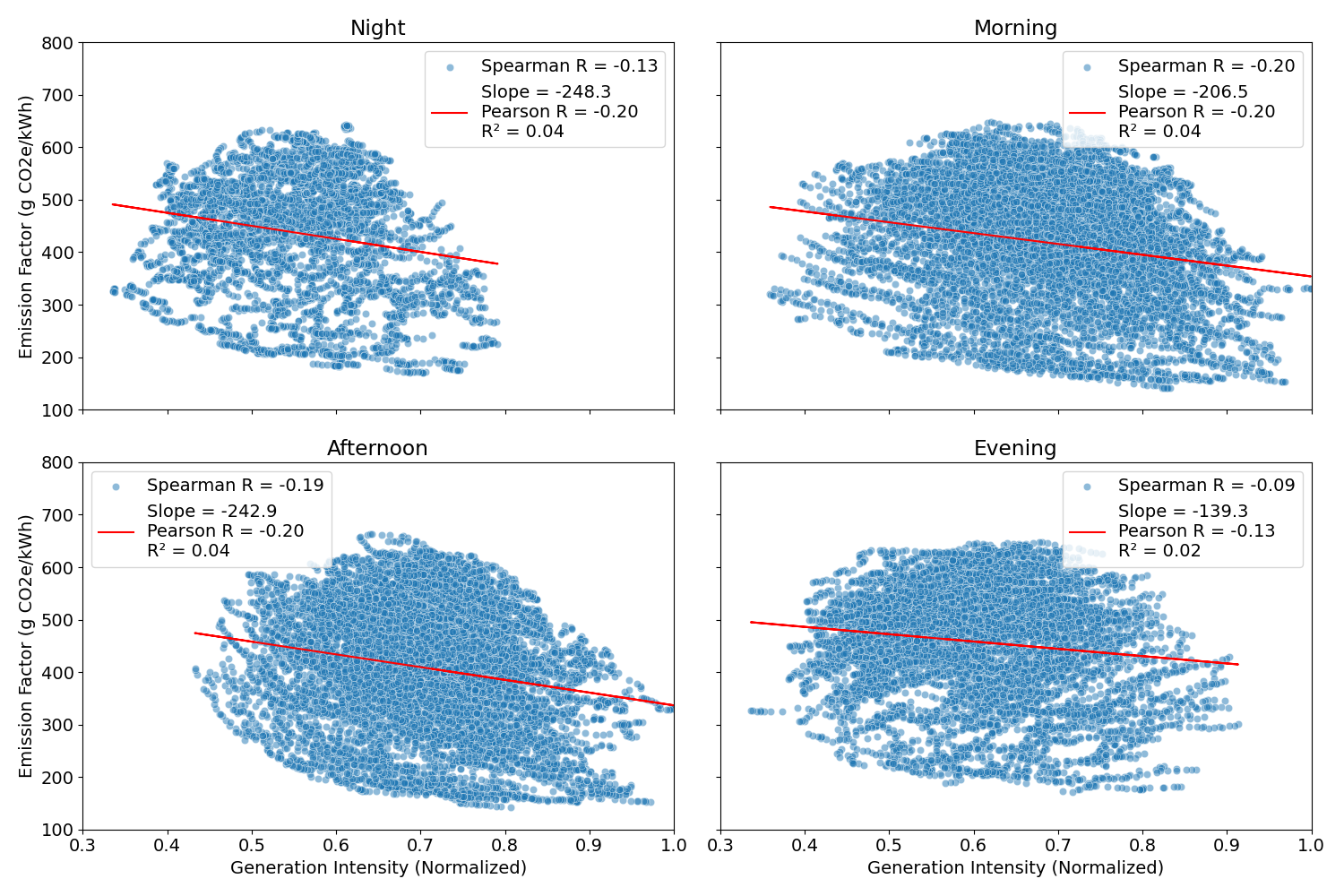}
    \caption{2021}
    \label{fig:spearman_tod_2021}
    \end{subfigure}
    \begin{subfigure}[b]{0.45\textwidth}
    \includegraphics[keepaspectratio, width=\textwidth]{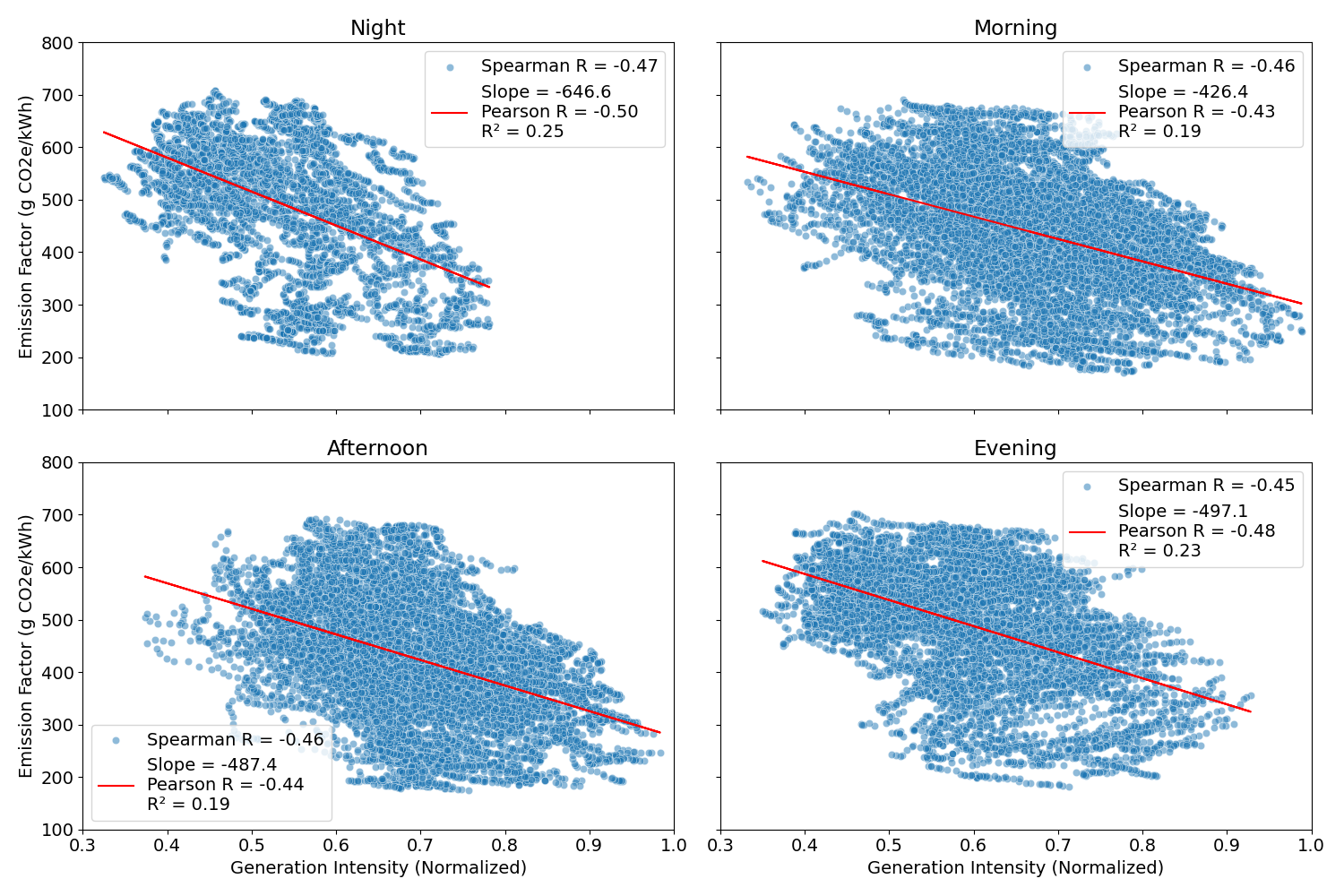}
    \caption{2022}
    \label{fig:spearman_tod_2022}
    \end{subfigure}

\caption{Correlation analysis for generation intensity and grid EF by time of day for years 2019-2022. The Spearman R indicates the degree to which the data can be represented by a strictly monotonous function, while the Pearson R indicates the degree to which the data can be represented by a linear function (in both cases, 0 describes no representation at all, and -1 or 1 a perfect representation).}
\label{fig:spearman_tod}
\end{figure*}

Finally, to illustrate differences throughout the course of a typical day, figure \ref{fig:spearman_tod} depicts the correlation of generation intensity and grid EF for the years 2019-2022 by time of day.`Night' is defined as the time from 00:00 h to 04:00 h, `Morning' from 04:00 h to 11:00 h, `Afternoon' from 11:00 h to 18:00 h, and `Night' from 18:00 h to 00:00 h.

The differentiation by time of day does not reveal any notable patterns. The correlation between generation intensity and grid EF is relatively weak overall (for both the Pearson and the Spearman Rank correlation), and strongest in the year 2022.

\section{Extended Discussion}
\label{sec:discussion_SI}

This section expands on the discussion covered in the main article.

\subsection{Extended Validation}
\label{sec:discussion_SI_validation}

\subsubsection{Extended Validation: Official Sources}
\label{sec:discussion_SI_validation_off}

Possible explanations for the relative differences between grid EF calculated in this study and those provided by official sources include differences in characterization factors used (from IPCC AR6 \cite{IPCC.2021} for this study, AR5 \cite{IPCC.2013} by the official sources). Also, all three official institutions assume zero emissions from the operational phase for renewable energy sources, while in this study, they are larger than zero. As the only institution, the UBA includes the emissions from flue gas desulfurization \cite{UBA.2023}.

An interesting effect that can be observed, and discussed in the main article discussion, is that the consumption-based EF by the UBA is larger than the comparable production-based EF, while for our own calculations, the opposite is true. This can be traced back to the way the UBA considers electricity trade: if we understand it correctly, the consumption based-EF is mere correction of the production-based EF using the net trade balance. The actual grid EF of the neighboring countries our not taken into account. Our approach, which we consider more sophisticated than the UBA approach (yet less sophisticated than the MRIO approaches discussed in Section \ref{sec:literature_results_detail}), does not only account for the net trade balance, but additionally takes into account the grid EF of the neighbors from which electricity is imported (and vice versa). Thus the opposing trends: for the UBA, electricity trade leads to a higher German grid EF, in our calculations, it leads to a lower EF.

\subsection{Extended Recommendations}
\label{sec:discussion_SI_recommendation}

\subsubsection{Extended Recommendations: Grid EF Calculation}
\label{sec:discussion_SI_recommendation_calculation}

\begin{figure*}[ht!]
\centering
    \includegraphics[keepaspectratio, width=\linewidth]{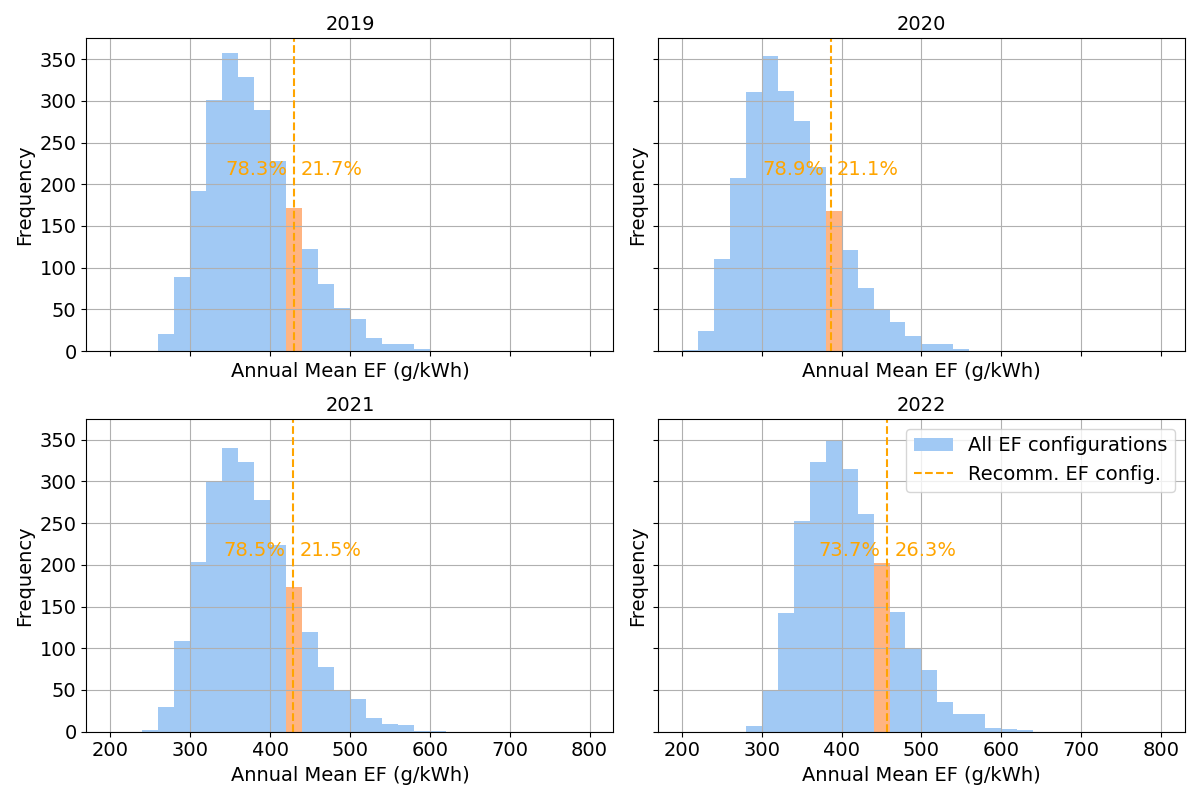}
    \caption{Annual mean of the recommended grid EF configuration (orange) compared to the annual means of all possible grid EF configurations (blue), by year. The recommended configuration is the one described in the discussion section of the main article. The percentages in the plot indicate the relative share of the 2304 configurations which result in a lower (left of the orange vertical line) and higher (right) annual mean EF than the recommended configuration.}
    \label{fig:disc_val_recom}
\end{figure*}

Referenced to all possible ways of calculating a grid EF with the methodology we propose, the recommended configuration (set of choices) is closer to the high end (see Figure \ref{fig:disc_val_recom}). About 21.1-26.3\% of the possible grid EF are larger than the recommended value, while the rest is smaller. The recommended configuration is therefore neither an extreme one nor the most typical (it's not around the median).

\subsection{Limitations}
\label{sec:discussion_lim}

Despite our best efforts to contribute in a comprehensive, complete and consistent manner to the standardization and harmonization of grid EF for Scope 2 emission accounting, some shortcomings remain. Besides some aspects already mentioned in the methodology section of the main article (e.g., omitting the technological and spatial resolution), more caveats apply to this study. We briefly list them here, distinguishing between general limitations and limitations that are specifically relevant for certain methodological aspects.

\subsubsection{General Limitations}
\label{sec:discussion_lim_gen}

Our study only covers Germany throughout the years 2019-2022. The results and conclusions may differ for other locations and time periods due to the different composition of the underlying energy system.

Also, we only assess the environmental impact with respect to GHG emissions. While this is the only aspect relevant for corporate GHG accounting and reporting, other types of impacts from electricity production and consumption (e.g., resource depletion, water use, impact on biodiversity) are of similar importance and should not be neglected.

Our calculations are based on primary-energy referenced EF for specific production types from UBA \cite{UBA.2021}. However, other approaches exist as well, such as the two approaches demonstrated by \citeauthor{Unnewehr.2022} that are based on total emission data from the European Emissions Trading System (ETS) and the UNFCCC national emissions inventory, respectively. Since the data sources are unlikely to match completely, the outcome will differ between approaches.

The mapping process described in the methodology section of the main article, which we include because of non-matching data categories, is likely to introduce some error. If the data providers were to use matching data categories, this could be avoided.

Additionally, for some production types, the primary-energy referenced EF are not provided by the data sources and had to be guessed (see Section \ref{sec:methodology_input_uba}). Furthermore, these EF are only applicable to Germany, and refer to the year 2020. For other countries and years, these EF are probably not the best representation of the operational and upstream emissions of the production types in question.

\subsubsection{Aspect-specific Limitations}
\label{sec:discussion_lim_spec}

The following limitations apply to specific methodological aspects and choices covered in this study.

\paragraph{Impact Metric.}
The GWP values used in this study are based on the CF from the IPCC AR6 \cite{IPCC.2021}, and will therefore differ from other publications that rely on EF from the IPCC  AR5 \cite{IPCC.2013}. A more detailed analysis could also take into account probabilistic instead of deterministic CF, as demonstrated by \citeauthor{Pereira.2020} \cite{Pereira.2020}.

\paragraph{System Boundaries.}
While we only distinguish between operational and life cycle (operational \& upstream) emissions, one could go further and distinguish between the feedstock and the infrastructure life cycle, as well as the up- and downstream emissions.

\paragraph{Co-generation of Heat.}
We cover six different allocation methods to account for electricity and heat co-generation in CHP units. However, there are several more that could be included, and no consensus appears to exist which one is the "correct" or most suitable one. The reference efficiencies used by the UBA (0.8 and 0.4) may be considered quite low for state-of-the-art heat and power generators. The temperature levels assumed for the allocation by exergy (T$_0$: 282 K, T: 363 K) are probably not entirely unrealistic, but should be verified and/or varied using a sensitivity analysis in future studies.

\paragraph{Auxiliary Consumption.}
There are no limitations of this study with respect to auxiliary consumption that we are aware of.

\paragraph{Auto-producers.}
Our definition of auto-producer rests on the assumption that these units do not feed electricity into the grid. We did not find information about this in our data sources. Should this assumption be wrong, then the calculations based on this assumption should be reviewed and updated.

\paragraph{Temporal Resolution.}
We only assess the temporal resolution levels of 15 minutes and one year. Additional, intermediate resolution levels may provide further relevant insights. Also, as mentioned in the main article's discussion, 15 min data may not be available in every region, especially outside of Europe and North America.

\paragraph{Electricity Trading.}
Our approach, while perhaps more sophisticated than the one used by the UBA (see main article's discussion section), is based on the SFOT approach, not on the more complex (and probably more accurate) MRIO approach. This means that our study only considers electricity exchange with direct neighbors, but ignores additional, network-wide interactions.

\paragraph{Storage Cycling.}
The method to calculate virtual emissions for the electricity flows going into and coming out of pumped hydro storage units in this study is a simplified one. It assumes a constant grid EF for these electricity flows, instead of calculating a grid EF at a higher temporal resolution. Should the inflows and outflows occur at times when the grid EF is systematically higher or lower than the annual average grid EF, this approach will introduce an error.

\paragraph{Transformation \& Distribution.}
Similarly, our calculation of T\&D losses does not temporally (or spatially) disaggregate losses in the grid. Also, the losses at different voltage levels of the grid are not analyzed separately.

\subsection{Future Work}
\label{sec:discussion_fut}

Besides the obvious opportunities for future work, i.e. addressing the limitations mentioned in Section \ref{sec:discussion_lim} (to the extent possible), we want to briefly highlight some additional promising avenues for research and applications.

To find out if and how it applies to a different context, we would welcome it if researchers were to use the methodology demonstrated in this study for calculating grid EF for other countries and time periods. Some data sources may have to be adjusted (e.g., for the primary-energy referenced EF \cite{UBA.2021}). The results could be compared to the grid EF calculated by official institutions (IEA, EEA, national institutions).

Another interesting approach would be to engage in a type of global sensitivity analysis (GSA). While in this study, we only investigated single effects in isolation, it may be worth mapping the entire solution space more systematically. Existing approaches from mathematics \cite{Sobol.2001, Saltelli.1995}, some of which have been already applied to life cycle assessment \cite{Kim.2022, Jolivet.2021, Groen.2017}, may be a good starting point. The findings may e.g., provide additional insight into how individual aspects (effects) interact with one another.
 
Moving away from research, and towards application, it may be worth developing a user interface (UI) for the methodology. This UI could e.g., be an interactive, browser based solution which requires no installations by the user. This UI would lower the barriers to using the methodology and applying it to Scope 2 emission accounting.



\balance


\bibliography{rsc} 
\bibliographystyle{rsc} 